\documentclass{article}
\usepackage{arxiv}
\usepackage[utf8]{inputenc} 
\usepackage[T1]{fontenc}    
\usepackage{hyperref}       
\usepackage{url}            
\usepackage{booktabs}       
\usepackage{amsfonts}       
\usepackage{amssymb}       
\usepackage{nicefrac}       
\usepackage{microtype}      
\usepackage{lipsum}		
\usepackage{graphicx}
\usepackage{xspace}
\usepackage{fourier}
\usepackage{mathtools}
\usepackage[numbers,compress,square]{natbib}
\usepackage{caption}
\usepackage{subcaption}
\usepackage{doi}
\usepackage[nameinlink,capitalize,noabbrev]{cleveref}
\usepackage[linesnumbered,ruled]{algorithm2e}
\usepackage[shortlabels]{enumitem}
\usepackage{wrapfig}
\usepackage{setspace}
\usepackage{multirow}
\usepackage{tabularx,booktabs}
\usepackage[table,xcdraw]{xcolor}
\usepackage[export]{adjustbox}
\usepackage{wasysym}
\usepackage{enumitem}

\SetCommentSty{mycommfont}
\Crefname{algocf}{Algorithm}{Algorithms}

\usepackage{tensorNotation}
\usepackage{finiteVolumeNotation}

\newcommand{\OF}[0]{OpenFOAM}

\crefname{equation}{Eq.}{Eqs.}

\graphicspath{
    {./figures/numerical_scheme/}
    {./figures/interface_advection/}
    {./figures/results/}
    {./figures/results/plic_surfaces_in_advection/}
    {./figures/results/plic_surfaces_in_reconstruction/}
}

\hypersetup{
    colorlinks=true,
    linkcolor=teal,
    filecolor=magenta,      
    urlcolor=cyan,
    citecolor=orange
}

\newcommand{\volume}{{\ooalign{\hfil$V$\hfil\cr\kern0.08em--\hfil\cr}}}

\title{A Geometric VOF Method for Interface Flow Simulations}

\date{} 					

\author{
    \parbox{\linewidth}{
        \centering
        Dezhi Dai\textsuperscript{1,*},
        Haomin Yuan\textsuperscript{1},
        Albert Y. Tong\textsuperscript{2},
        and Adrian Tentner\textsuperscript{1}
        \bigskip\\
        \normalfont{\textit{\textsuperscript{1} Nuclear Science and Engineering Division, Argonne National Laboratory, Lemont, IL 60439}}
        \\
        \normalfont{\textit{\textsuperscript{2} Department of Mechanical and Aerospace Engineering, University of Texas at Arlington, Arlington, TX 76019}}
        \bigskip\\
        \textsuperscript{*} Corresponding email: \texttt{daid@anl.gov}
    }
}

\renewcommand{\shorttitle}{A Geometric VOF Method for Interface Flow Simulations}

\hypersetup{
    pdftitle={A Geometric VOF Method for Interface Flow Simulations},
    pdfsubject={CFD, VOF, PLIC, SimPLIC, OpenFOAM},
    pdfauthor={Dezhi Dai, Haomin Yuan, Albert Y. Tong, Adrian Tentner},
    pdfkeywords={VOF, PLIC, SimPLIC, Simpson's Rule, Unstructured Meshes, \OF},
}

\begin{document}

\maketitle

\begin{abstract}
    A novel numerical technique designed for interface flow simulations using the Volume of Fluid (VOF) method on arbitrary unstructured meshes has been introduced. The method is called SimPLIC, which seamlessly integrates Piecewise Linear Interface Calculation (PLIC) and Simpson's rule. The main focus of the proposed method is to compute the volume of the primary phase that moves across a mesh face within a single time step. This is achieved by reconstructing the interface and assessing how the submerged face area evolves over time. Simpson's rule is employed to integrate the time evolution of this submerged face area, ensuring an accurate estimation of the volume of the transported primary phase. The method's robustness was validated by solving a spherical interface advection problem in a non-uniform three-dimensional flow across unstructured meshes with diverse cell types and dimensions. Key metrics such as volume conservation, shape retention, friction boundedness and solving efficiency were meticulously monitored and juxtaposed. Numerical outcomes underscored the precision and adequacy of the PLIC-VOF technique when complemented with Simpson's rule in advecting the interface. Furthermore, the SimPLIC method has been integrated into \OF\,v2312 as an unofficial extension and is now accessible to the community.
\end{abstract}


\section{Introduction}\label{s_introduction}

Interface flows are critical in a wide range of engineering applications, encompassing phenomena such as tank sloshing, fuel spray and atomization dynamics, jet breakups, dam breaks, interactions with waves, water entry or exit events, and phase changes. Despite their widespread occurrence, the substantial potential of numerical simulations for improving interface capturing and tracking is yet to be fully exploited. Leveraging this potential can greatly diminish risks and lower expenses in various predefined scenarios. Consequently, the development of an efficient and highly accurate numerical method for interface advection is essential for the research and analysis of multiphase flows.

The Volume of Fluid (VOF) method \cite{nichols1975methods,hirt1981volume} is a fundamental technique for capturing interfaces in multiphase flow simulations. It uses the primary phase volume fractions, symbolized as $\alpha$, to implicitly represent the interface. In a computational cell, if the cell is completely occupied by the primary phase, $\alpha$ is equal to one. If the cell is wholly filled with the secondary phase, $\alpha$ is zero. For cells containing both phases, known as mixed cells, $\alpha$ ranges between zero and one. The advection of the interface is tackled by solving the governing equation for $\alpha$, commonly referred to as the VOF equation.

In solving the VOF equation to advect the interface, both algebraic and geometric methods are employed. Algebraic approaches are generally more straightforward to implement, offering efficiency and adaptability across various mesh types \cite{roenby2016computational,maric2020unstructured}.  However, they encounter issues with numerical diffusion at interface cells, a consequence of the discontinuous nature of the fraction field $\alpha$. The common-used algebraic methods, Multidimensional Universal Limiter with Explicit Solution (MULES) \cite{deshpande2012evaluating}, High Resolution Interface Capturing (HRIC) \cite{muzaferija1999two}, and Compressive Interface Capturing Scheme for Arbitrary Meshes (CICSAM) \cite{ubbink1999method}, have been compared with the isoAdvector, a new efficient geometric VOF method for general polyhedral meshes in \cite{roenby2016computational}. The results indicated that the geometric VOF method is superior at shape preservation, volume conservation, fraction boundedness and interface sharpness. On the other hand, the geometric schemes maintain a sharp interface while preserving mass conservation, but this comes at the cost of an additional reconstruction step \cite{roenby2016computational,dai2019analytical,maric2020unstructured}. These carefully reconstructed interfaces play a crucial role in calculating the transition of the primary phase volume across the faces of the cells.

The determination of the interface location within a mixed cell relies on a predefined interface shape, such as a plane \cite{dai2019analytical} or an isosurface \cite{roenby2016computational}, along with the volume fraction $\alpha$. A variety of algorithms have been developed for interface reconstruction, applicable to both hexahedral \cite{noh2005slic,youngs1982time,ashgriz1991flair,rider1998reconstructing,pilliod2004second,zhang2008new,vignesh2013noniterative} and arbitrary polyhedral \cite{yang2006analytic,huang2012piecewise,ito2013volume} meshes. Among these, the Piecewise Linear Interface Calculation (PLIC) method \cite{yang2006analytic} stands out for its effective representation of the interface as an oriented plane, which is expressed as $\vec{n} \cdot \vec{x} + D = 0$. This plane is mathematically described by the equation $\vec{n} \cdot \vec{x} + D = 0$, where $\vec{n}$ is the unit outward normal vector, $\vec{x}$ denotes the position vector, and $D$ represents the signed distance from the origin. In the interface reconstruction step of the PLIC-VOF method, accurately determining $\vec{n}$, and either $\vec{x}$ or $D$, is crucial for precisely locating the interface plane within a mixed cell.

The existing methods for calculating $\vec{n}$ include several notable algorithms:
\begin{itemize}
    \item Young's algorithm \cite{youngs1982time}: Utilizes the fraction gradient. The accuracy of Young's algorithm and its variants in unstructured meshes largely depends on mesh types and gradient calculation methods \cite{dai2019numerical}. Common gradient calculation models for unstructured meshes include the Green-Gauss and least-square (LS) methods. However, the LS method's accuracy and convergence can significantly deteriorate in unstructured meshes with poor quality \cite{maric2020unstructured}, and the Green-Gauss method's effectiveness is sensitive to the face interpolation schemes used. Detailed formulations of these methods are discussed in \cref{sss_interface_orientation}.
    \item Mosso-Swartz algorithm \cite{swartz1989second,mosso1996recent}: Begins by computing orientations using Young's algorithm, then iteratively refines the accuracy through a least-squares minimization between the modified and estimated normal vectors in all mixed cells.
    \item Least squares Volume-of-fluid Interface Reconstruction Algorithm (LVIRA) \cite{pilliod2004second}: Determines orientation vectors by iteratively minimizing the discrete $L^{\infty}$ or $L^{1}$ errors between the true and approximated interfaces' volume fractions. LVIRA achieves second-order accuracy in reconstructing smooth stationary interfaces.
    \item Least Squares Fit (LSF) algorithm \cite{scardovelli2003interface,aulisa2007interface}: Initially uses Young's algorithm to compute orientations, then iteratively enhances accuracy by minimizing the distances between interface planes and surrounding cell centroids.
    \item Moment of Fluid (MoF) algorithm \cite{dyadechko2005moment}: Adds an extra dataset containing centroids of cell fractions, which are also advected by the flow velocity field.
    \item Conservative Level Contour Interface Reconstruction (CLCIR) \cite{lopez2008new}: Similar to the Mosso-Swartz algorithm, it averages the normal vectors from surrounding reconstructed interface polygons to evaluate the interface normals.
    \item Reconstructed Distance Function (RDF) model \cite{cummins2005estimating,scheufler2019accurate}: Reconstructs the signed distance function from the fraction field, then computes orientation vectors from the gradient of this function. It includes iterations to minimize the average difference of normal vectors across successive iterations \cite{scheufler2019accurate}. This method is also known as Coupled Level-Set and Volume of Fluid (CLSVOF).
\end{itemize}

A comprehensive comparison of the convergence orders and relative computational costs of these orientation schemes are available in Table 1 of \cite{maric2020unstructured}. The Young's algorithm and RDF model are employed in the present study.

Once $\vec{n}$ is established, the determination of either $\vec{x}$ or $D$ is unique for a specified fraction value, regardless of the chosen interface locating method. The primary distinction among different interface locating methods lies in their computational efficiency. For arbitrary polyhedral meshes, Ahn and Shashkov \cite{ahn2008geometric} introduced an iterative method for three-dimensional generalized polyhedral meshes, which is based on an algorithm for finding the intersection between a plane and polyhedral cells. L{\'o}pez and Hern{\'a}ndez \cite{lopez2008analytical} proposed an analytical approach for general grids, featuring more efficient geometric operations and the centered sequential bracketing (CSB) algorithm. Diot and Fran{\c{c}}ois \cite{diot2016interface} developed a noniterative interface reconstruction method for 3D arbitrary convex cells, offering higher overall efficiency. L{\'o}pez et al. \cite{lopez2016new} introduced an improved analytical method, the Coupled Interpolation Bracketing Analytical Volume Enforcement (CIBRAVE) algorithm, along with a new bracketing procedure, known as interpolation bracketing (IB), which demonstrated significant reductions in relative CPU time. Skarysz et al. \cite{skarysz2018iterative} presented an iterative approach for interface reconstruction in general convex cells based on tetrahedral decomposition. Dai and Tong proposed an analytical interface locating algorithm to calculate $D$ for two-dimensional polygonal meshes \cite{dai2018analytical} and later extended it to three-dimensional general polyhedral meshes \cite{dai2019analytical}. Chen and Zhang \cite{chen2019predicted} introduced an iterative algorithm using the Newton method, achieving much faster convergence compared to Ahn and Shashkov's approach \cite{ahn2008geometric}. The analytical method developed in \cite{dai2019analytical} coupled with the IB algorithm from \cite{lopez2016new} (as detailed in \cref{sss_interface_location}) is employed in the present study.

In the review paper of the unstructured un-split geometrical VOF methods \cite{maric2020unstructured}, the interface advection methods for unstructured meshes have been categorized into two families: the fully geometric and geometric/algebraic methods. When calculating the transition of the primary phase volume across a specific face, the fully geometric methods first generate a polyhedral volume by tracking the face points backward in the time step and then truncate this polyhedral volume by the PLIC interface. The resulting submerged volume is treated as the primary phase volume across this face. The drawbacks of the fully geometric methods are complexity implementation, low computational efficiency and issues of overlapped polyhedral volumes of backward-tracked faces. A comprehensive review of the fully geometric methods is available in Section 4.1 of \cite{maric2020unstructured}.

In the review paper on unstructured un-split geometrical VOF methods \cite{maric2020unstructured}, interface advection methods for unstructured meshes are classified into two main categories: fully geometric and geometric/algebraic methods. The fully geometric methods commence by creating a polyhedral volume by tracking the face points backwards within the time step, and then slicing this volume with the PLIC interface. The resulting submerged portion is considered the primary phase volume traversing this face. However, these methods are hampered by complex implementation, reduced computational efficiency, and issues with overlapping polyhedral volumes created by the backward tracking of face points. A detailed review of fully geometric methods can be found in Section 4.1 of \cite{maric2020unstructured}.

The geometric/algebraic methods employ an algebraic approach to calculate the primary phase volume across a specific face. Roenby et al. \cite{roenby2016computational} introduced the isoAdvector scheme, which estimates the primary phase volume traversing a particular face by tracking the intersection line between the interface and face within the time step. The isoAdvector method reduced three-dimensional geometric operations and improves computational efficiency. When combined with the RDF scheme, the isoAdvector method has been shown to match the performance of contemporary un-split geometrical VOF methods while significantly reducing computational costs. Xie and Xiao \cite{xie2017toward} developed the Tangent of Hyperbola Interface Capturing with Quadratic surface representation and Gaussian Quadrature (THINC/QQ) algorithm. This approach relies on Gaussian quadrature to approximate the integration of a cell-wise multi-dimensional hyperbolic tangent reconstruction function, which is then utilized to reconstruct the interface from the volume fraction value of the target cell. Additionally, THINC/QQ incorporates a third-order accurate explicit Runge-Kutta scheme for temporal integration.

In the present study, a new method called SimPLIC is introduced. The computation of the primary phase volume crossing a specific face is done using the PLIC method with Simpson’s rule \citep{burden2001numerical}, which is also known as the 3-point closed Newton-Cotes formula. This approach was initially introduced by the authors in \cite{dai2022adaptive} and has since been re-implemented in \OF\,v2312, specifically targeting efficiency improvements. This method is distinguished by its zero truncation error in the time integration of the submerged area of a face. To ascertain the accuracy and efficacy of the SimPLIC method, a classic benchmark problem involving three-dimensional interface advection was simulated across four different unstructured mesh types. Additionally, the newly proposed SimPLIC method is compared with the officially released PLIC-VOF methods in \OF\,v2312 in terms of accuracy and efficiency. A brief overview of the SimPLIC method is presented next followed by the benchmark problem descriptions and testing results.

\section{Numerical Formulations}\label{s_numerical_formulations}


The VOF method delineates the interface by addressing the transportation equation for the volume fraction. This equation is expressed as:
\begin{equation}
    \ddt{\alpha} + \nabla \cdot \left( \alpha \vec{U} \right) = 0,
    \label{eq_vof}%
\end{equation}
where $\vec{U}$ denotes the velocity field.

The Finite Volume Method (FVM) in \OF\,subdivides the computational domain into a finite number of contiguous control volumes or cells. Integrating \cref{eq_vof} over an arbitrary polyhedral cell $P$ with $N_f$ faces, from $t$ to $t + \deltaT$, yields the following relationship \cite{dai2022adaptive}:
\begin{equation}
    \timeInt{\left( \int_{V_P} \ddt{\alpha} \, dV \right)} + \timeInt{\left( \int_{V_P} \nabla \cdot \left( \alpha \vec{U} \right) \, dV \right)} = 0.
    \label{eq_vof_integration}%
\end{equation}

Under the presumption that the fraction value $\alpha$ remains consistent throughout the cell and the velocity $\vec{U}$ is invariant within the time interval, the updated value of $\alpha$ in the subsequent time step can be determined as:
\begin{equation}
    \alpha_{P}^{n+1} = \frac{V_{P}^n}{V_{P}^{n+1}} \alpha_{P}^{n} - \frac{1}{V_{P}^{n+1}} \sum_{f=1}^{N_f}{\left( \frac{\q_f^n}{|\Sf|^{n+1}} \timeInt{A_f(t)} \right)},
    \label{eq_updated_alpha_general}%
\end{equation}
where $V_{P}$ represents the volume of cell $P$, $\q_f$ signifies the the volumetric flux across face $f$, $\Sf$ is the face outward area vector, and $A_f(t)$ designates the submerged area. In the context of a stationary mesh, \cref{eq_updated_alpha_general} can be distilled to:
\begin{equation}
    \alpha_{P}^{n+1} = \alpha_{P}^{n} - \frac{1}{V_{P}} \sum_{f=1}^{N_f} {\left(\frac{\q_f^n}{\left| \Sf \right|} \timeInt{A_f(t)}\right)}.
    \label{eq_updated_alpha}%
\end{equation}

Consequently, the crucial task in solving the VOF equation lies in computing the time integration term $\timeInt{A_f(t)}$ as indicated in \cref{eq_updated_alpha}. Yet, the submerged area $A_f(t)$ is not smoothly defined, making its expression challenging to delineate \cite{roenby2016computational,dai2022adaptive}. In the present study, the $A_f(t)$ and $\timeInt{A_f(t)}$ are determined utilizing the PLIC method and Simpson's rule, respectively.

\subsection{Interface reconstruction}\label{ss_interface_reconstruction}

\begin{figure}[htbp]
    \centering
    \includegraphics[width=0.35\textwidth]{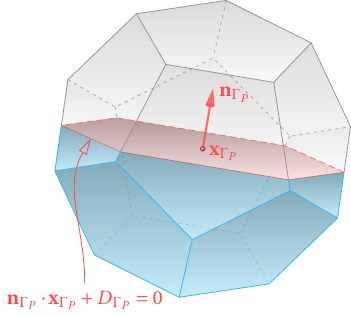}
    \caption{Approximated interface plane $\Gamma_P$.}
    \label{fig_poly_cell_with_interface}
\end{figure}

The submerged area $A_f(t)$ is derived using an approximated interface plane situated within its upwind neighboring cell \cite{roenby2016computational}, from which the face receives fluid during the time step. Numerically, cell $P$ is considered as mixed if
\begin{equation}
    \epsilon < \alpha_P < 1-\epsilon,
    \label{eq_mixed_cell_condition}%
\end{equation}
where $\epsilon$ is a numerical tolerance and a common value of $10^{-8}$ is used. As illustrated in \cref{fig_poly_cell_with_interface}, the interface in a mixed cell $P$ is approximated as a plane $\Gamma_P$ which is expressed by:
\begin{equation}
    \vec{n}_{\Gamma_P} \cdot \vec{x}_{\Gamma_P} + D_{\Gamma_P} = 0.
    \label{eq_plic_plane}%
\end{equation}
Determining the orientation vector $\vec{n}_{\Gamma_P}$ (\cref{sss_interface_orientation}) and the signed distance $D_{\Gamma_P}$ (\cref{sss_interface_location}) is essential to pinpoint the location of this interface plane.

\subsubsection{Interface orientation}\label{sss_interface_orientation}

\paragraph{Fraction Gradient} The fraction-gradient-based methods calculate $\vec{n}_{\Gamma_P}$ as:
\begin{equation}
    \vec{n}_{\Gamma_P} = - \frac{\left( \fvcgrad{\alpha} \right)_{P}}{\left| \left( \fvcgrad{\alpha} \right)_{P} \right|},
    \label{eq_vec_n_grad}%
\end{equation}
where the negative sign indicates that the vector $\vec{n}_{\Gamma_P}$ is oriented from the liquid towards the gas, as illustrated in \cref{fig_poly_cell_with_interface}. Within the \OF\,framework, the methods of evaluating the fraction gradient $\left( \fvcgrad{\alpha} \right)_P$ are outlined below:
\begin{itemize}
\item \textbf{CAG}. The fraction gradient $\left( \fvcgrad{\alpha} \right)_P$ is typically derived using the Green-Gauss method, which is expressed as:
\begin{equation}
    \left( \fvcgrad{\alpha} \right)_P = \frac{1}{V_P} \sum_{f=1}^{N_f}{\alpha_f \Sf},
    \label{eq_grad_alpha_green}%
\end{equation}
where $\alpha_f$ represents the fraction value of face $f$. One straightforward approach to determine $\alpha_f$ is the cell-averaged scheme, in which $\alpha_f$ is evaluated as a weighted average of the two adjacent cells $P$ and $N$:
\begin{equation}
    \alpha_f = w_f \alpha_P + (1 - w_f) \alpha_N.
    \label{eq_alpha_f_cag}
\end{equation}

Here, the face weighting factor $w_f$ is ascertained based on mesh geometry:
\begin{equation}
    w_f = \frac{\left| \vec{C}_{f} - \x_{P} \right|}{\left| \x_{N} - \x_{P} \right|},
    \label{eq_face_weighting_factor}
\end{equation}
where $\vec{C}_{f}$ is the weighted center of face $f$.

\item \textbf{NAG}. The cell-averaged Gauss (CAG) gradient method can exhibit skewness errors in the presence of suboptimal mesh quality. A superior approach leverages nodal values. The fraction field at cell centres is interpolated to mesh points using an inverse distance weighting method:
\begin{equation}
    \alpha_p = \frac{\sum_{c=1}^{N_{c,p}}{\frac{\alpha_{c}}{\left| \vec{x}_{p} - \x_{c} \right|}}}{\sum_{c=1}^{N_{c,p}}{\frac{1}{\left| \vec{x}_{p} - \x_{c} \right|}}},
    \label{eq_alpha_p_idw}
\end{equation}
where $N_{c,p}$ represents the number of cells $c$ adjacent the mesh point $p$. Subsequently, the face value $\alpha_f$ is determined by averaging its associated points:
\begin{equation}
    \alpha_f = \frac{1}{N_{p,f}} \sum_{p=1}^{N_{p,f}} \alpha_{p},
    \label{eq_alpha_f_nag}
\end{equation}
with $N_{p,f}$ being the number of points $p$ associated with face $f$. While the node-averaged Gauss (NAG) gradient method yields more accurate results, it demands additional computational effort, especially when determining node values.

\item \textbf{LS}. The accuracy of $\left( \fvcgrad{\alpha} \right)_P$ can be significantly enhanced using the LS gradient method, employing a cell-point-cell stencil \cite{scheufler2021twophaseflow}. This LS gradient approach is particularly advantageous for situations involving highly irregular meshes and for determining the orientation and curvature of interfaces in VOF simulations \cite{of221_2013_release_notes}. This method incorporates neighboring cells across both cell faces and points. A comprehensive overview of the LS gradient method using a cell-point-cell stencil can be found in Section 3.3.1 of \cite{maric2014openfoam}.
\end{itemize}

\paragraph{RDF} The idea of RDF model was first presented in \cite{cummins2005estimating} and then extended to general polyhedral meshes \cite{scheufler2019accurate}. In this model, the signed distance function, denoted as $\psi$, is numerically reconstructed from the fraction field $\alpha$ and the associated reconstructed interfaces. Subsequently, the orientation vector $\vec{n}_{\Gamma_P}$ is derived as:
\begin{equation}
    \vec{n}_{\Gamma_P} = \frac{\left( \fvcgrad{\psi} \right)_{P}}{\left| \left( \fvcgrad{\psi} \right)_{P} \right|}.
    \label{eq_vec_n_RDF}%
\end{equation}

In \OF\,v2312, the RDF model\footnote{\href{https://www.openfoam.com/documentation/guides/latest/api/classFoam_1_1reconstruction_1_1plicRDF.html}{https://www.openfoam.com/documentation/guides/latest/api/classFoam\_1\_1reconstruction\_1\_1plicRDF.html}} calculates the orientation vectors as follows \cite{scheufler2019accurate}:
\begin{enumerate}
    \small
    \item Initialize the orientation vectors $\vec{n}_{\Gamma}$ in all mixed cells using the LS method.
    \item Reconstruct the interface planes in all mixed cells.
    \item Calculate the distance function $\psi$ in all mixed cells and their point neighbours.
    \item Compute the gradient of distance function $\fvcgrad{\psi}$ utilizing the LS scheme.
    \item Update the orientation vectors $\vec{n}_{\Gamma}$ using \cref{eq_vec_n_RDF}.
    \item Update the RDF residuals $res$ and $res_{curv}$ (their definitions could be found in Eqs. (17) and (20) of \cite{scheufler2019accurate}).
    \item Repeat Steps 2-6 for a maximum of $I_{max}^{RDF}$ iterations or until either $res$ or $res_{curv}$ falls below the predefined tolerances.
\end{enumerate}

The default values of $I_{max}^{RDF}$, $res$ and $res_{curv}$ are $5$, $10^{-6}$ and $0.1$, respectively. A detailed explanation of the RDF method is available in \cite{scheufler2019accurate,scheufler2021twophaseflow}.




\subsubsection{Interface location}\label{sss_interface_location}

\begin{figure}[htbp]
    \centering
    \includegraphics[width=\textwidth]{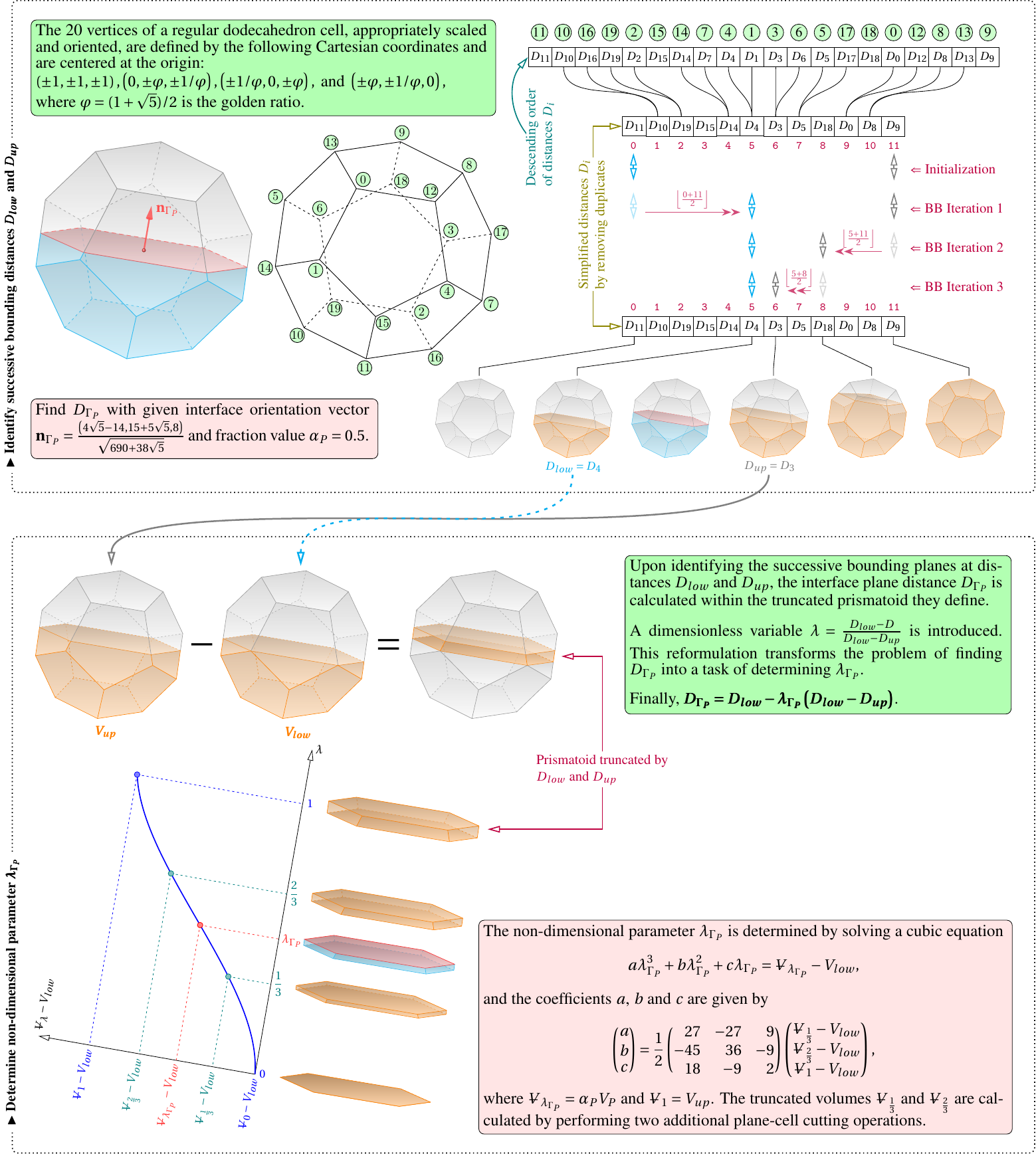}
    \caption{An illustration of locating the interface plane in a regular dodecahedron cell, with given orientation vector $\vec{n}_{\Gamma_P}$ and fraction value $\alpha_{P}$.}
    \label{fig_interface_location_illustration}
\end{figure}

The analytical distance-finding algorithm introduced in \cite{dai2019analytical} is employed in the present study. A general convex polyhedral cell with the presence of an interface is shown in \cref{fig_interface_location_illustration}. The cell consists of $N_f$ faces and $N_p$ vertices. For illustrative purposes and without loss of generality, a regular dodecahedron comprising 20 vertices is used to explicate the methodology.

The preliminary step involves determining the prismatoid formed by truncating the cell with two bounding planes. These planes pass through the cell vertices and are associated with the distances $D_{low}$ and $D_{up}$. The vertex indices are first sorted in descending order of distance $D_i\,(i=0, 1, \cdots, N_p-1)$ and stored in a label array to ensure a unique solution to the distance-finding problem. As illustrated in \cref{fig_interface_location_illustration}, the vertex indices are organized following the direction of $\vec{n}_{\Gamma_P}$. Subsequently, the distance list is filtered by eliminating duplicate values, resulting in $N_D$ distinct elements.

The next step is to identify the two bounding distances $D_{low}$ and $D_{up}$ using the Binary Bracketing (BB) procedure \cite{lopez2008analytical}. Initiate the process with the two ending elements of the uniquely sorted descending distance array, marked by the indices $k_{min}=0$ and $k_{max}=N_D$. Compute the central index $k_c$ as $\left\lfloor\left({k_{min}+{k_{max}}}\right) / 2\right\rfloor$, which is the greatest integer less than or equal to $\left(\left({k_{min}+{k_{max}}}\right) / 2\right)$, and use the corresponding plane as the cutting tool to derive the next truncated volume $\volume_{k_c}$ \cite{lopez2016new}. If $\volume_{k_c} < \alpha_{P} V_{P}$, update $k_{min}$ to $k_c$; otherwise, adjust $k_{max}$ to $k_c$. Persist with these steps until $k_{max}-k_{min}$ equals 1. The value of $k_{max}-k_{min}$ is halved after each iteration, converging to 1 at a rate of $O\left(\log_2{N_D}\right)$ \cite{lopez2016new}. As depicted in \cref{fig_interface_location_illustration}, the BB procedure reaches convergence after only three iterations.

In \cref{fig_interface_location_illustration}, it is shown how the identification of the two bounding distances $D_{low}$ and $D_{up}$ leads to the establishment of a prismatoid between these two planes. To facilitate the calculation, a dimensionless variable $\lambda$ has been introduced in \cite{dai2019analytical}. It is defined as $\lambda = \left(D_{low} - D\right) / \left(D_{low} - D_{up}\right)$, where $D$ falls within the range of $D_{low}$ to $D_{up}$, allowing $\lambda$ to vary from zero to one. This reformulation transforms the problem of finding $D_{\Gamma_P}$ into the task of determining $\lambda_{\Gamma_P}$. Consequently, the interface distance $D_{\Gamma_P}$ can be calculated as \cite{dai2019analytical}:
\begin{equation}
    D_{\Gamma_P} = D_{low} - \lambda_{\Gamma_P} \left(D_{low} - D_{up}\right).
    \label{eq_D_i}
\end{equation}

In accordance with the method outlined in \cite{dai2019analytical}, the non-dimensional parameter $\lambda_{\Gamma_P}$ is obtained by solving a cubic equation
\begin{equation}
    a \lambda_{\Gamma_P}^3 + b \lambda_{\Gamma_P}^2 + c \lambda_{\Gamma_P} = \volume_{\lambda_{\Gamma_P}}-V_{low},
\end{equation}
and the coefficients $a$, $b$ and $c$ are determined as follows:
\begin{equation}
\begin{pmatrix*}
a \\
b \\
c
\end{pmatrix*}
= \frac{1}{2}
\begin{pmatrix*}[r]
27 & -27 & 9 \\
-45 & 36 & -9 \\
18 & -9 & 2
\end{pmatrix*}
\begin{pmatrix*}
\volume_{\frac{1}{3}}-V_{low} \\
\volume_{\frac{2}{3}}-V_{low} \\
\volume_{1}-V_{low}
\end{pmatrix*},
\end{equation}
where $\volume_{\lambda_{\Gamma_P}} = \alpha_{P} V_{P}$ and $\volume_{1} = V_{up}$. To calculate the truncated volumes $\volume_{\frac{1}{3}}$ and $\volume_{\frac{2}{3}}$, two supplementary plane-cell cutting operations are required, as illustrated in \cref{fig_interface_location_illustration}.

\subsection{Interface advection}\label{ss_interface_advection}

\begin{figure}[htbp]
    \centering
    \includegraphics[scale=1]{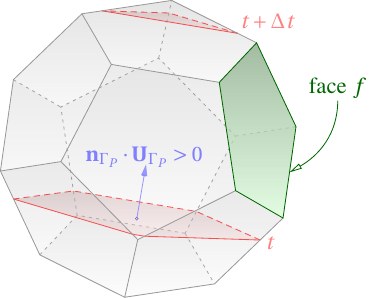}
    \caption{A mixed cell $P$ that is filling up.}
    \label{fig_interface_moving_and_face_f}
\end{figure}

For a clear and general demonstration of the interface advection, a mixed cell $P$ that is filling up is used to illustrate the methodology. As shown in \cref{fig_interface_moving_and_face_f}, the face $f$ is empty at time $t$ and then becomes fully submerged by time $t + \Delta t$. The acceleration of the moving interface is disregarded, implying that its velocity $\vec{U}_{\Gamma_P}$ remains constant throughout the time step. The interface velocity $\vec{U}_{\Gamma_P}$ is determined by interpolating velocities from the center and vertices of cell $P$, following these steps:
\begin{itemize}
    \item Decomposing each face into triangles.
    \item Constructing tetrahedra with these triangles and the center of the cell.
    \item Iterating through all tetrahedra to identify the one containing the interface's center.
    \item Employing inverse distance weighting for linear interpolation.
\end{itemize}

\begin{figure}[htbp]
    \centering
    \begin{subfigure}[b]{0.45\textwidth}
        \centering
        \includegraphics[scale=1]{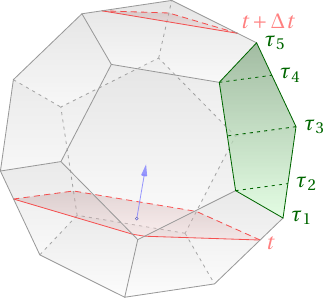}
        \caption{Calculations of sub-times $\tau_j \, \left(j=1,\cdots,N_{p,f}\right)$.}
        \label{fig_sub_vertex_times}
    \end{subfigure}
    \begin{subfigure}[b]{0.45\textwidth}
        \centering
        \includegraphics[scale=1]{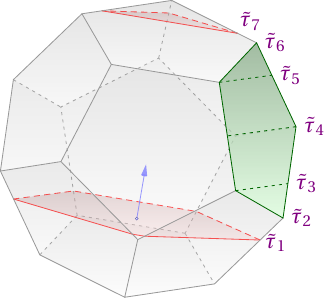}
        \caption{Determinations of key instants $\tilde{\tau}_k \, \left(k=1,\cdots,N_{\tilde{\tau}}\right)$.}
        \label{fig_sub_times}
    \end{subfigure}
    \caption{Illustrations of the key time instants $\tilde{\tau}_k$.}
    \label{fig_sub_time_intervals}
\end{figure}

Adopting the approach detailed in \cite{roenby2016computational}, the intersection line between the interface and face $f$ sweeps across the face within the sub-time interval $\left[\tau_j, \tau_{j+1}\right]$. Here, $\tau_j \, \left(j=1,\cdots,N_{p,f}\right)$ represents the moment when the interface plane intersects the $j$-th vertex of face $f$, sorted in the direction of the interface's movement (refer to \cref{fig_sub_vertex_times}). This is approximated by:
\begin{equation}
    \tau_j \approx t + \frac{\vec{n}_{\Gamma_P} \cdot \vec{x}_j + D_{\Gamma_P}}{\vec{n}_{\Gamma_P} \cdot \vec{U}_{\Gamma_P}},
    \label{eq_tau_j}%
\end{equation}
where $\vec{x}_j$ denotes the coordinates of the $j$-th vertex in the sorted order. Subsequently, the time integration of the submerged area $A_f(t)$ is evaluated as
\begin{equation}
    \int_{t}^{t+\Delta t}{A_f(t)\,dt} = \sum_{k=1}^{N_{\tilde{\tau}}-1}{\left(\int_{\tilde{\tau}_k}^{\tilde{\tau}_{k+1}}{A_f(t)\,dt}\right)},
    \label{eq_t_inte_sub_area_sum}%
\end{equation}
where $N_{\tilde{\tau}}$ signifies the count of key time instants (including the sorted times $\tau_j$, $t$ and $t + \Delta t$) between $t$ and $t + \Delta t$. As illustrated in \cref{fig_sub_times}, the key instants are $\tilde{\tau}_1 = t$, $\tilde{\tau}_7 = t + \Delta t$ and for $k=2, \cdots, 6$, $\tilde{\tau}_k = \tau_j \, \left(j=1, \cdots, 5\right)$.

Given that face $f$ is empty in the interval $\left[\tilde{\tau}_1, \tilde{\tau}_2\right]$, then $\int_{\tilde{\tau}_1}^{\tilde{\tau}_{2}}{A_f(t)\,dt} = 0$. Similarity, since face $f$ is fully submerged in $\left[\tilde{\tau}_6, \tilde{\tau}_7\right]$, it follows that $\int_{\tilde{\tau}_6}^{\tilde{\tau}_{7}}{A_f(t)\,dt} = \left| \vec{S}_f \right| \left(\tilde{\tau}_{7} - \tilde{\tau}_{6}\right)$. For other key intervals where the interface intersects face $f$, the time integration $\int_{\tilde{\tau}_k}^{\tilde{\tau}_{k+1}}{A_f(t)\,dt}$ could be calculated by deriving the exact expression of $A_f(t)$ as in Eq. (3.7) of \cite{roenby2016computational}. However, this approach increases coding complexity. Thus, the integration is computed using the Simpson's rule in the present study, namely:
\begin{equation}
    \int_{\tilde{\tau}_k}^{\tilde{\tau}_{k+1}}{A_f(t)\,dt} = \frac{\Delta \tilde{\tau}_k}{3} \left(A_f\left(\tilde{\tau}_k\right) + 4A_f\left(\tilde{\tau}_m\right) + A_f\left(\tilde{\tau}_{k+1}\right)\right) - \frac{\left(\Delta \tilde{\tau}_k\right)^5}{90} A_f^{(4)}\left(\xi_k\right),
    \label{eq_t_inte_sub_area_simpson_with_error}%
\end{equation}
where $\Delta \tilde{\tau}_k = \frac{\tilde{\tau}_{k+1}-\tilde{\tau}_{k}}{2}$, $\tilde{\tau}_m = \frac{\tilde{\tau}_k+\tilde{\tau}_{k+1}}{2}$ and $\xi_k \in \left(\tilde{\tau}_k, \tilde{\tau}_{k+1}\right)$. The error term in \cref{eq_t_inte_sub_area_simpson_with_error} involves the fourth derivative of $A_f(t)$, which is a quadratic function in the sub-interval $\left[\tilde{\tau}_k, \tilde{\tau}_{k+1}\right]$ \cite{roenby2016computational,dai2022adaptive}. Consequently, Simpson's rule yields the exact value of the time integration $\int_{\tilde{\tau}_k}^{\tilde{\tau}_{k+1}}{A_f(t)\,dt}$, expressed as:
\begin{equation}
    \int_{\tilde{\tau}_k}^{\tilde{\tau}_{k+1}}{A_f(t)\,dt} = \frac{\Delta \tilde{\tau}_k}{3} \left(A_f\left(\tilde{\tau}_k\right) + 4A_f\left(\tilde{\tau}_m\right) + A_f\left(\tilde{\tau}_{k+1}\right)\right).
    \label{eq_t_inte_sub_area_simpson}%
\end{equation}
This formulation allows for the evaluation of the three submerged areas using the same function, thereby simplifying the implementation.


\subsection{Bounding}\label{ss_bounding}

The updated volume fraction for cell $P$ in the new time step is determined using by \cref{eq_updated_alpha}. However, the resultant fraction value $\alpha_{P}^{n+1}$ might surpass its physical bounds of $0 \leq \alpha_{P}^{n+1} \leq 1$. Therefore, implementing a bounding procedure is essential to ensure strict adherence to these bounds.

A straightforward approach is to correct unphysical values by clipping any undershoots ($\alpha_{P}^{n+1} < 0$) and overshoots to ($\alpha_{P}^{n+1} > 1$) $\alpha_{P}^{n+1} = 0$ and $\alpha_{P}^{n+1} = 1$, respectively. This method is effective in cases with minimal unboundedness, such as simulations with very small time steps. However, this clipping process can inadvertently add or remove liquid in unbounded cells, thereby disrupting strict mass/volume conservation.

The study by \cite{roenby2016computational} introduced a mass-conservative bounding scheme that neither adds nor removes liquid from the domain, a method also adopted in the present study. This approach effectively redistributes any surplus or shortage of liquid in unbounded cells to maintain mass conservation.

In cases where cell $P$ is overfilled with liquid, indicated by $\alpha_{P}^{n+1} > 1$, the excess liquid is distributed to the downwind neighboring cells. Suppose cell $P$ has a liquid surplus $V^+_P$ and there are $N_d$ downwind neighbors, each with a corresponding face volumetric fluxes $\phi_f$ (for $f = 1, \cdots, N_d$). If cell $P$ is filled with liquid at time $t^*$ (where $t < t^* < t + \Delta t$), the liquid volume transported across face $f$ is $\Delta V_f^*$. The surplus volume $V^+_P$ transported across the $j$-th downwind face is then apportioned as:
\begin{equation}
    \Delta V^+_j = \min{\left(\phi_j \Delta t - \Delta V_f^*, V^+_P \frac{\phi_j}{\sum_{f=1}^{N_d}{\phi_f}}\right)}.
\end{equation}
This redistribution process continues until all of the excess liquid $V^+_P$ is appropriately allocated to the downwind neighbors.

Conversely, when $\alpha_{P}^{n+1} < 0$, cell $P$ is overfilled with gas, which equates to $\beta_{P}^{n+1} \equiv 1-\alpha_{P}^{n+1} > 1$. In this scenario, the same bounding method used for $\alpha_{P}^{n+1} > 1$ is applied to redistribute the excess gas.

Ultimately, to guarantee adherence to the strict physical boundaries of $\alpha$, a fraction clipping step is employed after redistributing any surplus or shortage of liquid.

\subsection{Warped face treatment}\label{ss_warped_faces}

\OF employs a mesh composed of arbitrary polyhedral cells, each bounded by polygonal faces with no restrictions on the number of faces per cell, the number of edges per face, or their alignment. Notably, the faces can be warped, further enhancing the adaptability of the mesh. This highly flexible mesh architecture offers considerable versatility in generating and manipulating meshes, proving particularly beneficial for domains with complex geometries. However, the geometric-dependent nature of the PLIC-VOF method makes it highly sensitive to the quality of face flatness in the mesh. Therefore, adequately addressing warped faces is crucial in the implementation of the PLIC-VOF method.

\begin{figure}[htbp]
    \centering
    \includegraphics[scale=1]{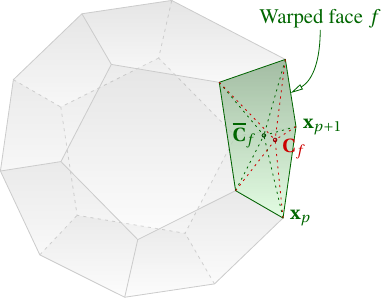}
    \caption{An illustration of the warped face $f$.}
    \label{fig_warped_face}
\end{figure}

As illustrated in \cref{fig_warped_face}, the vertices of face $f$ are not coplanar, with their average denoted as $\overline{\vec{C}}_f \equiv \sum_{p=1}^{N_{p,f}} \vec{x}_p$. In \OF, the face center $\vec{C}_f$ and area vector $\vec{S}_f$ are calculated as follows:
\begin{subequations}
    \begin{align}
        \vec{C}_f &= \frac{\sum_{p=1}^{N_{p,f}} \left|\overline{\vec{S}}^{\bigtriangleup}_p\right| \overline{\vec{C}}^{\bigtriangleup}_p}{\sum_{p=1}^{N_{p,f}} \left|\overline{\vec{S}}^{\bigtriangleup}_p\right|}, \\
        \vec{S}_f &= \sum_{p=1}^{N_{p,f}} \overline{\vec{S}}^{\bigtriangleup}_p,
    \end{align}
\end{subequations}
where $\overline{\vec{S}}^{\bigtriangleup}_p$ and $\overline{\vec{C}}^{\bigtriangleup}_p$ are the area vector and center of the triangle formed by the average center $\overline{\vec{C}}_f$, and the $p$-th and $(p+1)$-th vertices, given by:
\begin{subequations}
    \begin{align}
        \overline{\vec{S}}^{\bigtriangleup}_p &= \frac{1}{2} \left(\left(\vec{x}_{p+1}-\vec{x}_{p}\right) \times \left(\overline{\vec{C}}_f-\vec{x}_{p}\right)\right), \\
        \overline{\vec{C}}^{\bigtriangleup}_p &= \frac{1}{3} \left(\vec{x}_{p+1}+\vec{x}_{p}+\overline{\vec{C}}_f\right).
    \end{align}
\end{subequations}

The flatness of face $f$ is quantified as:
\begin{equation}
    \zeta_f = \frac{\left|\sum_{p=1}^{N_{p,f}} \overline{\vec{S}}^{\bigtriangleup}_p\right|}{\sum_{p=1}^{N_{p,f}} \left|\vec{S}^{\bigtriangleup}_p\right|},
\end{equation}
where the area vector $\vec{S}^{\bigtriangleup}_p$ is calculated using $\vec{C}_f$:
\begin{equation}
    \vec{S}^{\bigtriangleup}_p = \frac{1}{2} \left(\left(\vec{x}_{p+1}-\vec{x}_{p}\right) \times \left(\vec{C}_f-\vec{x}_{p}\right)\right).
\end{equation}

For flat faces, $\zeta_f = 1$, while for warped faces, $\zeta_f < 1$. In the case of faces with a flatness measure $\zeta_f < 1$, a triangular decomposition using $\vec{C}_f$ is implemented in both interface reconstruction and advection steps.

\section{Interface Advection in a Non-uniform Flow}\label{s_interface_advection}


A benchmark three-dimensional interface advection problem is considered, as described in various studies \cite{roenby2016computational,leveque1996high,ubbink1999method,shin2011local,liovic20063d,enright2005fast}. This test employs a predefined non-uniform velocity field and is crucial for evaluating the proposed advection method, especially its competence in managing highly distorted interfaces.

\begin{figure}[htbp]
    \centering
    \includegraphics[width=0.45\textwidth]{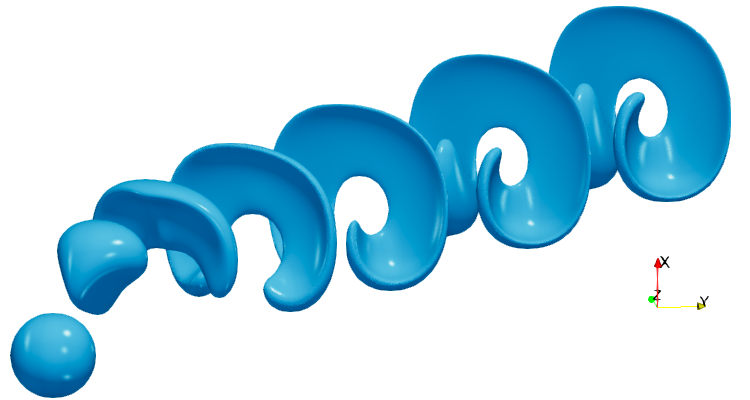}
    \caption{An illustration of the advected interface profile evolutions from $t=0\,s$ to $t=1.5\,s$ ($\Delta t = 0.25\,s$).}
    \label{fig_illustration_interface_evolutions}
\end{figure}

Initially, a spherical interface with a radius of $R=0.15\,m$ and center at $\vec{x}_0=(0.35\,m,0.35\,m,0.35\,m)$ is positioned inside a unit cube with its center at $(0.5\,m,0.5\,m,0.5\,m)$. The velocity field governing the advection is given by:
\begin{equation}
    \vec{U}(\vec{x}(t), t) = \frac{d \vec{x}(t)}{d t} = \cos{\left(\frac{2 \pi t}{T}\right)}
    \begin{pmatrix}
    {2 \sin^2{(\pi x)}  \sin{(2 \pi y)} \sin{(2 \pi z)}} \\
    {-\sin{(2 \pi x)} \sin^2{(\pi y)} \sin{(2 \pi z)}} \\
    {-\sin{(2 \pi x)} \sin{(2 \pi y)} \sin^2{(\pi z)}}
    \end{pmatrix},
    \label{eq_U_field}%
\end{equation}
where $\vec{x}(t) = (x(t), y(t), z(t))$ represents the position vector and $T = 6\,s$ is the period. The interface experiences its most pronounced deformation at $t = 1.5\,s$, a point at which the velocity field begins to reverse direction. By $t = 3\,s$, the distorted interface reverts to its original shape and location. The evolution of the interface profiles for this benchmark problem is depicted in \cref{fig_illustration_interface_evolutions}.

\subsection{Fraction field initialization}\label{ss_alpha_initialization}

The fraction field $\alpha$ is initialized using a geometric method. To represent the spherical interface shape, a 7$^{th}$-order icosphere mesh is employed (refer to \cref{fig_icosphere_ref_shape_back_view_t_0}). By performing a Boolean operation of intersection with the mixed cells using this icosphere mesh, the fraction value is derived by normalizing the ascertained intersection volume with respect to the total cell volume.

\subsection{Solution quantification}\label{ss_sol_quantification}

To quantify the numerical solutions derived from the PLIC-VOF method combined with Simpson's rule, the following error metrics are defined:

\begin{itemize}
    \item \textbf{Shape preservation}.

    \begin{itemize}
        \item \textbf{Symmetric difference error}. The symmetric difference error $E_{sd}$ has traditionally been used solely to assess the accuracy of interface reconstruction \cite{dyadechko2005moment,maric2013vofoam}. This metric leverages the symmetric difference between a surface mesh (\textit{which represents the precise interface}) and the reconstructed interfaces, providing insight into the fidelity of shape preservation. The dynamic symmetric difference error $E_{sd}(t)$ is defined as:
        \begin{equation}
            E_{sd}(t) \equiv \frac{\sum_{i \in \mathbb{M}}{\left(S_i(t) \Delta R^c_i(t)\right)} + \sum_{i \notin \mathbb{M}}{V_i \left|\alpha_i(t)-\alpha_i^{ext}(t)\right|}}{\sum_{i=1}^{N_c}{\alpha_i^{ext}(t) V_i}},
            \label{eq_E_sd_definition}
        \end{equation}
        where
        \begin{itemize}
            \item $\mathbb{M}$ denotes the set of all mixed cells.
            \item $S_i(t)$ represents the shape of the tool surface mesh $S(t)$ (\textit{which represents the precise interface at the time of $t$}) clipped by the $i$-th cell.
            \item $R^c_i(t)$ is the reconstructed shape of the primary phase within the $i$-th cell at the time of $t$.
            \item $\alpha_i^{ext}(t)$ is the exact solution for the $i$-th cell at the time of $t$, which is derived using the tool surface mesh $S(t)$ in accordance with the fraction field initialization method (\cref{ss_alpha_initialization}).
            \item $V_i$ is the volume of the $i$-th cell.
            \item $N_c$ is the total number of cells.
        \end{itemize}

        The tool surface mesh $S(t)$ is obtained by integrating the velocity field $\vec{U}(\vec{x}(t))$ from \cref{eq_U_field}. This integration utilizes the 4/5$^{th}$-order Dormand-Prince Runge–Kutta ODE solver \cite{dormand1980family,hairer1993solving} as incorporated in \OF v2212, under the implementation \texttt{Foam::RKDP45}. This solver operates on the points of an initial spherical surface mesh. A comprehensive description of the tool surface mesh $S(t)$ calculation can be found in \cref{ss_tool_mesh_calc}.

        The second term in the numerator of \cref{eq_E_sd_definition} accounts for contributions from unexpected empty or full cells during interface advection. This term becomes zero since $\alpha_i = \alpha_i^{ext}$ in the initial time $t=0\,s$. In the present study, this error is employed to examine the interface orientation schemes in the initial time $t=0\,s$ (\cref{ss_interface_orientation_schemes}), and it is simplified as:
        \begin{equation}
            E_{sd} \equiv \frac{\sum_{i \in \mathbb{M}}{\left(S_i(0) \Delta R^c_i(0)\right)}}{\sum_{i=1}^{N_c}{\alpha_i(0) V_i}}.
            \label{eq_E_sd_initial_time}
        \end{equation}

        \item \textbf{Shape error}. In addition, another quantitative measure of shape preservation, which is referred to as \textit{shape error} in the present study, is defined as \cite{roenby2016computational}
        \begin{equation}
            E_s(t) \equiv \frac{\sum_{i=1}^{N_c}{V_i \left|\alpha_i(t)-\alpha_i^{ext}(t)\right|}}{\sum_{i=1}^{N_c}{\alpha_i^{ext}(t) V_i}}.
            \label{eq_E_s_definition}
        \end{equation}
        $E_s(t)$ is used to quantify the shape preservation during the interface advection tests (\cref{ss_interface_advection}). It should be noted that $E_s(t)$ defined in \cref{eq_E_s_definition} equals to zero when $t=0\,s$ and can not evaluate the shape preservation in the initial state.
    \end{itemize}


    \item \textbf{Volume/mass conservation}. The volume conservation error $E_v(t)$ represents the relative deviation of the primary phase's total volume from its initial value. It is expressed as:
    \begin{equation}
        E_v(t) \equiv \frac{\sum_{i=1}^{N_c}{\alpha_i(t) V_i} - \sum_{i=1}^{N_c}{\alpha_i(0) V_i}}{\sum_{i=1}^{N_c}{\alpha_i(0) V_i}}.
        \label{eq_E_v_definition}
    \end{equation}

    \item \textbf{Boundedness}. To be physically meaningful, the fraction field $\alpha(t)$ must strictly satisfy $0 \leq \alpha(t) \leq 1$. The relevant measures are the minimum value $\min{(\alpha(t))}$ and the complementary maximum value $\max{(1-\alpha(t))}$ across all mesh cells.

    \item \textbf{Efficiency}. The \OF\,v2312 was compiled using GCC 11.4.0 on a Red Hat Enterprise Linux 8.8 production cluster named \texttt{Improv} of the Laboratory Computing Resource Center (LCRC) in Argonne National Laboratory. Each of the computing nodes is powered by dual 2.0-GHz AMD 7713 64-core processors. All computations were executed serially. The CPU times required for both interface reconstruction ($T_{rec}\,[s]$) and advection ($T_{adv}\,[s]$) steps were monitored using the \texttt{Foam::Time::elapsedCpuTime()}\footnote{\href{https://www.openfoam.com/documentation/guides/latest/api/classFoam_1_1Time.html}{https://www.openfoam.com/documentation/guides/latest/api/classFoam\_1\_1Time.html}} function. What's more, the entire simulation time $T_{calc}\,[s]$ is also tracked.
\end{itemize}

The definitions of $E_s(t)$, $E_p(t)$ and $E_v(t)$ correspond to those in \cite{roenby2016computational}, where they are denoted as $E_1(t)$, $\delta W_{rel}(t)$ and $\delta V_{rel}(t)$, respectively. It's worth noting that that $E_1(t)$, $\delta W_{rel}(t)$ and $\delta V_{rel}(t)$ were assessed only at the end of a simulation in \cite{roenby2016computational}.

\subsection{Tool surface mesh calculation}\label{ss_tool_mesh_calc}

For a given surface mesh consisting of $n_v$ vertices and $n_f$ triangular faces, the initial-value problem $\frac{d \vec{x}(t)}{d t} = \vec{U}(\vec{x}(t), t), \vec{x}(t_0)=\vec{x}_0$ as defined in \cref{eq_U_field}, can be solved for each vertex. Maintaining the vertex-edge-face connections produces the exact shape of the surface mesh advected by the velocity field $\vec{U}(\vec{x}(t), t)$ at the time of $t$. The 4$^{th}$- and 5$^{th}$-order Runge–Kutta approximations for $\vec{x}(t_{n+1})$, denoted as $\vec{x}_{n+1}$ and $\hat{\vec{x}}_{n+1}$, are expressed as:
\begin{equation}
\begin{aligned}
    \vec{x}_{n+1} &= \vec{x}_{n} + \frac{35}{384} k_1 + \frac{500}{1113} k_3 + \frac{125}{192} k_4 - \frac{2187}{6784} k_5 + \frac{11}{84} k_6, \\
    \hat{\vec{x}}_{n+1} &= \vec{x}_{n} + \frac{5179}{57600} k_1 + \frac{7571}{16695} k_3 + \frac{393}{640} k_4 - \frac{92097}{339200} k_5 + \frac{187}{2100} k_6 + \frac{1}{40} k_7,
\end{aligned}
\label{eq_RK_approximations}
\end{equation}
where the coefficients $k_i(i = 1, 2, \cdots, 7)$ are given by:
\begin{equation}
\begin{aligned}
    k_1 &= h \vec{U}\left(\vec{x}_n, t_n\right), \\
    k_2 &= h \vec{U}\left(\vec{x}_n+\frac{1}{5}k_1, t_n+\frac{1}{5}h\right), \\
    k_3 &= h \vec{U}\left(\vec{x}_n+\frac{3}{40}k_1+\frac{9}{40}k_2, t_n+\frac{3}{10}h\right), \\
    k_4 &= h \vec{U}\left(\vec{x}_n+\frac{44}{45}k_1-\frac{56}{15}k_2+\frac{32}{9}k_3, t_n+\frac{4}{5}h\right), \\
    k_5 &= h \vec{U}\left(\vec{x}_n+\frac{19372}{6561}k_1-\frac{25360}{2187}k_2+\frac{64448}{6561}k_3-\frac{212}{729}k_4, t_n+\frac{8}{9}h\right), \\
    k_6 &= h \vec{U}\left(\vec{x}_n+\frac{9017}{3168}k_1-\frac{355}{33}k_2+\frac{46732}{5247}k_3+\frac{49}{176}k_4-\frac{5103}{18656}k_5, t_n+h\right), \\
    k_7 &= h \vec{U}\left(\vec{x}_n+\frac{35}{384}k_1+\frac{500}{1113}k_3+\frac{125}{192}k_4-\frac{2187}{6784}k_5+\frac{11}{84}k_6, t_n+h\right),
\end{aligned}
\label{eq_RK_coeffs}
\end{equation}
where $h$ represents the step size.

The discrepancy between the 4$^{th}$- and 5$^{th}$-order approximations is characterized by the global error
\begin{equation}
    \tau_{n+1} = \hat{\vec{x}}_{n+1} - \vec{x}_{n+1} = -\frac{71}{57600} k_1 + \frac{71}{16695} k_3 - \frac{71}{1920}k_4 + \frac{17253}{339200}k_5 - \frac{22}{525}k_6 + \frac{1}{40}k_7,
    \label{eq_error_tau}
\end{equation}
then $\tau_{n+1}$ is considered as the error of the 4$^{th}$-order approximation $\vec{x}_{n+1}$.

The method outlined above is referred to as the 4/5$^{th}$-order Dormand-Prince Runge–Kutta (DPRK45) scheme \cite{dormand1980family,hairer1993solving}. A detailed procedure for calculating the position of a surface mesh vertex at $t = t_{end}$ from its initial position $\vec{x}_0$ at $t_0$ is provided in \cref{alg_tool_surface_mesh}. It should be noted that the DPRK45 scheme is versatile and can be applied to other interface advection benchmarks with non-uniform, pre-defined velocity fields.

\begin{algorithm}[htbp]
\setstretch{0.5}
\caption{Tool surface mesh calculation algorithm}
\label{alg_tool_surface_mesh}
\DontPrintSemicolon
\SetKwInOut{Input}{Input}
\SetKwInOut{Output}{Output}
\SetNoFillComment
\SetKwRepeat{Do}{do}{while}

    \Input{$t_0 = 0$, $t_{end}$, $h=10^{-6}$, $\vec{x}_0$, $\varepsilon_{abs}=10^{-15}$, $\varepsilon_{rel}=10^{-9}$ and $I_{max}=10000$}
    \Output{$\vec{x}_{end}$}

    $t \gets t_0$

    $\vec{x} \gets \vec{x}_0$

    \For{($iter = 0;\ iter < I_{max};\ iter$++)}
    {
        \tcc{Check if this is a truncated step and set $h$ to integrate to $t_{end}$}
        \If{$(t+h - t_{end})(t+h - t_0) > 0$} 
        {
            $h \gets t_{end}-t$
        }

        \tcc{Integrate up to $t_{end}$}
        $\tau \gets 0$
        
        $\vec{x}_n \gets \vec{x}$

        \Do{$\tau > 1$}
        {
            Compute $k_i(i = 1, 2, \cdots, 7)$ \tcp*[l]{\cref{eq_RK_coeffs}}
            
            Compute $\vec{x}_{n+1}$ \tcp*[l]{\cref{eq_RK_approximations}}

            Compute $\hat{\vec{x}}_{n+1}$ \tcp*[l]{\cref{eq_RK_approximations}}

            $\varepsilon \gets \varepsilon_{abs}+\varepsilon_{rel} \max{\left(|\vec{x}_{n}|, |\vec{x}_{n+1}|\right)}$

            $\tau \gets \max{\left(|\hat{\vec{x}}_{n+1} - \vec{x}_{n+1}| / \varepsilon\right)}$
            
            $s = \max{\left(0.9 \tau^{-0.25}, 0.2\right)}$ \tcp*[l]{safe scaling}

            $h \gets hs$
        }

        \tcc{Update $t$ and $\vec{x}$}
        $t \gets t + h$

        $\vec{x} \gets \vec{x}_{n+1}$

        \tcc{If the error is small increase $h$}
        \eIf{$\tau > (10/0.9)^{-5}$}
        {
            $h \gets \min{\left(\max{\left(0.9 \tau^{-0.2}, 0.2\right)}, 10\right)} h$
        }
        {
            $h \gets 9h$
        }

        \tcc{Check if reached $t_{end}$}
        \If{$(t - t_{end})(t_{end} - t_0) >= 0$} 
        {
            $\vec{x}_{end} \gets \vec{x}$

            \Return
        }
    }
\end{algorithm}

\begin{figure}[htbp]
    \centering
    \begin{subfigure}[b]{0.49\textwidth}
        \centering
        \includegraphics[width=0.6\textwidth]{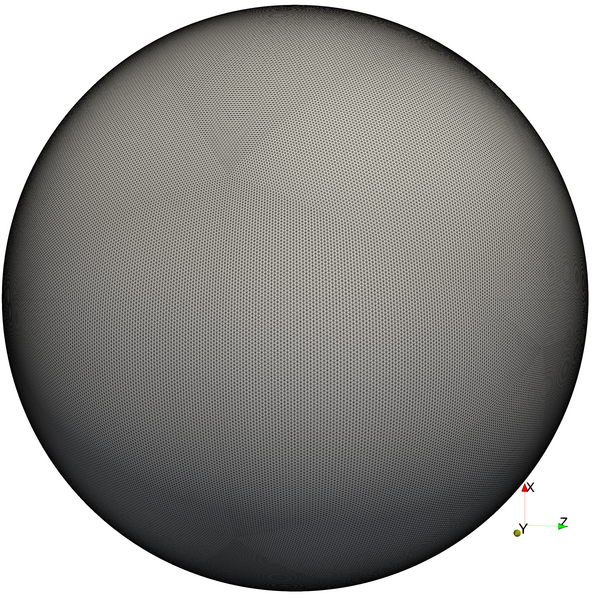}
        \caption{Icosphere ($n_v=163842, n_f=327680$)}
        \label{fig_icosphere_ref_shape_back_view_t_0}
    \end{subfigure}
    \begin{subfigure}[b]{0.49\textwidth}
        \centering
        \includegraphics[width=0.6\textwidth]{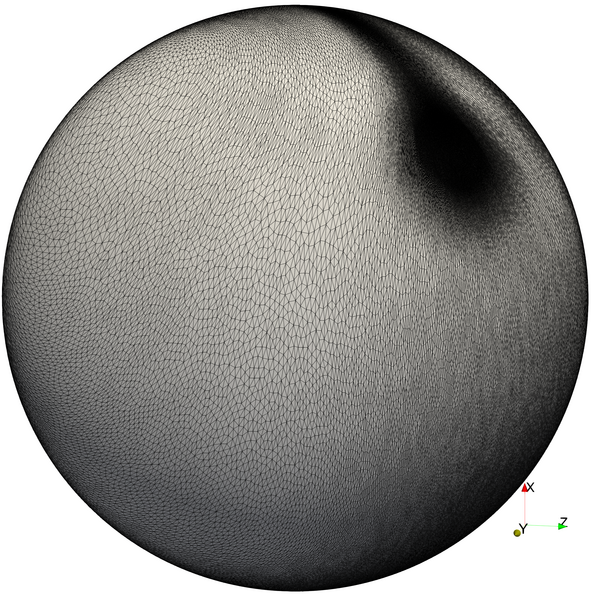}
        \caption{Non-uniform surface ($n_v=88659, n_f=177314$)}
        \label{fig_nonuniform_ref_shape_back_view_t_0}
    \end{subfigure}
    \caption{Initial tool surface meshes ($t = 0\,s$).}
    \label{fig_ref_shape_back_view_t_0}
\end{figure}

\begin{figure}[htbp]
    \centering
    \begin{subfigure}[b]{0.49\textwidth}
        \centering
        \includegraphics[width=0.6\textwidth]{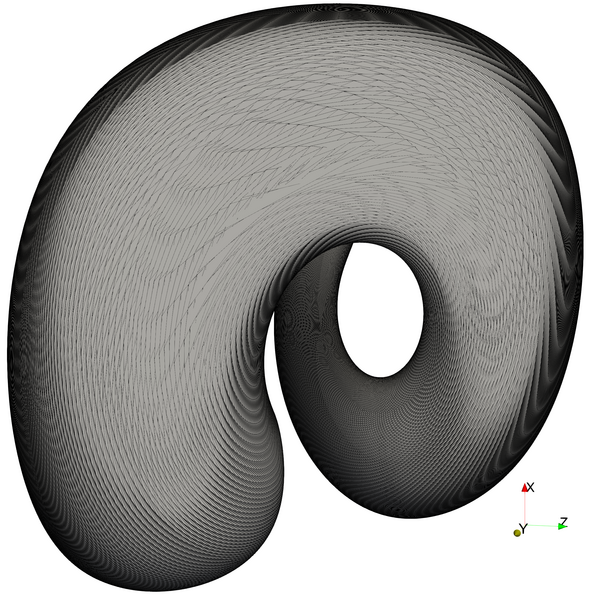}
        \caption{Icosphere ($n_v=163842, n_f=327680$)}
        \label{fig_icosphere_ref_shape_back_view_t_1_5}
    \end{subfigure}
    \begin{subfigure}[b]{0.49\textwidth}
        \centering
        \includegraphics[width=0.6\textwidth]{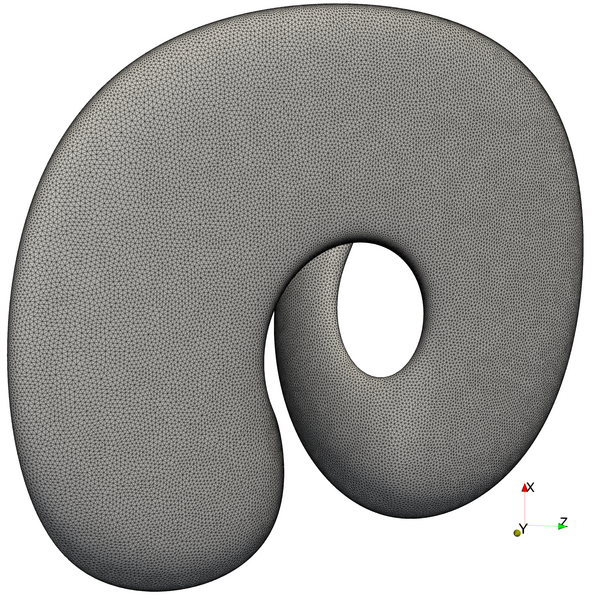}
        \caption{Non-uniform surface ($n_v=88659, n_f=177314$)}
        \label{fig_nonuniform_ref_shape_back_view_t_1_5}
    \end{subfigure}
    \caption{Maximum deformed tool surface meshes ($t = 1.5\,s$).}
    \label{fig_ref_shape_back_view_t_1_5}
\end{figure}

A 7$^{th}$-order icosphere surface mesh, used to represent the initial spherical interface shape, is depicted in \cref{fig_icosphere_ref_shape_back_view_t_0}. This mesh also initializes the $\alpha$ field in \cref{ss_alpha_initialization}. The shape of this mesh at $t=1.5\,s$, as determined by the DPRK45 method, is presented in \cref{fig_icosphere_ref_shape_back_view_t_1_5}. Notably, this resulted in certain triangular elements becoming highly deformed or overlapping. To address this, a non-uniform surface mesh was introduced (see \cref{fig_nonuniform_ref_shape_back_view_t_0}). Its shape at $t=1.5\,s$ is illustrated in \cref{fig_nonuniform_ref_shape_back_view_t_1_5}. Both meshes were evolved from their initial state until $t_{end}=3\,s$, with snapshots taken every $0.005\,s$.

The aspect ratio $AR$ is employed to assess the surface mesh quality. For a given triangular element, the $AR$ is defined as:
\begin{equation}
    AR = \frac{\underset{1 \leq i \leq 3}{\max}{l_i} \sum_{i=1}^{3}{l_i}}{4 \sqrt{3} A},
\end{equation}
where $l_i(i=1,2,3)$ represents the lengths of the three edges and $A$ denotes the area of the triangle.

\begin{figure}[htbp]
    \centering
    \begin{subfigure}[b]{0.45\textwidth}
        \centering
        \includegraphics[width=0.85\textwidth]{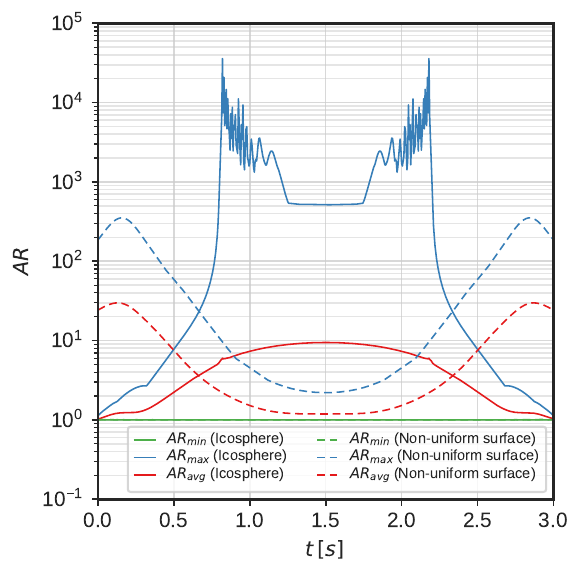}
        \caption{$AR$}
        \label{fig_ref_shape_AR}
    \end{subfigure}
    \begin{subfigure}[b]{0.45\textwidth}
        \centering
        \includegraphics[width=0.85\textwidth]{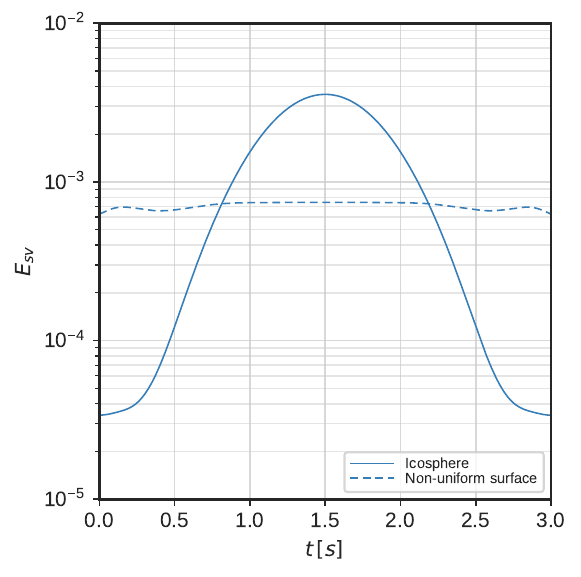}
        \caption{$E_{sv}$}
        \label{ref_shape_Ev}
    \end{subfigure}
    \caption{Time histories of $AR$ and $E_{sv}$.}
    \label{fig_ref_shape_qualities}
\end{figure}

The temporal evolutions of the minimum, maximum, and average aspect ratios for the two distinct surface meshes are depicted in \cref{fig_ref_shape_AR}. Remarkably, all these distributions are symmetric. The maximum aspect ratio $AR_{max}$ for the icosphere surface mesh escalates swiftly from $t=0\,s$ and surpasses $10^3$ within intervals $0.805\,s \leq t \leq 1.21\,s$ and $1.79\,s \leq t \leq 2.195\,s$. Furthermore, between $1.21\,s$ and $1.79\,s$, $AR_{max}$ for the icosphere exceeds $500$. In contrast, the non-uniform surface mesh starts with a high $AR_{max}$ but declines to $22.1$ by $t=1.5\,s$. Notably, for the interval $0.665\,s \leq t \leq 2.335\,s$, the non-uniform surface mesh consistently maintains a lower average aspect ratio than its icosphere counterpart.

The absolute relative errors, $E_{sv}$, of the volume enclosed by the surface meshes are shown in \cref{ref_shape_Ev}. The non-uniform mesh exhibits a steady distribution across different time instances. Conversely, the icosphere mesh encounters pronounced errors within the span $0.815\,s \leq t \leq 2.185\,s$.

Consequently, in \cref{ss_sol_quantification}, the tool surface meshes $S(t)$ derived from the DPRK45 algorithm originate from:
\begin{enumerate}[(a)]
    \item the non-uniform surface mesh for $0.665\,s \leq t \leq 2.335\,s$,
    \item the icosphere surface for time instances where $t<0.665\,s$ or $t>2.335\,s$.
\end{enumerate}

\subsection{Volume meshes used in tests}\label{ss_meshes}

\begin{figure}[htbp]
    \centering
    \includegraphics[scale=0.89]{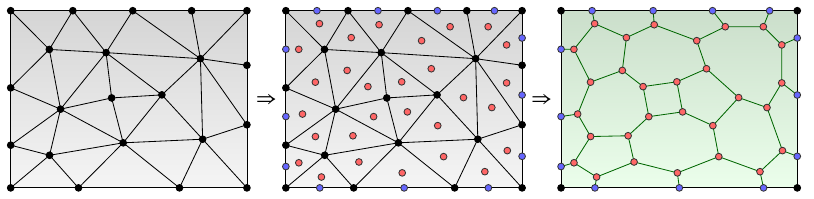}
    \caption{An illustration of general polyhedral mesh generation.}
    \label{fig_general_polyhedral_mesh}
\end{figure}

\begin{figure}[htbp]
    \centering
    \includegraphics[scale=0.89]{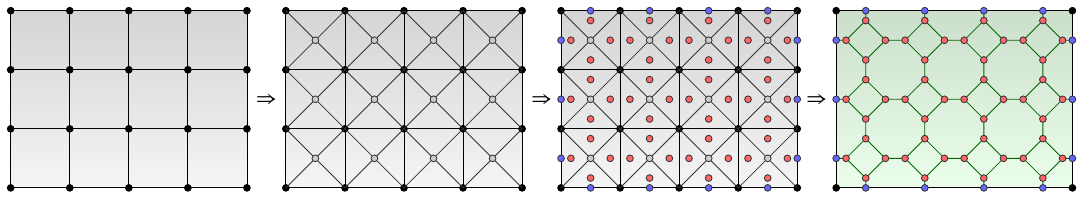}
    \caption{An illustration of structured polyhedral mesh generation.}
    \label{fig_structured_polyhedral_mesh}
\end{figure}

The interface advection problem outlined in the beginning is addressed across four distinct mesh types, each characterized by differing mesh sizes denoted as $\Delta s$. The meshers and utilities employed for mesh generation are outlined as follows:
\begin{enumerate}[label=(\alph*)]
    \item \textbf{Tetrahedral meshes}. The \texttt{Tetrahedral Mesher} in Simcenter STAR-CCM+ 2306 \cite{ccm} generates tetrahedral meshes with a base size of $\Delta s$ and a volume growth rate of unity, ensuring uniform mesh density across the domain. It employs the Delaunay method, which iteratively introduces points into the domain and constructs high-quality tetrahedra \cite{ccmUG}. The generated tetrahedral meshes are also used for the general polyhedral mesh conversions, as detailed in (c).
    
    \item \textbf{Hexahedral meshes}. The hexahedral meshes are created using the \texttt{blockMesh} utility in \OF\,v2312. Here, the block is constructed with $1 / \Delta s$ cells in each dimension and expansion ratios set to unity in all directions. This configuration results in meshes comprising a total of $1 / (\Delta s)^{3}$ cells.
    

    \item \textbf{General polyhedral meshes}. The general polyhedral meshes employed in the tests are generated using the \texttt{polyDualMesh} converter in \OF\,v2312. As depicted in \cref{fig_general_polyhedral_mesh}, a dualization scheme is implemented in the tetrahedral meshes. This involves marking the centroids of the tetrahedral cells (\textit{red dots}) and boundary faces (\textit{blue dots}), followed by constructing polyhedral cells through connections between centroids of cells sharing a common vertex. It should be noted that \texttt{polyDualMesh} doesn't attempt to improve the face flatness quality of the converted polyhedral meshes.
    
    \item \textbf{Structured polyhedral meshes}. The structured polyhedral meshes are produced using \texttt{blockPolyMesh} \cite{blockPolyMeshCode2023}, a structured polyhedral mesh generator that builds upon the \OF\,\texttt{blockMesh} utility. Each cell in the structured hexahedral mesh is first split into $24$ tetrahedrons, which are then transformed into polyhedra following the algorithm in the \texttt{polyDualMesh} utility. \cref{fig_structured_polyhedral_mesh} provides an illustration of this structured polyhedral mesh generation process. In comparison to the method used for general polyhedral meshes (refer to \cref{fig_general_polyhedral_mesh}), this approach includes an additional step of decomposing hexahedral cells before applying the dualization scheme.
\end{enumerate}

\begin{figure}[htbp]
    \centering
    \begin{subfigure}[b]{0.4\textwidth}
        \centering
        \includegraphics[width=0.9\textwidth]{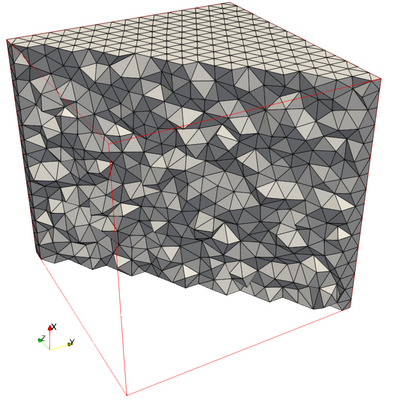}
        \caption{Tetrahedral mesh}
    \end{subfigure}
    \begin{subfigure}[b]{0.4\textwidth}
        \centering
        \includegraphics[width=0.9\textwidth]{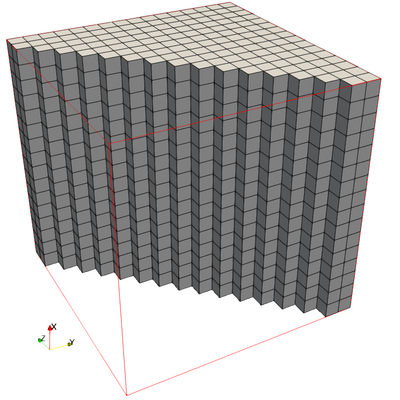}
        \caption{Hexahedral mesh}
    \end{subfigure}
    \begin{subfigure}[b]{0.4\textwidth}
        \centering
        \includegraphics[width=0.9\textwidth]{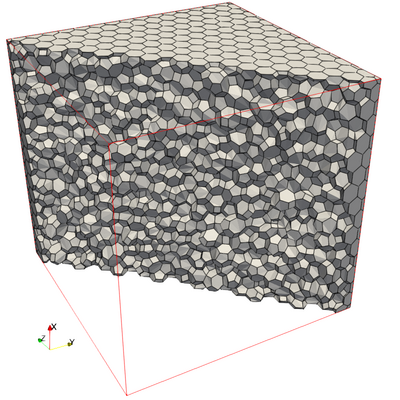}
        \caption{General polyhedral mesh}
    \end{subfigure}
    \begin{subfigure}[b]{0.4\textwidth}
        \centering
        \includegraphics[width=0.9\textwidth]{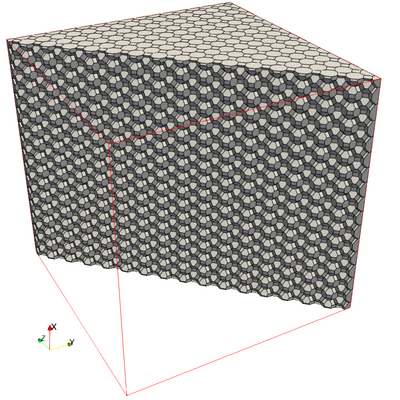}
        \caption{Structured polyhedral mesh}
    \end{subfigure}
    \caption{Various mesh types with $\Delta s = 2^{-4}\,m$.}
    \label{fig_various_meshes}
\end{figure}

\cref{fig_various_meshes} illustrates these four mesh types with $\Delta s = 1/16\,m$. The statistics of all meshes used are summarized in \cref{tab_stat_of_all_meshes}. The general polyhedral meshes yield approximately 5 times fewer cells and roughly 1.6 times fewer faces compared to their tetrahedral sources. Conversely, the structured polyhedral meshes result in about 5 times more cells and 9.7 times more faces compared to their hexahedral counterparts.

\begin{table}[htbp]
\footnotesize
\centering
\caption{Statistics of all meshes.}
\label{tab_stat_of_all_meshes}
\renewcommand{\arraystretch}{1.15}
\small
\begin{tabular}{c|rrr|rrr}
\hline
\rowcolor[HTML]{EFEFEF} 
\cellcolor[HTML]{EFEFEF} &
  \multicolumn{3}{c|}{\cellcolor[HTML]{EFEFEF}Tetrahedral meshes} &
  \multicolumn{3}{c}{\cellcolor[HTML]{EFEFEF}Hexahedral meshes} \\ \cline{2-7} 
\rowcolor[HTML]{EFEFEF} 
\multirow{-2}{*}{\cellcolor[HTML]{EFEFEF}\begin{tabular}[c]{@{}c@{}}$\Delta s$\\ $[m]$\end{tabular}} &
  \multicolumn{1}{c|}{\cellcolor[HTML]{EFEFEF}\texttt{nPoints}} &
  \multicolumn{1}{c|}{\cellcolor[HTML]{EFEFEF}\texttt{nCells}} &
  \multicolumn{1}{c|}{\cellcolor[HTML]{EFEFEF}\texttt{nFaces}} &
  \multicolumn{1}{c|}{\cellcolor[HTML]{EFEFEF}\texttt{nPoints}} &
  \multicolumn{1}{c|}{\cellcolor[HTML]{EFEFEF}\texttt{nCells}} &
  \multicolumn{1}{c}{\cellcolor[HTML]{EFEFEF}\texttt{nFaces}} \\ \hline
$2^{-5}$ &
  \multicolumn{1}{r|}{\texttt{32,288}} &
  \multicolumn{1}{r|}{\texttt{172,779}} &
  \texttt{351,606} &
  \multicolumn{1}{r|}{\texttt{35,937}} &
  \multicolumn{1}{r|}{\texttt{32,768}} &
  \texttt{101,376} \\
$2^{-6}$ &
  \multicolumn{1}{r|}{\texttt{244,556}} &
  \multicolumn{1}{r|}{\texttt{1,376,270}} &
  \texttt{2,776,924} &
  \multicolumn{1}{r|}{\texttt{274,625}} &
  \multicolumn{1}{r|}{\texttt{262,144}} &
  \texttt{798,720} \\
$2^{-7}$ &
  \multicolumn{1}{r|}{\texttt{1,905,268}} &
  \multicolumn{1}{r|}{\texttt{11,001,507}} &
  \texttt{22,100,934} &
  \multicolumn{1}{r|}{\texttt{2,146,689}} &
  \multicolumn{1}{r|}{\texttt{2,097,152}} &
  \texttt{6,340,608} \\
$2^{-8}$ &
  \multicolumn{1}{r|}{\texttt{15,030,359}} &
  \multicolumn{1}{r|}{\texttt{87,932,754}} &
  \texttt{176,257,956} &
  \multicolumn{1}{r|}{\texttt{16,974,593}} &
  \multicolumn{1}{r|}{\texttt{16,777,216}} &
  \texttt{50,528,256} \\ \hline
\rowcolor[HTML]{EFEFEF} 
\cellcolor[HTML]{EFEFEF} &
  \multicolumn{3}{c|}{\cellcolor[HTML]{EFEFEF}General polyhedral meshes} &
  \multicolumn{3}{c}{\cellcolor[HTML]{EFEFEF}Structured polyhedral meshes} \\ \cline{2-7} 
\rowcolor[HTML]{EFEFEF} 
\multirow{-2}{*}{\cellcolor[HTML]{EFEFEF}\begin{tabular}[c]{@{}c@{}}$\Delta s$\\ $[m]$\end{tabular}} &
  \multicolumn{1}{c|}{\cellcolor[HTML]{EFEFEF}\texttt{nPoints}} &
  \multicolumn{1}{c|}{\cellcolor[HTML]{EFEFEF}\texttt{nCells}} &
  \multicolumn{1}{c|}{\cellcolor[HTML]{EFEFEF}\texttt{nFaces}} &
  \multicolumn{1}{c|}{\cellcolor[HTML]{EFEFEF}\texttt{nPoints}} &
  \multicolumn{1}{c|}{\cellcolor[HTML]{EFEFEF}\texttt{nCells}} &
  \multicolumn{1}{c}{\cellcolor[HTML]{EFEFEF}\texttt{nFaces}} \\ \hline
$2^{-5}$ &
  \multicolumn{1}{r|}{\texttt{185,267}} &
  \multicolumn{1}{r|}{\texttt{32,288}} &
  \texttt{217,552} &
  \multicolumn{1}{r|}{\texttt{811,400}} &
  \multicolumn{1}{r|}{\texttt{170,081}} &
  \texttt{981,478} \\
$2^{-6}$ &
  \multicolumn{1}{r|}{\texttt{1,425,814}} &
  \multicolumn{1}{r|}{\texttt{244,556}} &
  \texttt{1,670,367} &
  \multicolumn{1}{r|}{\texttt{6,390,536}} &
  \multicolumn{1}{r|}{\texttt{1,335,489}} &
  \texttt{7,726,022} \\
$2^{-7}$ &
  \multicolumn{1}{r|}{\texttt{11,198,891}} &
  \multicolumn{1}{r|}{\texttt{1,905,268}} &
  \texttt{13,104,156} &
  \multicolumn{1}{r|}{\texttt{50,726,408}} &
  \multicolumn{1}{r|}{\texttt{10,584,449}} &
  \texttt{61,310,854} \\
$2^{-8}$ &
  \multicolumn{1}{r|}{\texttt{88,720,730}} &
  \multicolumn{1}{r|}{\texttt{15,030,359}} &
  \texttt{103,751,086} &
  \multicolumn{1}{r|}{\texttt{404,229,128}} &
  \multicolumn{1}{r|}{\texttt{84,280,065}} &
  \texttt{488,509,190} \\ \hline
\end{tabular}
\end{table}

In tetrahedral, hexahedral, and structured polyhedral meshes, warped faces are non-existent, with the minimum face flatness values being unity. Therefore, triangular decompositions for warped faces are not required in these three types of meshes. However, warped faces frequently occur in general polyhedral meshes, which are derived from tetrahedral meshes. The maximum, minimum, and area-averaged face flatness values for these general polyhedral meshes are detailed in \cref{tab_face_flatness_general_poly_meshes}.

\begin{table}[htbp]
\footnotesize
\centering
\caption{Face flatness qualities in all general polyhedral meshes.}
\label{tab_face_flatness_general_poly_meshes}
\renewcommand{\arraystretch}{1.15}
\small
\begin{tabular}{c|c|c|c}
\hline
\rowcolor[HTML]{EFEFEF} 
$\Delta s\,[m]$ & $\zeta_{max}$ & $\zeta_{min}$ & $\zeta_{avg}$ \\ \hline
$2^{-5}$ & 1 & 0.890 & 0.992 \\
$2^{-6}$ & 1 & 0.880 & 0.992 \\
$2^{-7}$ & 1 & 0.879 & 0.991 \\
$2^{-8}$ & 1 & 0.861 & 0.991 \\ \hline
\end{tabular}
\end{table}

\section{Numerical Tests}\label{s_results}

\subsection{Interface orientation schemes}\label{ss_interface_orientation_schemes}

To assess the performance of the four orientation methods detailed in \cref{sss_interface_orientation}, the interface reconstruction at the initial state is analyzed. The mixed cell tolerance $\epsilon$ is fixed to $10^{-8}$. The parameters of RDF scheme, $I_{max}^{RDF}$, $res$ and $res_{curv}$, are using the default values in \OF\,v2312, which are $5$, $10^{-6}$ and $0.1$, respectively. These schemes have been tested across four different types of meshes, each with varying resolutions. Initially, the study focuses on evaluating the impacts of triangular decomposition on warped cell faces in general polyhedral meshes, followed by comparative analyses across various mesh types.

\subsubsection{Warped face decomposition in general polyhedral meshes}

The impacts of triangular decomposition of warped faces within general polyhedral meshes have been investigated. \cref{tab_effects_of_tri_decomposition} summarizes the symmetric difference errors $E_{sd}$ and the convergence orders $\mathcal{O}\left(E_{sd}\right)$ of various orientation schemes in the initial interface reconstruction, with and without warped face decomposition in general polyhedral meshes. The order of convergence $\mathcal{O}\left(E_{sd}\right)$ is calculated as:
\begin{equation}
    \mathcal{O}\left(E_{sd}\right) = \ln{\frac{E_{sd}(\Delta s)}{E_{sd}(\Delta s/2)}} \bigg/ \ln{2}.
    \label{eq_convergence_order}
\end{equation}

The triangular decomposition of warped cell faces leads to a reduction in $E_{sd}$ across all orientation schemes. However, this approach only marginally improves the $\mathcal{O}\left(E_{sd}\right)$ in fraction-gradient-based methods. For RDF scheme, employing the warped face decomposition technique elevates $\mathcal{O}\left(E_{sd}\right)$ from approximately unity to 1.6.

\newlength\MAX
\setlength\MAX{4.425mm}
\newcommand*\Chart[1]{#1~\rlap{\textcolor{blue!20}{\rule{2.26\MAX}{1.5ex}}}\textcolor{blue!70}{\rule{#1\MAX}{1.5ex}}}

\begin{table}[htbp]
\footnotesize
\centering
\caption{$E_{sd}$ and $\mathcal{O}\left(E_{sd}\right)$ of various orientation schemes in general polyhedral meshes.}
\label{tab_effects_of_tri_decomposition}
\renewcommand{\arraystretch}{1.15}
\begin{tabular}{c|c|c|l|c|l}
\hline
\rowcolor[HTML]{EFEFEF} 
\cellcolor[HTML]{EFEFEF} &
  \cellcolor[HTML]{EFEFEF} &
  \multicolumn{2}{c|}{\cellcolor[HTML]{EFEFEF}Without decomposition} &
  \multicolumn{2}{c}{\cellcolor[HTML]{EFEFEF}With decomposition} \\ \cline{3-6}
\rowcolor[HTML]{EFEFEF}
\multirow{-2}{*}{\cellcolor[HTML]{EFEFEF}\begin{tabular}[c]{@{}c@{}}Orientation\\ Scheme\end{tabular}} &
  \multirow{-2}{*}{\cellcolor[HTML]{EFEFEF}\begin{tabular}[c]{@{}c@{}}$\Delta s$\\ $[m]$\end{tabular}} &
  $E_{sd}$ &
  \multicolumn{1}{c|}{\cellcolor[HTML]{EFEFEF}$\mathcal{O}\left(E_{sd}\right)$} &
  $E_{sd}$ &
  \multicolumn{1}{c}{\cellcolor[HTML]{EFEFEF}$\mathcal{O}\left(E_{sd}\right)$} \\ \hline
                       & $2^{-5}$ & $2.71 \times 10^{-2}$ &              & $2.00 \times 10^{-2}$ &              \\
                       & $2^{-6}$ & $8.48 \times 10^{-3}$ & \Chart{1.67} & $6.84 \times 10^{-3}$ & \Chart{1.55} \\
                       & $2^{-7}$ & $4.53 \times 10^{-3}$ & \Chart{0.90} & $3.40 \times 10^{-3}$ & \Chart{1.01} \\
\multirow{-4}{*}{CAG}  & $2^{-8}$ & $2.05 \times 10^{-3}$ & \Chart{1.15} & $1.53 \times 10^{-3}$ & \Chart{1.15} \\ \hline
                       & $2^{-5}$ & $1.79 \times 10^{-2}$ &              & $1.79 \times 10^{-2}$ &              \\
                       & $2^{-6}$ & $7.56 \times 10^{-3}$ & \Chart{1.25} & $5.77 \times 10^{-3}$ & \Chart{1.64} \\
                       & $2^{-7}$ & $3.66 \times 10^{-3}$ & \Chart{1.04} & $2.83 \times 10^{-3}$ & \Chart{1.03} \\
\multirow{-4}{*}{NAG}  & $2^{-8}$ & $1.71 \times 10^{-3}$ & \Chart{1.10} & $1.23 \times 10^{-3}$ & \Chart{1.20} \\ \hline
                       & $2^{-5}$ & $2.53 \times 10^{-2}$ &              & $1.80 \times 10^{-2}$ &              \\
                       & $2^{-6}$ & $6.94 \times 10^{-3}$ & \Chart{1.87} & $5.98 \times 10^{-3}$ & \Chart{1.59} \\
                       & $2^{-7}$ & $3.31 \times 10^{-3}$ & \Chart{1.07} & $2.61 \times 10^{-3}$ & \Chart{1.19} \\
\multirow{-4}{*}{LS}   & $2^{-8}$ & $1.64 \times 10^{-3}$ & \Chart{1.01} & $1.29 \times 10^{-3}$ & \Chart{1.02} \\ \hline
                       & $2^{-5}$ & $2.14 \times 10^{-2}$ &              & $1.41 \times 10^{-2}$ &              \\
                       & $2^{-6}$ & $5.61 \times 10^{-3}$ & \Chart{1.93} & $3.25 \times 10^{-3}$ & \Chart{2.12} \\
                       & $2^{-7}$ & $2.57 \times 10^{-3}$ & \Chart{1.13} & $9.96 \times 10^{-4}$ & \Chart{1.71} \\
\multirow{-4}{*}{RDF}  & $2^{-8}$ & $1.22 \times 10^{-3}$ & \Chart{1.07} & $3.56 \times 10^{-4}$ & \Chart{1.48} \\ \hline
\end{tabular}
\end{table}

\begin{figure}[htbp]
    \centering
    \begin{subfigure}[b]{0.4\textwidth}
        \centering
        \includegraphics[width=\textwidth]{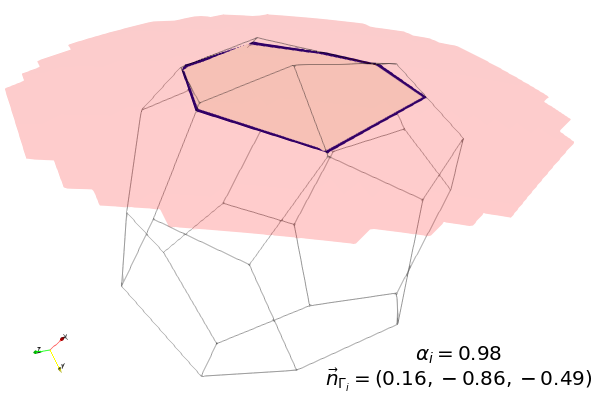}
        \caption{Without triangular decomposition}
        \label{fig_plicSurfaceNoSplitCell}
    \end{subfigure}\hspace{1cm}
    \begin{subfigure}[b]{0.4\textwidth}
        \centering
        \includegraphics[width=\textwidth]{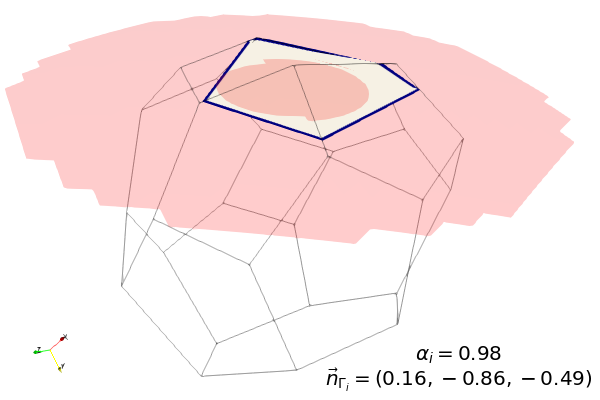}
        \caption{With triangular decomposition}
        \label{fig_plicSurfaceWithSplitCell}
    \end{subfigure}
    \caption{Different reconstructed interface planes in a polyhedral cell (red surface is the exact interface).}
    \label{fig_plicSurfac_split_comparison}
\end{figure}

\cref{fig_plicSurfac_split_comparison} displays reconstructed interface planes with and without warped face decomposition in a polyhedral cell characterized by $\alpha=0.98$ and $\vec{n}_{\Gamma} = (0.16, -0.86, -0.49)$. The exact interface is indicated in red. It is clearly observable that the interface plane reconstructed with warped face decomposition aligns more closely with the exact solution, whereas the plane reconstructed without these modifications falls inside the exact interface. Additionally, \cref{tab_reconstruction_plic_surfraces_warped_face_decomp} offers a comparative view of the reconstructed interface planes using warped face decomposition in the general polyhedral meshes. There are no significant deviations between reconstructed interfaces with and without warped face decomposition, except in the scenario of RDF scheme with $\Delta s = 2^{-8}$, the inclusion of warped face decomposition notably enhances the quality of the poor interface reconstructions.

\begin{table}[htbp]
\footnotesize
\centering
\caption{Reconstructed interfaces at $t=0\,s$ with ($\mathrlap{\hexstar}\hexagon$) and without ($\hexagon$) warped face decomposition in general polyhedral meshes.}
\label{tab_reconstruction_plic_surfraces_warped_face_decomp}
\renewcommand{\arraystretch}{1.15}
\begin{tabular}{c|c|c|c|c|c}
\hline
\rowcolor[HTML]{EFEFEF} 
$\Delta s\,[m]$ &
  $\hexagon$/$\mathrlap{\hexstar}\hexagon$ &
  CAG &
  NAG &
  LS &
  RDF \\ \hline
 &
  $\hexagon$ &
  \includegraphics[height=0.1525\textwidth,valign=c]{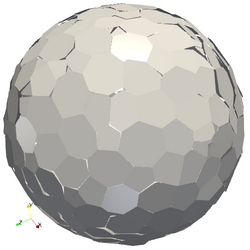} &
  \includegraphics[height=0.1525\textwidth,valign=c]{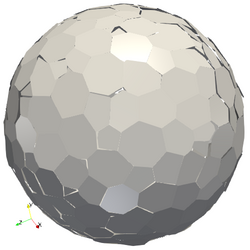} &
  \includegraphics[height=0.1525\textwidth,valign=c]{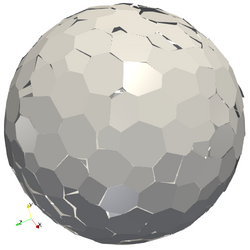} &
  \includegraphics[height=0.1525\textwidth,valign=c]{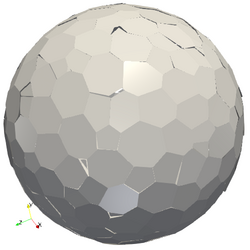} \\
 \multirow{-2}{*}[4.5em]{$2^{-5}$} &
  $\mathrlap{\hexstar}\hexagon$ &
  \includegraphics[height=0.1525\textwidth,valign=c]{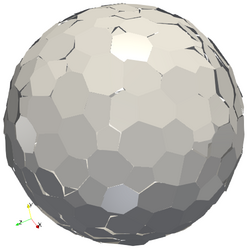} &
  \includegraphics[height=0.1525\textwidth,valign=c]{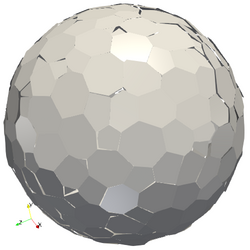} &
  \includegraphics[height=0.1525\textwidth,valign=c]{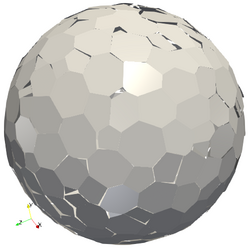} &
  \includegraphics[height=0.1525\textwidth,valign=c]{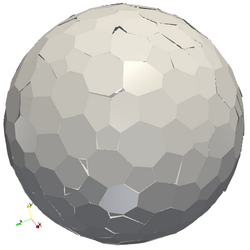} \\ \hline
 &
  $\hexagon$ &
  \includegraphics[height=0.1525\textwidth,valign=c]{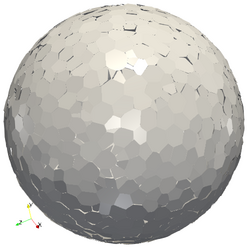} &
  \includegraphics[height=0.1525\textwidth,valign=c]{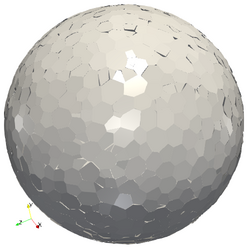} &
  \includegraphics[height=0.1525\textwidth,valign=c]{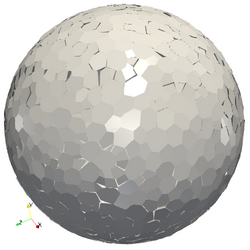} &
  \includegraphics[height=0.1525\textwidth,valign=c]{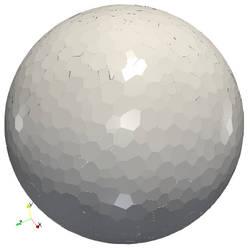} \\
 \multirow{-2}{*}[4.5em]{$2^{-6}$} &
  $\mathrlap{\hexstar}\hexagon$ &
  \includegraphics[height=0.1525\textwidth,valign=c]{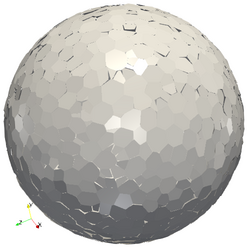} &
  \includegraphics[height=0.1525\textwidth,valign=c]{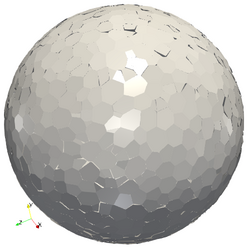} &
  \includegraphics[height=0.1525\textwidth,valign=c]{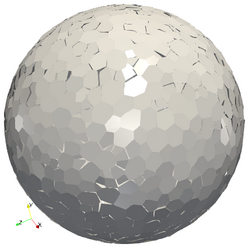} &
  \includegraphics[height=0.1525\textwidth,valign=c]{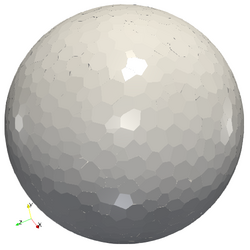} \\ \hline
 &
  $\hexagon$ &
  \includegraphics[height=0.1525\textwidth,valign=c]{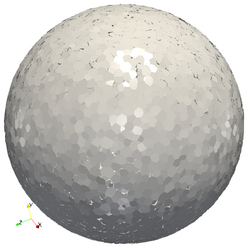} &
  \includegraphics[height=0.1525\textwidth,valign=c]{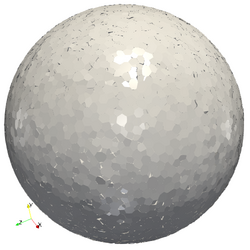} &
  \includegraphics[height=0.1525\textwidth,valign=c]{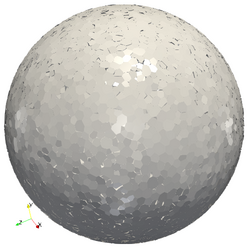} &
  \includegraphics[height=0.1525\textwidth,valign=c]{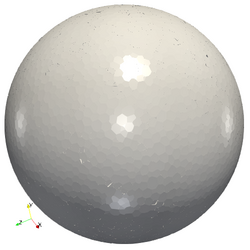} \\
 \multirow{-2}{*}[4.5em]{$2^{-7}$} &
  $\mathrlap{\hexstar}\hexagon$ &
  \includegraphics[height=0.1525\textwidth,valign=c]{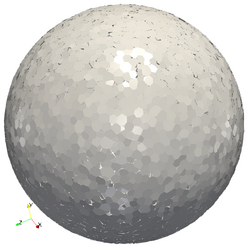} &
  \includegraphics[height=0.1525\textwidth,valign=c]{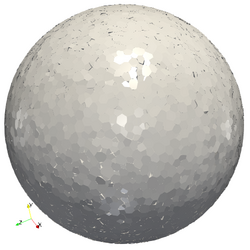} &
  \includegraphics[height=0.1525\textwidth,valign=c]{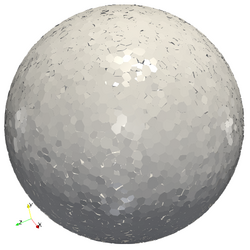} &
  \includegraphics[height=0.1525\textwidth,valign=c]{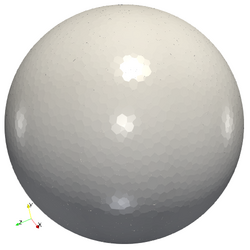} \\ \hline
 &
  $\hexagon$ &
  \includegraphics[height=0.1525\textwidth,valign=c]{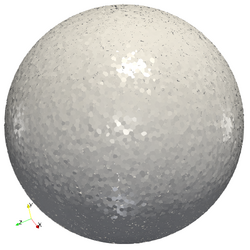} &
  \includegraphics[height=0.1525\textwidth,valign=c]{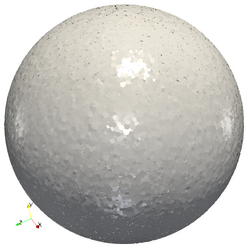} &
  \includegraphics[height=0.1525\textwidth,valign=c]{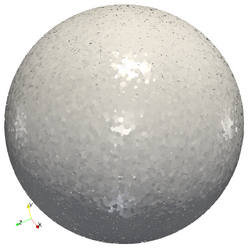} &
  \includegraphics[height=0.1525\textwidth,valign=c]{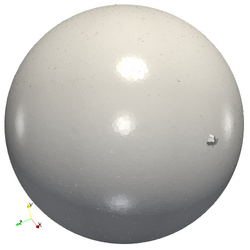} \\
 \multirow{-2}{*}[4.5em]{$2^{-8}$} &
  $\mathrlap{\hexstar}\hexagon$ &
  \includegraphics[height=0.1525\textwidth,valign=c]{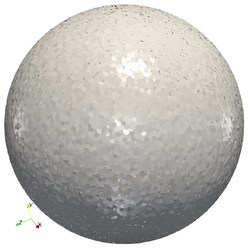} &
  \includegraphics[height=0.1525\textwidth,valign=c]{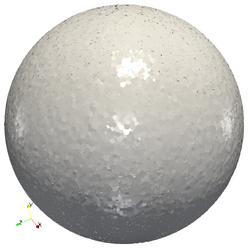} &
  \includegraphics[height=0.1525\textwidth,valign=c]{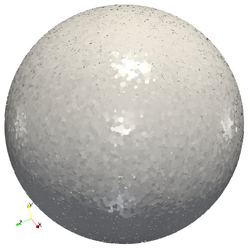} &
  \includegraphics[height=0.1525\textwidth,valign=c]{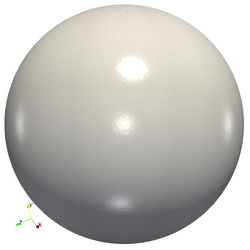} \\ \hline
\end{tabular}%
\end{table}

\subsubsection{Orientation schemes in various mesh types}

The reconstructed interface planes for these four orientation schemes across different mesh types are showcased in \cref{tab_reconstruction_plic_surfraces_tet_meshes,tab_reconstruction_plic_surfraces_hex_meshes,tab_reconstruction_plic_surfraces_blockPoly_meshes}. The CAG method, due to its suboptimal orientation evaluations, results in unsmoothed interface planes in both tetrahedral and hexahedral meshes. As the number of faces per cell increases, the CAG scheme begins to produce interface planes akin to those generated by the NAG and LS methods, observable in both general and structured polyhedral meshes. The NAG and LS schemes consistently yield similar interface planes across all mesh types. In contrast, the RDF method demonstrates a remarkable capability to produce exceptionally smooth interface planes.

\begin{table}[htbp]
\footnotesize
\centering
\caption{Reconstructed interfaces at $t=0\,s$ in tetrahedral meshes.}
\label{tab_reconstruction_plic_surfraces_tet_meshes}
\renewcommand{\arraystretch}{1.15}
\begin{tabular}{c|c|c|c|c}
\hline
\rowcolor[HTML]{EFEFEF} 
  $\Delta s\,[m]$ &
  CAG &
  NAG &
  LS &
  RDF \\ \hline
  $2^{-5}$ &
  \includegraphics[height=0.1525\textwidth,valign=c]{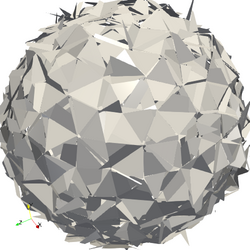} &
  \includegraphics[height=0.1525\textwidth,valign=c]{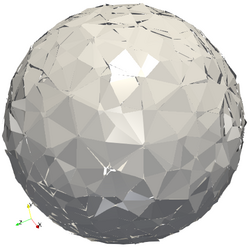} &
  \includegraphics[height=0.1525\textwidth,valign=c]{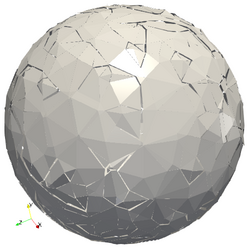} &
  \includegraphics[height=0.1525\textwidth,valign=c]{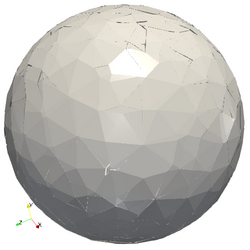} \\ \hline
  $2^{-6}$ &
  \includegraphics[height=0.1525\textwidth,valign=c]{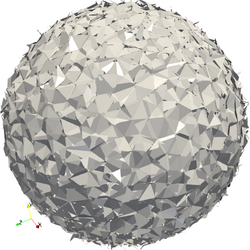} &
  \includegraphics[height=0.1525\textwidth,valign=c]{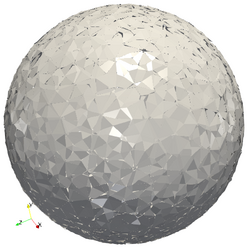} &
  \includegraphics[height=0.1525\textwidth,valign=c]{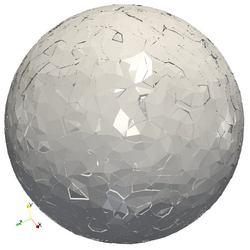} &
  \includegraphics[height=0.1525\textwidth,valign=c]{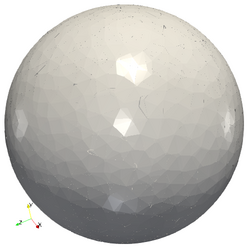} \\ \hline
  $2^{-7}$ &
  \includegraphics[height=0.1525\textwidth,valign=c]{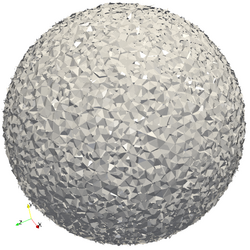} &
  \includegraphics[height=0.1525\textwidth,valign=c]{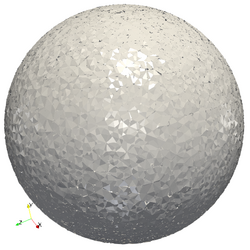} &
  \includegraphics[height=0.1525\textwidth,valign=c]{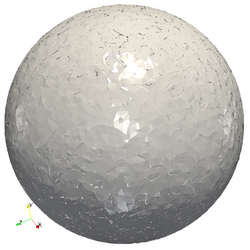} &
  \includegraphics[height=0.1525\textwidth,valign=c]{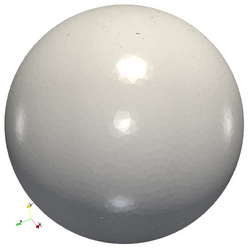} \\ \hline
  $2^{-8}$ &
  \includegraphics[height=0.1525\textwidth,valign=c]{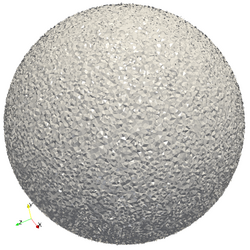} &
  \includegraphics[height=0.1525\textwidth,valign=c]{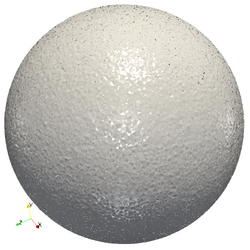} &
  \includegraphics[height=0.1525\textwidth,valign=c]{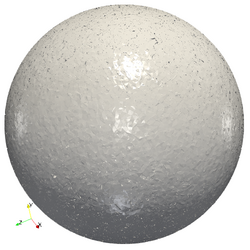} &
  \includegraphics[height=0.1525\textwidth,valign=c]{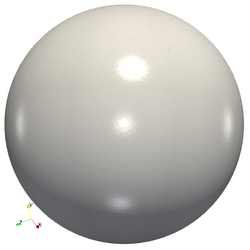} \\ \hline
\end{tabular}%
\end{table}

\begin{table}[htbp]
\footnotesize
\centering
\caption{Reconstructed interfaces at $t=0\,s$ in hexahedral meshes.}
\label{tab_reconstruction_plic_surfraces_hex_meshes}
\renewcommand{\arraystretch}{1.15}
\begin{tabular}{c|c|c|c|c}
\hline
\rowcolor[HTML]{EFEFEF} 
  $\Delta s\,[m]$ &
  CAG &
  NAG &
  LS &
  RDF \\ \hline
  $2^{-5}$ &
  \includegraphics[height=0.1525\textwidth,valign=c]{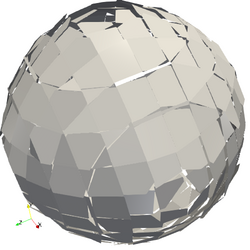} &
  \includegraphics[height=0.1525\textwidth,valign=c]{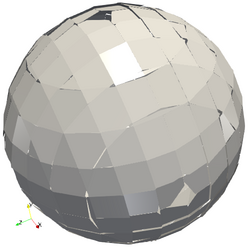} &
  \includegraphics[height=0.1525\textwidth,valign=c]{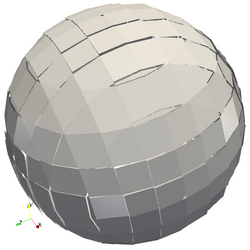} &
  \includegraphics[height=0.1525\textwidth,valign=c]{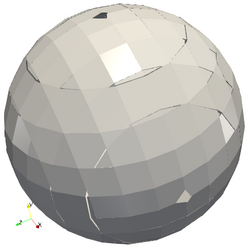} \\ \hline
  $2^{-6}$ &
  \includegraphics[height=0.1525\textwidth,valign=c]{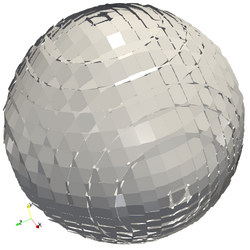} &
  \includegraphics[height=0.1525\textwidth,valign=c]{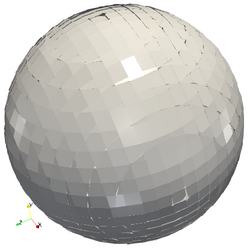} &
  \includegraphics[height=0.1525\textwidth,valign=c]{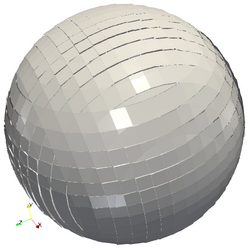} &
  \includegraphics[height=0.1525\textwidth,valign=c]{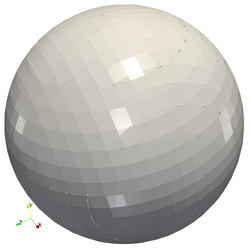} \\ \hline
  $2^{-7}$ &
  \includegraphics[height=0.1525\textwidth,valign=c]{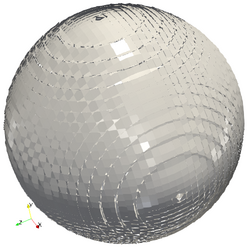} &
  \includegraphics[height=0.1525\textwidth,valign=c]{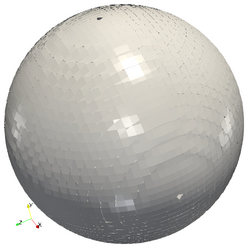} &
  \includegraphics[height=0.1525\textwidth,valign=c]{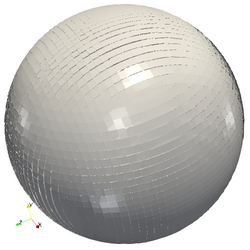} &
  \includegraphics[height=0.1525\textwidth,valign=c]{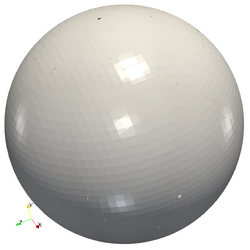} \\ \hline
  $2^{-8}$ &
  \includegraphics[height=0.1525\textwidth,valign=c]{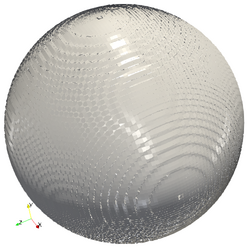} &
  \includegraphics[height=0.1525\textwidth,valign=c]{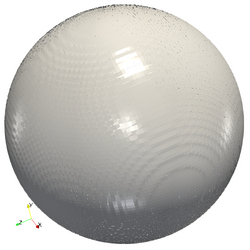} &
  \includegraphics[height=0.1525\textwidth,valign=c]{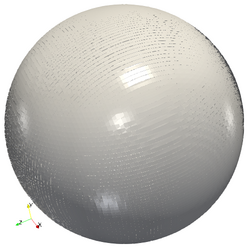} &
  \includegraphics[height=0.1525\textwidth,valign=c]{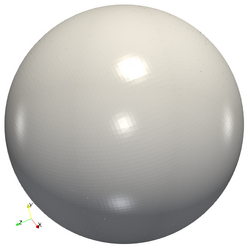} \\ \hline
\end{tabular}%
\end{table}

\begin{table}[htbp]
\footnotesize
\centering
\caption{Reconstructed interfaces at $t=0\,s$ in structured polyhedral meshes meshes.}
\label{tab_reconstruction_plic_surfraces_blockPoly_meshes}
\renewcommand{\arraystretch}{1.15}
\begin{tabular}{c|c|c|c|c}
\hline
\rowcolor[HTML]{EFEFEF} 
  $\Delta s\,[m]$ &
  CAG &
  NAG &
  LS &
  RDF \\ \hline
  $2^{-5}$ &
  \includegraphics[height=0.1525\textwidth,valign=c]{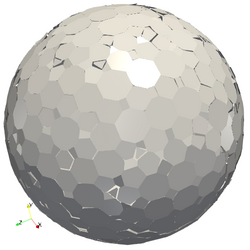} &
  \includegraphics[height=0.1525\textwidth,valign=c]{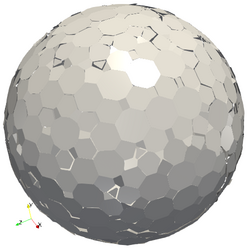} &
  \includegraphics[height=0.1525\textwidth,valign=c]{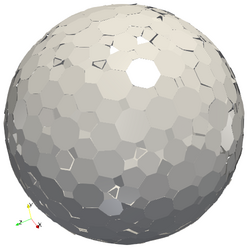} &
  \includegraphics[height=0.1525\textwidth,valign=c]{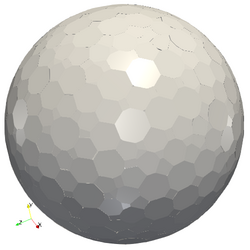} \\ \hline
  $2^{-6}$ &
  \includegraphics[height=0.1525\textwidth,valign=c]{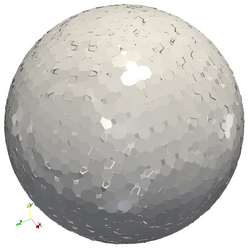} &
  \includegraphics[height=0.1525\textwidth,valign=c]{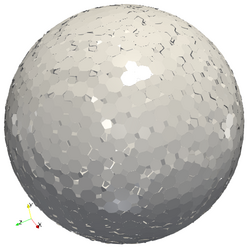} &
  \includegraphics[height=0.1525\textwidth,valign=c]{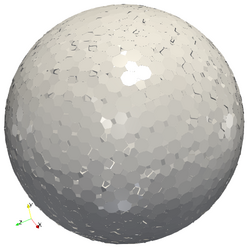} &
  \includegraphics[height=0.1525\textwidth,valign=c]{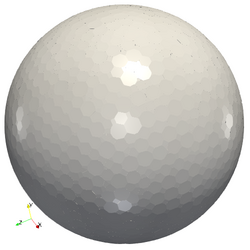} \\ \hline
  $2^{-7}$ &
  \includegraphics[height=0.1525\textwidth,valign=c]{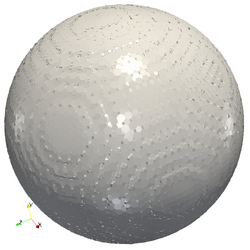} &
  \includegraphics[height=0.1525\textwidth,valign=c]{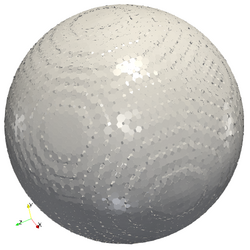} &
  \includegraphics[height=0.1525\textwidth,valign=c]{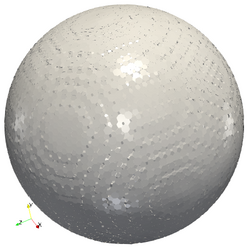} &
  \includegraphics[height=0.1525\textwidth,valign=c]{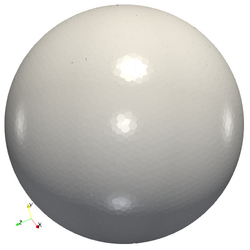} \\ \hline
  $2^{-8}$ &
  \includegraphics[height=0.1525\textwidth,valign=c]{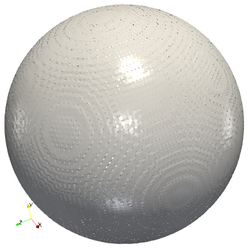} &
  \includegraphics[height=0.1525\textwidth,valign=c]{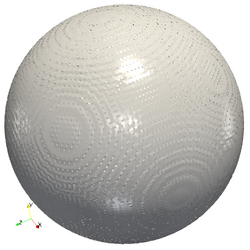} &
  \includegraphics[height=0.1525\textwidth,valign=c]{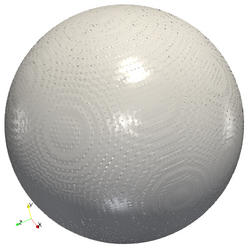} &
  \includegraphics[height=0.1525\textwidth,valign=c]{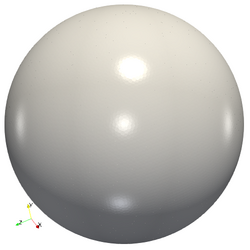} \\ \hline
\end{tabular}%
\end{table}

\begin{figure}[htbp]
    \centering
    \begin{subfigure}[b]{0.4\textwidth}
        \centering
        \includegraphics[width=\textwidth]{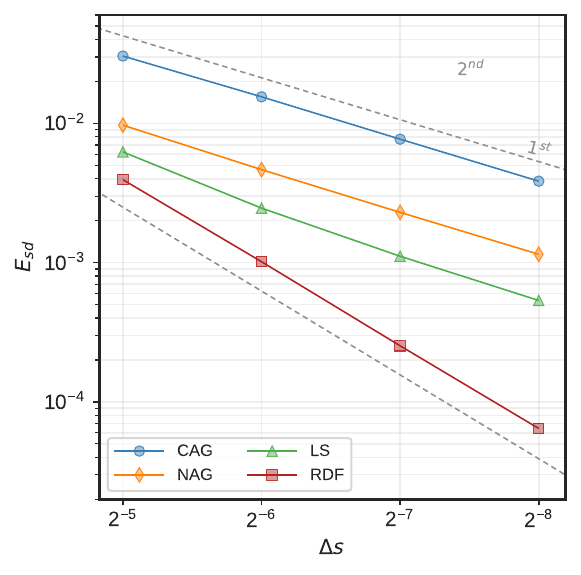}
        \caption{Tetrahedral meshes}
        \label{fig_err_symmDiff_orientation_schemes_tet_meshes}
    \end{subfigure}
    \begin{subfigure}[b]{0.4\textwidth}
        \centering
        \includegraphics[width=\textwidth]{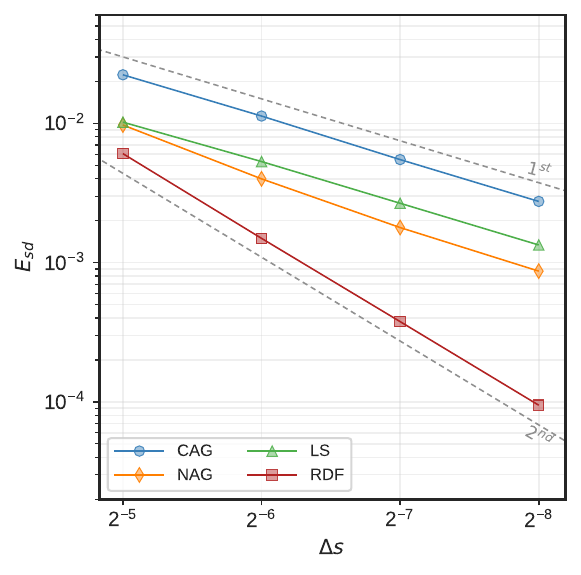}
        \caption{Hexahedral meshes}
        \label{fig_err_symmDiff_orientation_schemes_hex_meshes}
    \end{subfigure}
    \begin{subfigure}[b]{0.4\textwidth}
        \centering
        \includegraphics[width=\textwidth]{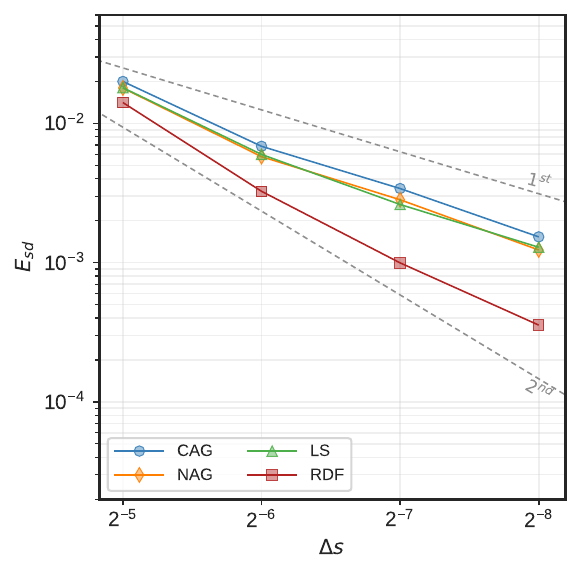}
        \caption{Polyhedral meshes}
        \label{fig_err_symmDiff_orientation_schemes_poly_meshes}
    \end{subfigure}
    \begin{subfigure}[b]{0.4\textwidth}
        \centering
        \includegraphics[width=\textwidth]{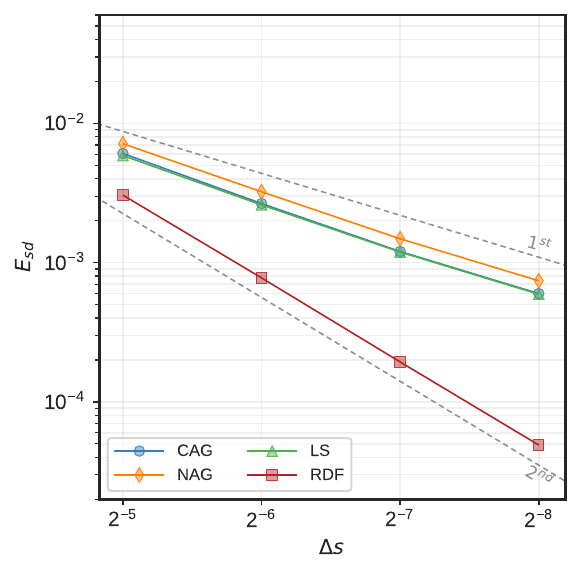}
        \caption{Structured polyhedral meshes}
        \label{fig_err_symmDiff_orientation_schemes_spoly_meshes}
    \end{subfigure}
    \caption{Symmetric difference errors $E_{sd}$ of various orientation schemes.}
    \label{fig_err_symmDiff_orientation_schemes}
\end{figure}

\cref{fig_err_symmDiff_orientation_schemes} provides a comparative analysis of $E_{sd}$ for the different orientation schemes applied to various mesh types. The results of polyhedral meshes, as seen in \cref{fig_err_symmDiff_orientation_schemes_poly_meshes}, are derived using triangular decomposition of warped cell faces. In addition, \cref{tab_average_convergence_orders} summarizes the average orders of convergence, denoted as $\bar{\mathcal{O}}\left(E_{sd}\right)$, which are evaluated as follows:
\begin{equation}
    \bar{\mathcal{O}}\left(E_{sd}\right) = \ln{\frac{E_{sd}(\Delta s=2^{-5})}{E_{sd}(\Delta s=2^{-8})}} \bigg/ \left(3 \ln{2}\right).
    \label{eq_avg_convergence_order}
\end{equation}

The RDF scheme invariably provides the most accurate orientation evaluations and reduces $E_{sd}$ with approximately second-order accuracy across all mesh types. The NAG scheme typically surpasses the CAG method in shape preservation, except in structured polyhedral meshes where the errors of the CAG and LS schemes are similar. Although the LS method performs well in tetrahedral meshes, its improvements are not as significant in polyhedral meshes, and it even falls behind the NAG method in hexahedral meshes. The CAG, NAG and LS schemes exhibit approximately first-order accuracy in $E_{sd}$.

\setlength\MAX{3.25mm}
\renewcommand*\Chart[1]{#1~\rlap{\textcolor{blue!20}{\rule{2.02\MAX}{1.5ex}}}\textcolor{blue!70}{\rule{#1\MAX}{1.5ex}}}

\begin{table}[htbp]
\footnotesize
\centering
\caption{Average convergence orders $\bar{\mathcal{O}}\left(E_{sd}\right)$ of all meshes.}
\label{tab_average_convergence_orders}
\renewcommand{\arraystretch}{1.15}
\begin{tabular}{c|m{0.07\textwidth}|m{0.07\textwidth}|m{0.07\textwidth}|m{0.07\textwidth}}
\rowcolor[HTML]{EFEFEF}
\hline
Meshes &
  \multicolumn{1}{c|}{\cellcolor[HTML]{EFEFEF}CAG} &
  \multicolumn{1}{c|}{\cellcolor[HTML]{EFEFEF}NAG} &
  \multicolumn{1}{c|}{\cellcolor[HTML]{EFEFEF}LS} &
  \multicolumn{1}{c}{\cellcolor[HTML]{EFEFEF}RDF} \\ \hline
Tetrahedral           & \Chart{0.99} & \Chart{1.03} & \Chart{1.18} & \Chart{1.98} \\ \hline
Hexahedral            & \Chart{1.01} & \Chart{1.16} & \Chart{0.98} & \Chart{2.00} \\ \hline
General polyhedral    & \Chart{1.24} & \Chart{1.29} & \Chart{1.27} & \Chart{1.77} \\ \hline
Structured polyhedral & \Chart{1.11} & \Chart{1.09} & \Chart{1.10} & \Chart{1.99} \\ \hline
\end{tabular}
\end{table}

\subsection{Interface advection}\label{ss_interface_advection_results}

The spherical interface is advected according to the velocity specified in \cref{eq_U_field}. For the testing, mesh resolutions of $\Delta s = 2^{-6}$ and $2^{-7}$ are employed across all mesh types. In these tests, the fraction clipping step is not activated and the mixed cell tolerance $\epsilon$ is fixed to $10^{-8}$. In the simulations, the time step values are regulated to ensure that both the global and interface Courant numbers do not exceed 0.5, denoted as $Co \leq 0.5$ and $Co_i \leq 0.5$, respectively. Initially, the effects of warped face decomposition in general polyhedral meshes are examined, followed by evaluations of different orientation schemes. Finally, the newly proposed SimPLIC method is benchmarked against the officially released PLIC-VOF methods in \OF\,v2312.

\subsubsection{Impacts of warped face decomposition in general polyhedral meshes}

As indicated in \cref{tab_advection_plic_surfraces_warped_face_decomp}, the implementation of warped face decomposition in general polyhedral meshes does not lead to significantly different reconstructed interfaces at $t=1.5\,s$ and $t=3\,s$. The volume conservation error $E_v(t)$, the extremal fraction bounding values $\alpha_{min}(t)$ and $1-\alpha_{max}(t)$, and the shape error $E_s(t)$ are sampled every $0.25\,s$. Their absolute maximum ("amax") and average ("avg") values are compiled in \cref{tab_advection_performance_data_warped_face_decomp}. This table also provides details on the interface reconstruction, advection, and total simulation times, denoted as $T_{rec}$, $T_{adv}$, and $T_{calc}$, respectively. Similar to the observations from the visualized interface planes, no notable deviations are detected, except for the fact that the warped face decomposition considerably increases both $T_{rec}$ and $T_{calc}$.

\begin{table}[htbp]
\footnotesize
\centering
\caption{Reconstructed interfaces at $t=1.5\,s$ (\textit{gray}) and $t=3\,s$ (\textit{red}) with ($\mathrlap{\hexstar}\hexagon$) and without ($\hexagon$) warped face decomposition in general polyhedral meshes.}
\label{tab_advection_plic_surfraces_warped_face_decomp}
\renewcommand{\arraystretch}{1.15}
\begin{tabular}{c|c|c|c|c|c}
\hline
\rowcolor[HTML]{EFEFEF} 
$\Delta s\,[m]$ &
  $\hexagon$/$\mathrlap{\hexstar}\hexagon$ &
  CAG &
  NAG &
  LS &
  RDF \\ \hline
 &
  $\hexagon$ &
  \includegraphics[height=0.16\textwidth,valign=c]{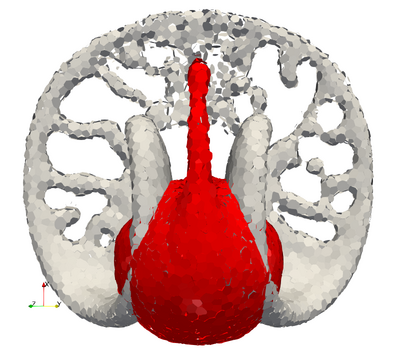} &
  \includegraphics[height=0.16\textwidth,valign=c]{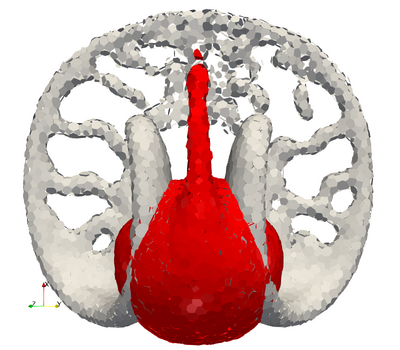} &
  \includegraphics[height=0.16\textwidth,valign=c]{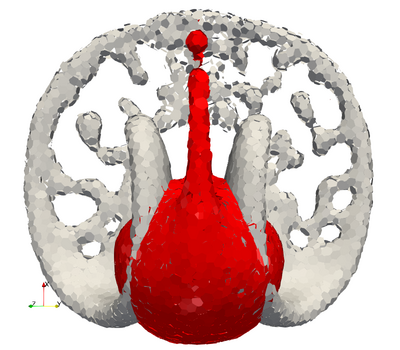} &
  \includegraphics[height=0.16\textwidth,valign=c]{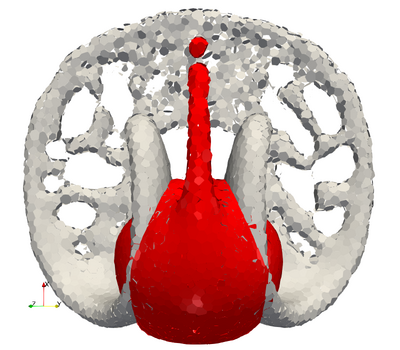} \\
 \multirow{-2}{*}[4.5em]{$2^{-6}$} &
  $\mathrlap{\hexstar}\hexagon$ &
  \includegraphics[height=0.16\textwidth,valign=c]{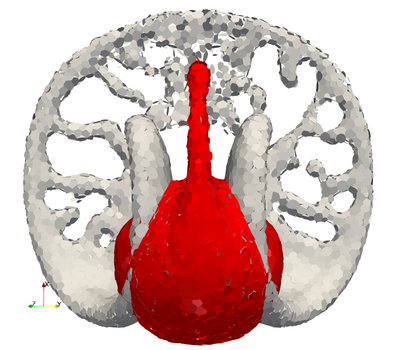} &
  \includegraphics[height=0.16\textwidth,valign=c]{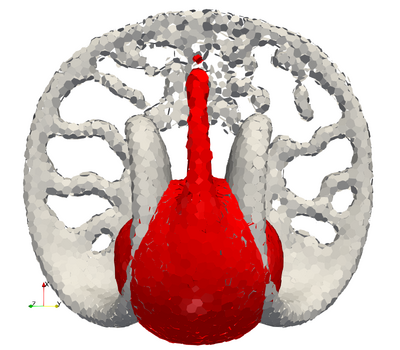} &
  \includegraphics[height=0.16\textwidth,valign=c]{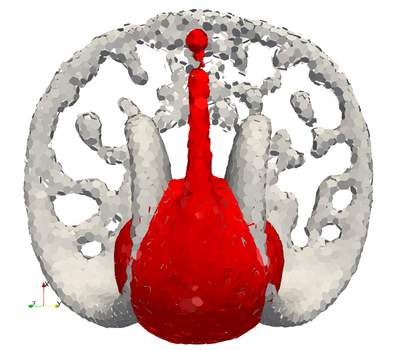} &
  \includegraphics[height=0.16\textwidth,valign=c]{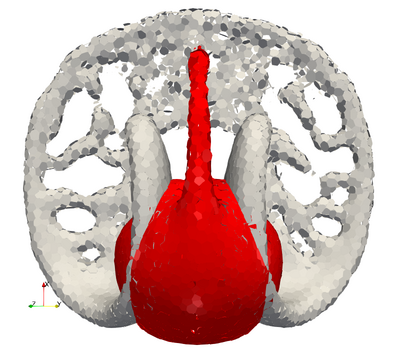} \\ \hline
 &
  $\hexagon$ &
  \includegraphics[height=0.16\textwidth,valign=c]{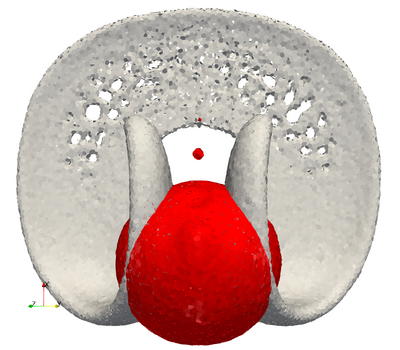} &
  \includegraphics[height=0.16\textwidth,valign=c]{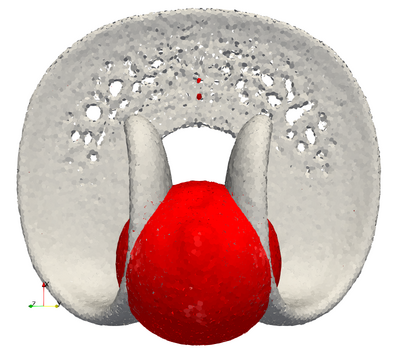} &
  \includegraphics[height=0.16\textwidth,valign=c]{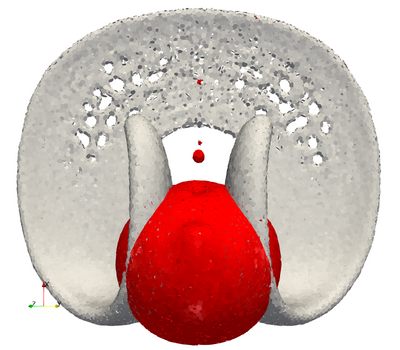} &
  \includegraphics[height=0.16\textwidth,valign=c]{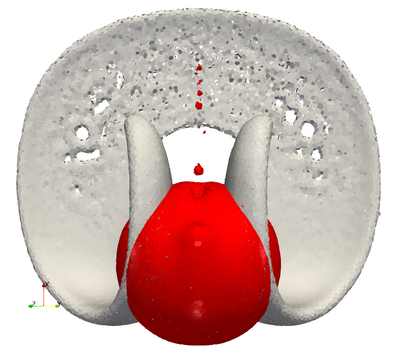} \\
 \multirow{-2}{*}[4.5em]{$2^{-7}$} &
  $\mathrlap{\hexstar}\hexagon$ &
  \includegraphics[height=0.16\textwidth,valign=c]{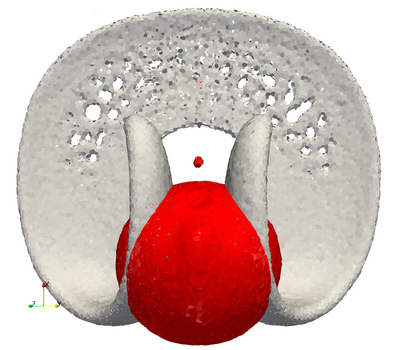} &
  \includegraphics[height=0.16\textwidth,valign=c]{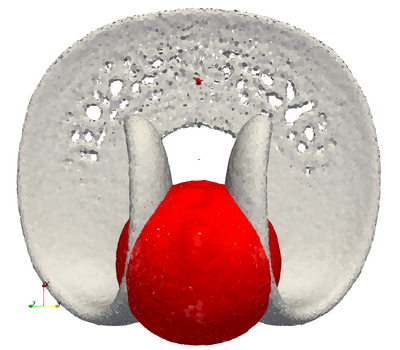} &
  \includegraphics[height=0.16\textwidth,valign=c]{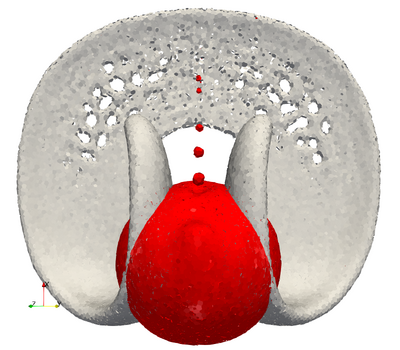} &
  \includegraphics[height=0.16\textwidth,valign=c]{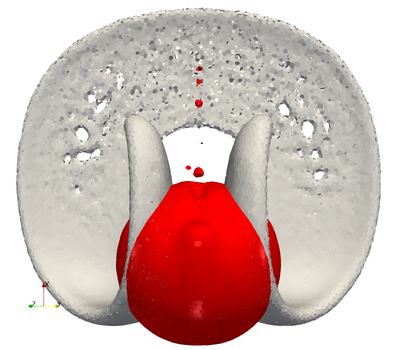} \\ \hline
\end{tabular}%
\end{table}

The data for $E_v(t)$ in \cref{tab_advection_performance_data_warped_face_decomp} shows that volume conservation is maintained in general polyhedral meshes, irrespective of whether warped face decomposition is implemented. The fraction field remains numerically bounded at its lower limit, but this is not the case for the upper limit. The additional process of warped face decomposition has minimal impact on $T_{adv}$, but it results in significant increases in $T_{rec}$ for different orientation schemes: approximately $165.4\%$, $62.8\%$, $205.5\%$, and $207.1\%$ for the CAG, NAG, LS, and RDF schemes, respectively. This outcome is anticipated since the face decomposition and the addition of extra cell vertices inevitably require additional CPU times of local memory copies and interface location determinations. The increases in $T_{calc}$ observed with warped face decomposition are primarily due to the corresponding increases in $T_{rec}$. Hence, while the warped face decomposition in general polyhedral meshes does not affect the accuracy of interface advections, it substantially reduces the computational efficiency.

\begin{table}[htbp]
\footnotesize
\centering
\caption{Errors and execution times at $t=3\,s$ with ($\mathrlap{\hexstar}\hexagon$) and without ($\hexagon$) warped face decomposition in general polyhedral meshes.}
\label{tab_advection_performance_data_warped_face_decomp}
\resizebox{\textwidth}{!}{%
\renewcommand{\arraystretch}{1.15}
\begin{tabular}{c|c|c|c|c|c|c|c|c|c|c|r|r|r}
\hline
\rowcolor[HTML]{EFEFEF} 
\multicolumn{1}{c|}{\cellcolor[HTML]{EFEFEF}} &
  \multicolumn{1}{c|}{\cellcolor[HTML]{EFEFEF}} &
  \multicolumn{1}{c|}{\cellcolor[HTML]{EFEFEF}} &
  \multicolumn{2}{c|}{\cellcolor[HTML]{EFEFEF}$E_v(t)$} &
  \multicolumn{2}{c|}{\cellcolor[HTML]{EFEFEF}$\alpha_{min}(t)$} &
  \multicolumn{2}{c|}{\cellcolor[HTML]{EFEFEF}1-$\alpha_{max}(t)$} &
  \multicolumn{2}{c|}{\cellcolor[HTML]{EFEFEF}$E_s(t)$} &
  \multicolumn{1}{c|}{\cellcolor[HTML]{EFEFEF}} &
  \multicolumn{1}{c|}{\cellcolor[HTML]{EFEFEF}} &
  \multicolumn{1}{c}{\cellcolor[HTML]{EFEFEF}} \\ \cline{4-11}
\rowcolor[HTML]{EFEFEF} 
\multicolumn{1}{c|}{\multirow{-2}{*}{\cellcolor[HTML]{EFEFEF}\begin{tabular}[c]{@{}c@{}}$\Delta s$\\ $[m]$\end{tabular}}} &
  \multicolumn{1}{c|}{\multirow{-2}{*}{\cellcolor[HTML]{EFEFEF}Scheme}} &
  \multicolumn{1}{c|}{\multirow{-2}{*}{\cellcolor[HTML]{EFEFEF}$\hexagon$/$\mathrlap{\hexstar}\hexagon$}} &
  \multicolumn{1}{c|}{\cellcolor[HTML]{EFEFEF}amax} &
  \multicolumn{1}{c|}{\cellcolor[HTML]{EFEFEF}avg} &
  \multicolumn{1}{c|}{\cellcolor[HTML]{EFEFEF}amax} &
  \multicolumn{1}{c|}{\cellcolor[HTML]{EFEFEF}avg} &
  \multicolumn{1}{c|}{\cellcolor[HTML]{EFEFEF}amax} &
  \multicolumn{1}{c|}{\cellcolor[HTML]{EFEFEF}avg} &
  \multicolumn{1}{c|}{\cellcolor[HTML]{EFEFEF}amax} &
  \multicolumn{1}{c|}{\cellcolor[HTML]{EFEFEF}avg} &
  \multicolumn{1}{c|}{\multirow{-2}{*}{\cellcolor[HTML]{EFEFEF}\begin{tabular}[c]{@{}c@{}}$T_{rec}$\\ $[s]$\end{tabular}}} &
  \multicolumn{1}{c|}{\multirow{-2}{*}{\cellcolor[HTML]{EFEFEF}\begin{tabular}[c]{@{}c@{}}$T_{adv}$\\ $[s]$\end{tabular}}} &
  \multicolumn{1}{c}{\multirow{-2}{*}{\cellcolor[HTML]{EFEFEF}\begin{tabular}[c]{@{}c@{}}$T_{calc}$\\ $[s]$\end{tabular}}} \\ \hline
 &
  &
   $\hexagon$ &
 $1.87 \times 10^{-11}$ &
 $1.85 \times 10^{-11}$ &
 $-3.44 \times 10^{-18}$ &
 $-4.09 \times 10^{-19}$ &
 $-3.97 \times 10^{-02}$ &
 $-1.94 \times 10^{-02}$ &
 0.295 &
 0.204 &
 125.6 &
 99.4 &
 630.9 \\
&
 \multirow{-2}{*}{CAG} &
 $\mathrlap{\hexstar}\hexagon$ &
 $1.87 \times 10^{-11}$ &
 $1.85 \times 10^{-11}$ &
 $-4.88 \times 10^{-19}$ &
 $-9.72 \times 10^{-20}$ &
 $-3.55 \times 10^{-02}$ &
 $-2.01 \times 10^{-02}$ &
 0.297 &
 0.204 &
 348.3 &
 98.9 &
 851.3 \\
&
  &
 $\hexagon$ &
 $1.87 \times 10^{-11}$ &
 $1.85 \times 10^{-11}$ &
 $-2.39 \times 10^{-18}$ &
 $-3.92 \times 10^{-19}$ &
 $-3.74 \times 10^{-02}$&
 $-2.02 \times 10^{-02}$&
 0.298 &
 0.205 &
 276.1 &
 97.4 &
 778.3 \\
&
 \multirow{-2}{*}{NAG} &
 $\mathrlap{\hexstar}\hexagon$ &
 $1.88 \times 10^{-11}$ &
 $1.85 \times 10^{-11}$ &
 $-2.32 \times 10^{-18}$ &
 $-2.73 \times 10^{-19}$ &
 $-3.60 \times 10^{-02}$ &
 $-2.00 \times 10^{-02}$ &
 0.299 &
 0.205 &
 493.4 &
 96.2 &
 994.8 \\
&
  &
 $\hexagon$ &
 $1.87 \times 10^{-11}$ &
 $1.85 \times 10^{-11}$ &
 $-4.61 \times 10^{-18}$ &
 $-5.29 \times 10^{-19}$ &
 $-3.80 \times 10^{-02}$&
 $-2.05 \times 10^{-02}$&
 0.311 &
 0.215 &
 108.4 &
 98.3 &
 603.1 \\
&
 \multirow{-2}{*}{LS} &
 $\mathrlap{\hexstar}\hexagon$ &
 $1.88 \times 10^{-11}$ &
 $1.85 \times 10^{-11}$ &
 $-6.30 \times 10^{-19}$ &
 $-1.52 \times 10^{-19}$ &
 $-3.38 \times 10^{-02}$ &
 $-1.98 \times 10^{-02}$ &
 0.312 &
 0.217 &
 335.4 &
 100.1 &
 831.5 \\
&
  &
 $\hexagon$ &
 $1.87 \times 10^{-11}$ &
 $1.85 \times 10^{-11}$ &
 $-1.06 \times 10^{-18}$ &
 $-2.07 \times 10^{-19}$ &
 $-3.18 \times 10^{-02}$&
 $-1.83 \times 10^{-02}$&
 0.305 &
 0.209 &
 325.4 &
 98.3 &
 818.5 \\
\multirow{-8}{*}{$2^{-6}$} &
 \multirow{-2}{*}{RDF} &
 $\mathrlap{\hexstar}\hexagon$ &
 $1.87 \times 10^{-11}$ &
 $1.85 \times 10^{-11}$ &
 $-1.10 \times 10^{-18}$ &
 $-2.86 \times 10^{-19}$ &
 $-2.97 \times 10^{-02}$ &
 $-1.83 \times 10^{-02}$ &
 0.301 &
 0.209 &
 1008.7 &
 99.0 &
 1503.0 \\ \hline
&
  &
 $\hexagon$ &
 $1.84 \times 10^{-11}$ &
 $1.70 \times 10^{-11}$ &
 $-8.67 \times 10^{-19}$ &
 $-2.41 \times 10^{-19}$ &
 $-9.65 \times 10^{-02}$&
 $-4.83 \times 10^{-02}$&
 0.092 &
 0.064 &
 1485.2 &
 1661.0 &
 11242.0 \\
&
 \multirow{-2}{*}{CAG} &
 $\mathrlap{\hexstar}\hexagon$ &
 $1.84 \times 10^{-11}$ &
 $1.70 \times 10^{-11}$ &
 $-4.40 \times 10^{-19}$ &
 $-1.48 \times 10^{-19}$ &
 $-8.95 \times 10^{-02}$ &
 $-4.63 \times 10^{-02}$ &
 0.095 &
 0.066 &
 3764.7 &
 1671.2 &
 13535.4 \\
&
  &
 $\hexagon$ &
 $1.84 \times 10^{-11}$ &
 $1.70 \times 10^{-11}$ &
 $-1.82 \times 10^{-18}$ &
 $-3.51 \times 10^{-19}$ &
 $-7.39 \times 10^{-02}$&
 $-4.72 \times 10^{-02}$&
 0.091 &
 0.064 &
 4855.1 &
 1660.0 &
 14581.6 \\
&
 \multirow{-2}{*}{NAG} &
 $\mathrlap{\hexstar}\hexagon$ &
 $1.84 \times 10^{-11}$ &
 $1.70 \times 10^{-11}$ &
 $-8.13 \times 10^{-19}$ &
 $-1.62 \times 10^{-19}$ &
 $-8.80 \times 10^{-02}$ &
 $-4.71 \times 10^{-02}$ &
 0.092 &
 0.065 &
 7131.7 &
 1668.1 &
 16866.0 \\
&
  &
 $\hexagon$ &
 $1.84 \times 10^{-11}$ &
 $1.70 \times 10^{-11}$ &
 $-1.13 \times 10^{-18}$ &
 $-3.37 \times 10^{-19}$ &
 $-9.58 \times 10^{-02}$&
 $-4.97 \times 10^{-02}$&
 0.091 &
 0.065 &
 1173.0 &
 1663.1 &
 10906.4 \\
&
 \multirow{-2}{*}{LS} &
 $\mathrlap{\hexstar}\hexagon$ &
 $1.84 \times 10^{-11}$ &
 $1.70 \times 10^{-11}$ &
 $-5.15 \times 10^{-19}$ &
 $-1.62 \times 10^{-19}$ &
 $-9.15 \times 10^{-02}$ &
 $-4.56 \times 10^{-02}$ &
 0.092 &
 0.066 &
 3537.9 &
 1691.0 &
 13315.8 \\
&
  &
 $\hexagon$ &
 $1.84 \times 10^{-11}$ &
 $1.70 \times 10^{-11}$ &
 $-1.79 \times 10^{-18}$ &
 $-4.33 \times 10^{-19}$ &
 $-8.08 \times 10^{-02}$&
 $-4.18 \times 10^{-02}$&
 0.098 &
 0.068 &
 3615.6 &
 1696.4 &
 13377.9 \\
\multirow{-8}{*}{$2^{-7}$} &
 \multirow{-2}{*}{RDF} &
 $\mathrlap{\hexstar}\hexagon$ &
 $1.84 \times 10^{-11}$ &
 $1.70 \times 10^{-11}$ &
 $-9.08 \times 10^{-19}$ &
 $-1.81 \times 10^{-19}$ &
 $-8.26 \times 10^{-02}$ &
 $-3.95 \times 10^{-02}$ &
 0.100 &
 0.070 &
 11004.0 &
 1705.6 &
 20768.0 \\ \hline
\end{tabular}
}
\end{table}

\subsubsection{Impacts of orientation schemes}

The reconstructed interface planes at $t=1.5\,s$ and $t=3\,s$, as well as the corresponding errors and execution times at $t=3\,s$ with various orientation schemes are presented in \cref{tab_advection_plic_surfraces_various_orientation_schemes,tab_advection_performance_data_various_orientation_schemes}, respectively. The simulations are conducted on the four mesh types, each with two different resolutions. It should be noted that the function of warped face decomposition is not activated in the cases involving general polyhedral meshes.

\cref{tab_advection_plic_surfraces_various_orientation_schemes} reveals that the reconstructed interface planes are inconsistent when using the CAG scheme in both tetrahedral and hexahedral meshes. This inconsistency stems from the poor accuracy of the CAG method in these mesh types, leading to the primary phase being transported to unintended cells. The NAG, LS, and RDF methods demonstrate an improvement in the quality of reconstructed interfaces for these two mesh types and do not exhibit significant differences in the resulting interfaces. In both general and structured polyhedral meshes, no notable differences in reconstructed interfaces are observed across these four orientation schemes.

\begin{table}[htbp]
\footnotesize
\centering
\caption{Reconstructed interfaces at $t=1.5\,s$ (\textit{gray}) and $t=3\,s$ (\textit{red}) with various orientation schemes.}
\label{tab_advection_plic_surfraces_various_orientation_schemes}
\renewcommand{\arraystretch}{1.15}
\begin{tabular}{c|c|c|c|c|c}
\hline
\rowcolor[HTML]{EFEFEF} 
$\Delta s\,[m]$ &
  Mesh &
  CAG &
  NAG &
  LS &
  RDF \\ \hline
 &
  \rotatebox[origin=c]{90}{Tetrahedral} &
  \includegraphics[height=0.16\textwidth,valign=c]{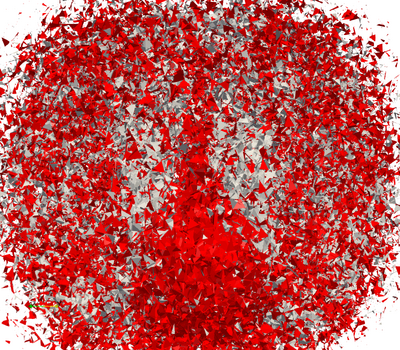} &
  \includegraphics[height=0.16\textwidth,valign=c]{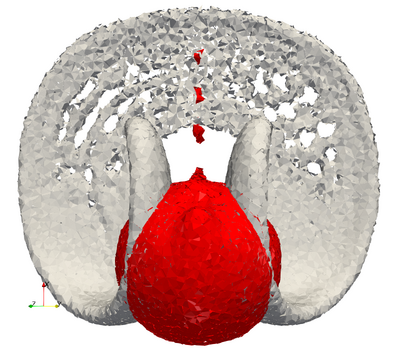} &
  \includegraphics[height=0.16\textwidth,valign=c]{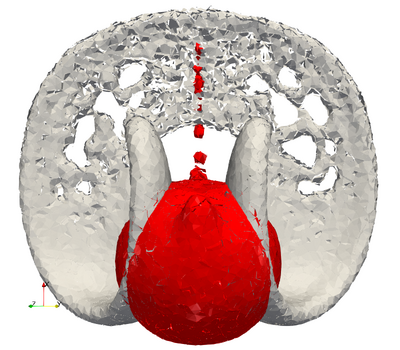} &
  \includegraphics[height=0.16\textwidth,valign=c]{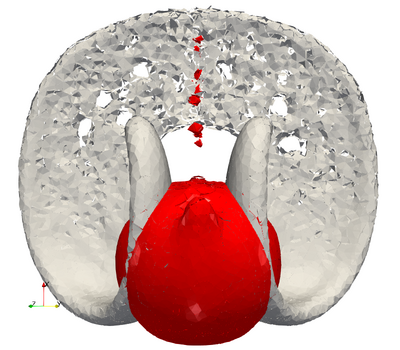} \\
 &
  \rotatebox[origin=c]{90}{Hexahedral} &
  \includegraphics[height=0.16\textwidth,valign=c]{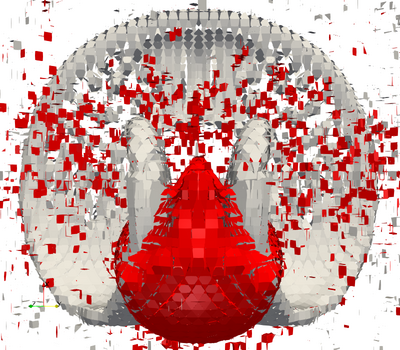} &
  \includegraphics[height=0.16\textwidth,valign=c]{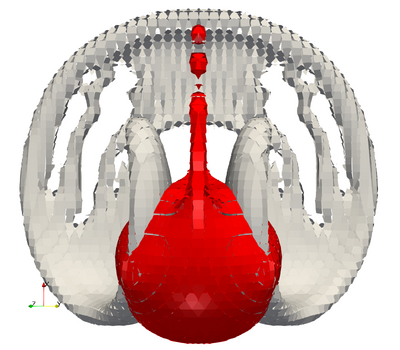} &
  \includegraphics[height=0.16\textwidth,valign=c]{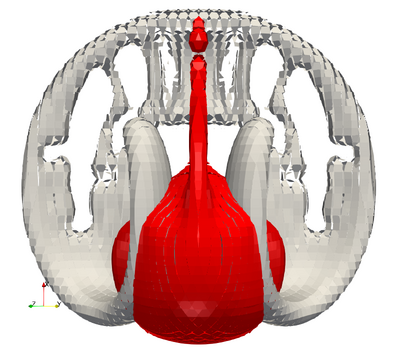} &
  \includegraphics[height=0.16\textwidth,valign=c]{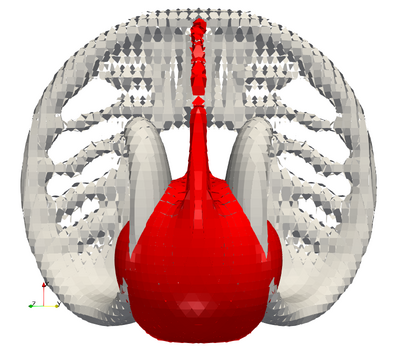} \\
 &
  \rotatebox[origin=c]{90}{\shortstack{General\\polyhedral}} &
  \includegraphics[height=0.16\textwidth,valign=c]{interface_polyMesh_plicVof_CAG_mesh_3.png} &
  \includegraphics[height=0.16\textwidth,valign=c]{interface_polyMesh_plicVof_NAG_mesh_3.png} &
  \includegraphics[height=0.16\textwidth,valign=c]{interface_polyMesh_plicVof_LS_mesh_3.png} &
  \includegraphics[height=0.16\textwidth,valign=c]{interface_polyMesh_plicVof_RDF_mesh_3.png} \\
\multirow{-4}{*}[13em]{$2^{-6}$} &
  \rotatebox[origin=c]{90}{\shortstack{Structured\\polyhedral}} &
  \includegraphics[height=0.16\textwidth,valign=c]{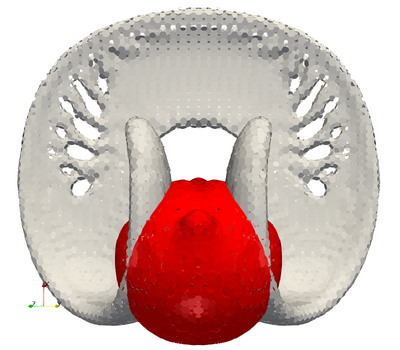} &
  \includegraphics[height=0.16\textwidth,valign=c]{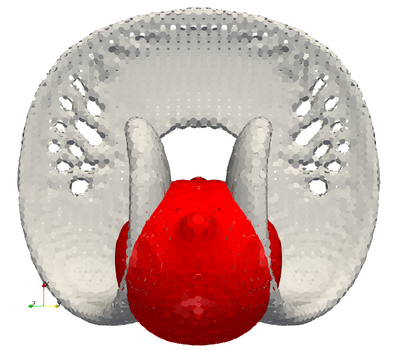} &
  \includegraphics[height=0.16\textwidth,valign=c]{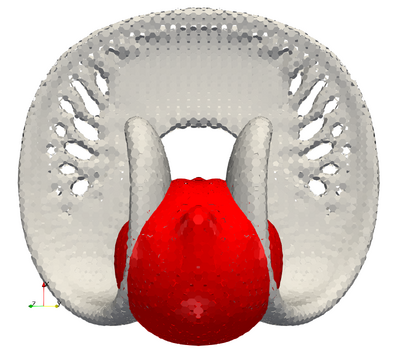} &
  \includegraphics[height=0.16\textwidth,valign=c]{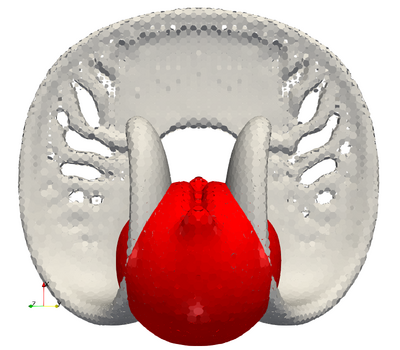} \\ \hline
 &
  \rotatebox[origin=c]{90}{Tetrahedral} &
  \includegraphics[height=0.16\textwidth,valign=c]{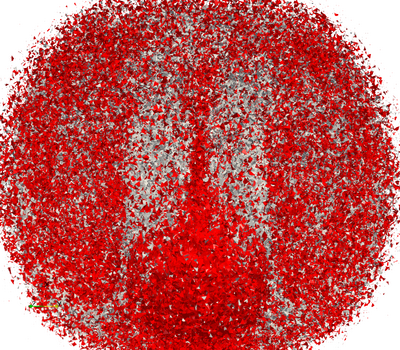} &
  \includegraphics[height=0.16\textwidth,valign=c]{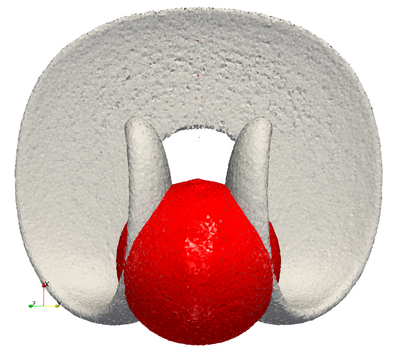} &
  \includegraphics[height=0.16\textwidth,valign=c]{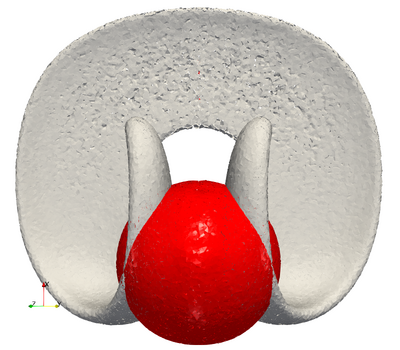} &
  \includegraphics[height=0.16\textwidth,valign=c]{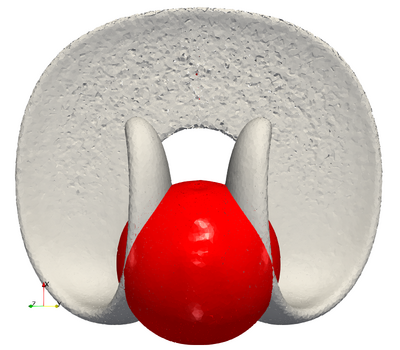} \\
 &
  \rotatebox[origin=c]{90}{Hexahedral} &
  \includegraphics[height=0.16\textwidth,valign=c]{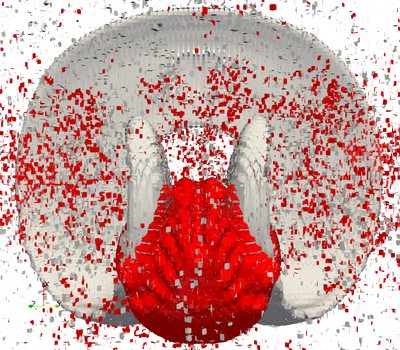} &
  \includegraphics[height=0.16\textwidth,valign=c]{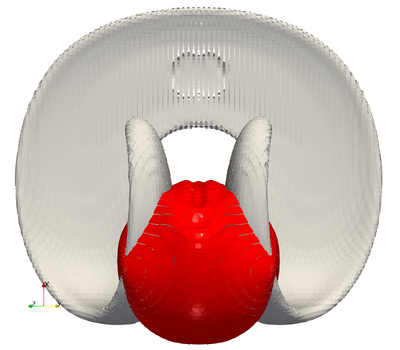} &
  \includegraphics[height=0.16\textwidth,valign=c]{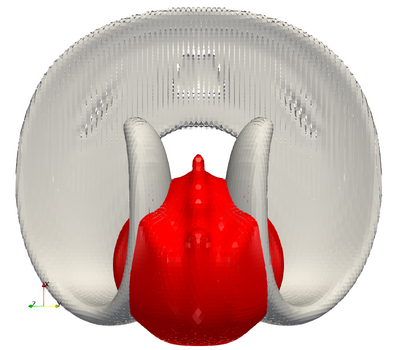} &
  \includegraphics[height=0.16\textwidth,valign=c]{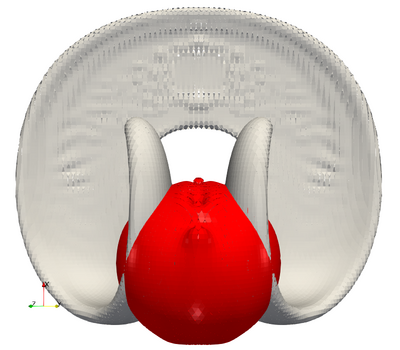} \\
 &
  \rotatebox[origin=c]{90}{\shortstack{General\\polyhedral}} &
  \includegraphics[height=0.16\textwidth,valign=c]{interface_polyMesh_plicVof_CAG_mesh_4.png} &
  \includegraphics[height=0.16\textwidth,valign=c]{interface_polyMesh_plicVof_NAG_mesh_4.png} &
  \includegraphics[height=0.16\textwidth,valign=c]{interface_polyMesh_plicVof_LS_mesh_4.png} &
  \includegraphics[height=0.16\textwidth,valign=c]{interface_polyMesh_plicVof_RDF_mesh_4.png} \\
\multirow{-4}{*}[13em]{$2^{-7}$} &
  \rotatebox[origin=c]{90}{\shortstack{Structured\\polyhedral}} &
  \includegraphics[height=0.16\textwidth,valign=c]{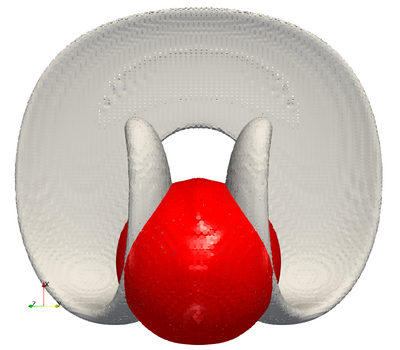} &
  \includegraphics[height=0.16\textwidth,valign=c]{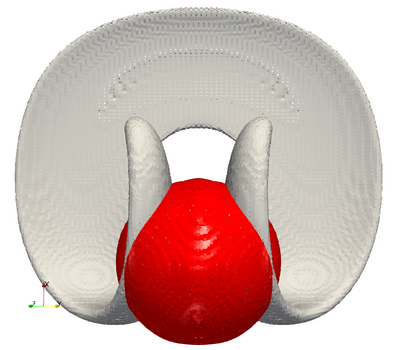} &
  \includegraphics[height=0.16\textwidth,valign=c]{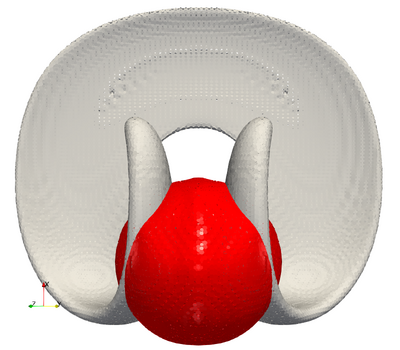} &
  \includegraphics[height=0.16\textwidth,valign=c]{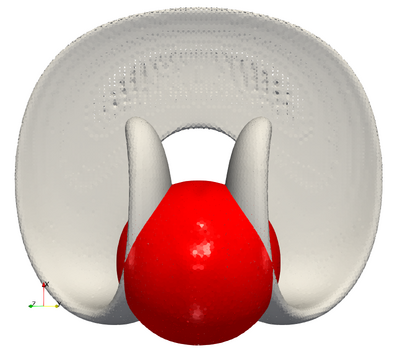} \\ \hline
\end{tabular}%
\end{table}

The $E_v(t)$ values presented in \cref{tab_advection_performance_data_various_orientation_schemes} suggest that all orientation schemes effectively preserve volume conservation. Across all schemes, the fraction field is consistently numerically bounded at the lower limit in general and structured polyhedral meshes. However, in tetrahedral and hexahedral meshes, there is an increase in the absolute values of the averaged $\alpha_{min}$. Similarly, while the fraction field is numerically bounded at the lower limit in hexahedral and structured polyhedral meshes, there is a significant rise in the averaged $1-\alpha_{max}$ values in tetrahedral and general polyhedral meshes. Notably, in general polyhedral meshes, the $1-\alpha_{max}$ values are on the order of $10^{-2}$.

Regarding shape errors, all orientation schemes demonstrate comparable performance in general and structured polyhedral meshes. In tetrahedral meshes, the CAG method shows the least effective shape preservation, with an average $E_s$ of $0.237$ for $\Delta s = 2^{-6}$ and $0.138$ for $\Delta s = 2^{-7}$. In hexahedral meshes, the LS and CAG schemes exhibit the highest shape errors for $\Delta s = 2^{-6}$ and $\Delta s = 2^{-7}$, respectively. Moreover, both the NAG and RDF methods display lower shape errors compared to the LS scheme in tetrahedral and hexahedral meshes.

The interface advection time, $T_{adv}$, does not exhibit significant differences across the various orientation schemes and mesh types, with the exception of the CAG method in tetrahedral and hexahedral meshes. In these cases, the CAG method creates additional mixed cells, leading to increased CPU time requirements for both interface reconstruction and advection. In contrast, the NAG method requires significantly more time for interface advection than the CAG method in both general and structured polyhedral meshes. The RDF scheme records the highest $T_{adv}$ in tetrahedral meshes. However, at a mesh resolution of $\Delta s = 2^{-7}$, the RDF method outperforms the NAG one in terms of speed in both hexahedral and polyhedral meshes. Across all mesh types, the LS scheme demonstrates the most efficient performance.

\begin{table}[htbp]
\footnotesize
\centering
\caption{Errors and execution times at $t=3\,s$ with various orientation schemes.}
\label{tab_advection_performance_data_various_orientation_schemes}
\resizebox{\textwidth}{!}{%
\renewcommand{\arraystretch}{1.15}
\begin{tabular}{l|c|c|l|l|l|l|l|l|l|l|r|r|r}
\hline
\rowcolor[HTML]{EFEFEF} 
\multicolumn{1}{c|}{\cellcolor[HTML]{EFEFEF}}                       & \cellcolor[HTML]{EFEFEF}                                                                             & \multicolumn{1}{c|}{\cellcolor[HTML]{EFEFEF}}                         & \multicolumn{2}{c|}{\cellcolor[HTML]{EFEFEF}$E_v(t)$}                                                & \multicolumn{2}{c|}{\cellcolor[HTML]{EFEFEF}$\alpha_{min}(t)$}                                       & \multicolumn{2}{c|}{\cellcolor[HTML]{EFEFEF}1-$\alpha_{max}(t)$}                                     & \multicolumn{2}{c|}{\cellcolor[HTML]{EFEFEF}$E_s(t)$}                                                & \multicolumn{1}{c|}{\cellcolor[HTML]{EFEFEF}}                                                                            & \multicolumn{1}{c|}{\cellcolor[HTML]{EFEFEF}}                                                                            & \multicolumn{1}{c}{\cellcolor[HTML]{EFEFEF}}                                                                             \\ \cline{4-11}
\rowcolor[HTML]{EFEFEF} 
\multicolumn{1}{c|}{\multirow{-2}{*}{\cellcolor[HTML]{EFEFEF}Mesh}} & \multirow{-2}{*}{\cellcolor[HTML]{EFEFEF}\begin{tabular}[c]{@{}c@{}}$\Delta s$\\ $[m]$\end{tabular}} & \multicolumn{1}{c|}{\multirow{-2}{*}{\cellcolor[HTML]{EFEFEF}Scheme}} & \multicolumn{1}{c|}{\cellcolor[HTML]{EFEFEF}amax} & \multicolumn{1}{c|}{\cellcolor[HTML]{EFEFEF}avg} & \multicolumn{1}{c|}{\cellcolor[HTML]{EFEFEF}amax} & \multicolumn{1}{c|}{\cellcolor[HTML]{EFEFEF}avg} & \multicolumn{1}{c|}{\cellcolor[HTML]{EFEFEF}amax} & \multicolumn{1}{c|}{\cellcolor[HTML]{EFEFEF}avg} & \multicolumn{1}{c|}{\cellcolor[HTML]{EFEFEF}amax} & \multicolumn{1}{c|}{\cellcolor[HTML]{EFEFEF}avg} & \multicolumn{1}{c|}{\multirow{-2}{*}{\cellcolor[HTML]{EFEFEF}\begin{tabular}[c]{@{}c@{}}$T_{rec}$\\ $[s]$\end{tabular}}} & \multicolumn{1}{c|}{\multirow{-2}{*}{\cellcolor[HTML]{EFEFEF}\begin{tabular}[c]{@{}c@{}}$T_{adv}$\\ $[s]$\end{tabular}}} & \multicolumn{1}{c}{\multirow{-2}{*}{\cellcolor[HTML]{EFEFEF}\begin{tabular}[c]{@{}c@{}}$T_{calc}$\\ $[s]$\end{tabular}}} \\ \hline
                                                                    &                                                                                                      & CAG                                                                   & $1.84 \times 10^{-11}$                                          & $1.82 \times 10^{-11}$                                         & $-1.14 \times 10^{-03}$                                         & $-2.33 \times 10^{-04}$                                        & $-4.18 \times 10^{-03}$                                         & $-1.56 \times 10^{-03}$                                        & 0.313                                             & 0.237                                            & 1628.8                                                                                                                   & 819.4                                                                                                                    & 4592.7                                                                                                                   \\
                                                                    &                                                                                                      & NAG                                                                   & $1.99 \times 10^{-11}$                                          & $1.92 \times 10^{-11}$                                         & $-3.75 \times 10^{-06}$                                         & $-7.37 \times 10^{-07}$                                        & $-7.22 \times 10^{-03}$                                         & $-3.03 \times 10^{-03}$                                        & 0.126                                             & 0.077                                            & 715.0                                                                                                                    & 311.5                                                                                                                    & 3169.4                                                                                                                   \\
                                                                    &                                                                                                      & LS                                                                    & $1.98 \times 10^{-11}$                                          & $1.91 \times 10^{-11}$                                         & $-6.85 \times 10^{-06}$                                         & $-2.48 \times 10^{-06}$                                        & $-7.23 \times 10^{-03}$                                         & $-3.59 \times 10^{-03}$                                        & 0.168                                             & 0.096                                            & 453.3                                                                                                                    & 313.4                                                                                                                    & 2909.0                                                                                                                   \\
                                                                    & \multirow{-4}{*}{$2^{-6}$}                                                                           & RDF                                                                   & $1.98 \times 10^{-11}$                                          & $1.91 \times 10^{-11}$                                         & $-1.73 \times 10^{-05}$                                         & $-4.49 \times 10^{-06}$                                        & $-7.22 \times 10^{-03}$                                         & $-3.24 \times 10^{-03}$                                        & 0.151                                             & 0.086                                            & 1824.1                                                                                                                   & 310.5                                                                                                                    & 4282.8                                                                                                                   \\ \cline{2-14} 
                                                                    &                                                                                                      & CAG                                                                   & $1.67 \times 10^{-11}$                                          & $1.33 \times 10^{-11}$                                         & $-2.26 \times 10^{-03}$                                         & $-5.01 \times 10^{-04}$                                        & $-2.08 \times 10^{-03}$                                         & $-8.62 \times 10^{-04}$                                        & 0.176                                             & 0.138                                            & 17911.4                                                                                                                  & 9933.4                                                                                                                   & 63281.7                                                                                                                  \\
                                                                    &                                                                                                      & NAG                                                                   & $2.52 \times 10^{-11}$                                          & $2.14 \times 10^{-11}$                                         & $-1.31 \times 10^{-05}$                                         & $-2.03 \times 10^{-06}$                                        & $-2.65 \times 10^{-03}$                                         & $-1.56 \times 10^{-03}$                                        & 0.030                                             & 0.022                                            & 9590.6                                                                                                                   & 4585.1                                                                                                                   & 49661.4                                                                                                                  \\
                                                                    &                                                                                                      & LS                                                                    & $2.53 \times 10^{-11}$                                          & $2.13 \times 10^{-11}$                                         & $-1.38 \times 10^{-04}$                                         & $-1.55 \times 10^{-05}$                                        & $-3.20 \times 10^{-03}$                                         & $-1.61 \times 10^{-03}$                                        & 0.032                                             & 0.023                                            & 4375.6                                                                                                                   & 4533.4                                                                                                                   & 44372.3                                                                                                                  \\
\multirow{-8}{*}{\rotatebox[origin=c]{90}{Tetrahedral}}                                       & \multirow{-4}{*}{$2^{-7}$}                                                                           & RDF                                                                   & $2.40 \times 10^{-11}$                                          & $2.09 \times 10^{-11}$                                         & $-1.36 \times 10^{-05}$                                         & $-1.91 \times 10^{-06}$                                        & $-2.51 \times 10^{-03}$                                         & $-1.55 \times 10^{-03}$                                        & 0.032                                             & 0.022                                            & 16918.4                                                                                                                  & 4533.3                                                                                                                   & 56880.6                                                                                                                  \\ \hline
                                                                    &                                                                                                      & CAG                                                                   & $1.88 \times 10^{-11}$                                          & $1.88 \times 10^{-11}$                                         & $-2.23 \times 10^{-03}$                                         & $-4.16 \times 10^{-04}$                                        & $-1.42 \times 10^{-06}$                                         & $-3.02 \times 10^{-07}$                                        & 0.246                                             & 0.164                                            & 73.2                                                                                                                     & 44.3                                                                                                                     & 202.1                                                                                                                    \\
                                                                    &                                                                                                      & NAG                                                                   & $1.90 \times 10^{-11}$                                          & $1.88 \times 10^{-11}$                                         & $-1.16 \times 10^{-05}$                                         & $-1.44 \times 10^{-06}$                                        & $-1.07 \times 10^{-07}$                                         & $-1.15 \times 10^{-08}$                                        & 0.237                                             & 0.146                                            & 41.8                                                                                                                     & 14.0                                                                                                                     & 139.9                                                                                                                    \\
                                                                    &                                                                                                      & LS                                                                    & $1.90 \times 10^{-11}$                                          & $1.88 \times 10^{-11}$                                         & $-1.91 \times 10^{-05}$                                         & $-1.62 \times 10^{-06}$                                        & $-1.08 \times 10^{-07}$                                         & $-1.68 \times 10^{-08}$                                        & 0.305                                             & 0.200                                            & 19.8                                                                                                                     & 14.7                                                                                                                     & 119.2                                                                                                                    \\
                                                                    & \multirow{-4}{*}{$2^{-6}$}                                                                           & RDF                                                                   & $1.90 \times 10^{-11}$                                          & $1.88 \times 10^{-11}$                                         & $-1.03 \times 10^{-03}$                                         & $-8.75 \times 10^{-05}$                                        & $-1.52 \times 10^{-04}$                                         & $-1.54 \times 10^{-05}$                                        & 0.275                                             & 0.172                                            & 61.8                                                                                                                     & 15.2                                                                                                                     & 161.8                                                                                                                    \\ \cline{2-14} 
                                                                    &                                                                                                      & CAG                                                                   & $1.86 \times 10^{-11}$                                          & $1.83 \times 10^{-11}$                                         & $-1.80 \times 10^{-03}$                                         & $-5.24 \times 10^{-04}$                                        & $-6.68 \times 10^{-04}$                                         & $-8.85 \times 10^{-05}$                                        & 0.111                                             & 0.074                                            & 931.6                                                                                                                    & 707.5                                                                                                                    & 3046.4                                                                                                                   \\
                                                                    &                                                                                                      & NAG                                                                   & $2.11 \times 10^{-11}$                                          & $1.95 \times 10^{-11}$                                         & $-2.72 \times 10^{-05}$                                         & $-4.02 \times 10^{-06}$                                        & $-2.09 \times 10^{-05}$                                         & $-1.88 \times 10^{-06}$                                        & 0.071                                             & 0.044                                            & 564.0                                                                                                                    & 216.3                                                                                                                    & 2187.7                                                                                                                   \\
                                                                    &                                                                                                      & LS                                                                    & $2.13 \times 10^{-11}$                                          & $1.94 \times 10^{-11}$                                         & $-2.61 \times 10^{-05}$                                         & $-4.11 \times 10^{-06}$                                        & $-4.24 \times 10^{-05}$                                         & $-4.41 \times 10^{-06}$                                        & 0.106                                             & 0.068                                            & 169.9                                                                                                                    & 210.6                                                                                                                    & 1783.1                                                                                                                   \\
\multirow{-8}{*}{\rotatebox[origin=c]{90}{Hexahedral}}                                        & \multirow{-4}{*}{$2^{-7}$}                                                                           & RDF                                                                   & $2.11 \times 10^{-11}$                                          & $1.95 \times 10^{-11}$                                         & $-4.80 \times 10^{-04}$                                         & $-5.97 \times 10^{-05}$                                        & $-1.70 \times 10^{-07}$                                         & $-2.66 \times 10^{-08}$                                        & 0.079                                             & 0.046                                            & 515.4                                                                                                                    & 215.0                                                                                                                    & 2136.8                                                                                                                   \\ \hline
                                                                    &                                                                                                      & CAG                                                                   & $1.87 \times 10^{-11}$                                          & $1.85 \times 10^{-11}$                                         & $-3.44 \times 10^{-18}$                                         & $-4.09 \times 10^{-19}$                                        & $-3.97 \times 10^{-02}$                                         & $-1.94 \times 10^{-02}$                                        & 0.295                                             & 0.204                                            & 125.6                                                                                                                    & 99.4                                                                                                                     & 630.9                                                                                                                    \\
                                                                    &                                                                                                      & NAG                                                                   & $1.87 \times 10^{-11}$                                          & $1.85 \times 10^{-11}$                                         & $-2.39 \times 10^{-18}$                                         & $-3.92 \times 10^{-19}$                                        & $-3.74 \times 10^{-02}$                                         & $-2.02 \times 10^{-02}$                                        & 0.298                                             & 0.205                                            & 276.1                                                                                                                    & 97.4                                                                                                                     & 778.3                                                                                                                    \\
                                                                    &                                                                                                      & LS                                                                    & $1.87 \times 10^{-11}$                                          & $1.85 \times 10^{-11}$                                         & $-4.61 \times 10^{-18}$                                         & $-5.29 \times 10^{-19}$                                        & $-3.80 \times 10^{-02}$                                         & $-2.05 \times 10^{-02}$                                        & 0.311                                             & 0.215                                            & 108.4                                                                                                                    & 98.3                                                                                                                     & 603.1                                                                                                                    \\
                                                                    & \multirow{-4}{*}{$2^{-6}$}                                                                           & RDF                                                                   & $1.87 \times 10^{-11}$                                          & $1.85 \times 10^{-11}$                                         & $-1.06 \times 10^{-18}$                                         & $-2.07 \times 10^{-19}$                                        & $-3.18 \times 10^{-02}$                                         & $-1.83 \times 10^{-02}$                                        & 0.305                                             & 0.209                                            & 325.4                                                                                                                    & 98.3                                                                                                                     & 818.5                                                                                                                    \\ \cline{2-14} 
                                                                    &                                                                                                      & CAG                                                                   & $1.84 \times 10^{-11}$                                          & $1.70 \times 10^{-11}$                                         & $-8.67 \times 10^{-19}$                                         & $-2.41 \times 10^{-19}$                                        & $-9.65 \times 10^{-02}$                                         & $-4.83 \times 10^{-02}$                                        & 0.092                                             & 0.064                                            & 1485.2                                                                                                                   & 1661.0                                                                                                                   & 11242.0                                                                                                                  \\
                                                                    &                                                                                                      & NAG                                                                   & $1.84 \times 10^{-11}$                                          & $1.70 \times 10^{-11}$                                         & $-1.82 \times 10^{-18}$                                         & $-3.51 \times 10^{-19}$                                        & $-7.40 \times 10^{-02}$                                         & $-4.72 \times 10^{-02}$                                        & 0.091                                             & 0.064                                            & 4855.1                                                                                                                   & 1660.0                                                                                                                   & 14581.6                                                                                                                  \\
                                                                    &                                                                                                      & LS                                                                    & $1.84 \times 10^{-11}$                                          & $1.70 \times 10^{-11}$                                         & $-1.13 \times 10^{-18}$                                         & $-3.37 \times 10^{-19}$                                        & $-9.58 \times 10^{-02}$                                         & $-4.97 \times 10^{-02}$                                        & 0.091                                             & 0.065                                            & 1173.0                                                                                                                   & 1663.1                                                                                                                   & 10906.4                                                                                                                  \\
\multirow{-8}{*}{\rotatebox[origin=c]{90}{\shortstack{General\\polyhedral}}}                                & \multirow{-4}{*}{$2^{-7}$}                                                                           & RDF                                                                   & $1.84 \times 10^{-11}$                                          & $1.70 \times 10^{-11}$                                         & $-1.79 \times 10^{-18}$                                         & $-4.33 \times 10^{-19}$                                        & $-8.08 \times 10^{-02}$                                         & $-4.18 \times 10^{-02}$                                        & 0.098                                             & 0.068                                            & 3615.6                                                                                                                   & 1696.4                                                                                                                   & 13377.9                                                                                                                  \\ \hline
                                                                    &                                                                                                      & CAG                                                                   & $1.85 \times 10^{-11}$                                          & $1.71 \times 10^{-11}$                                         & $-1.30 \times 10^{-17}$                                         & $-3.27 \times 10^{-18}$                                        & $-2.56 \times 10^{-05}$                                         & $-5.24 \times 10^{-06}$                                        & 0.122                                             & 0.073                                            & 577.7                                                                                                                    & 505.1                                                                                                                    & 3710.8                                                                                                                   \\
                                                                    &                                                                                                      & NAG                                                                   & $1.85 \times 10^{-11}$                                          & $1.71 \times 10^{-11}$                                         & $-1.07 \times 10^{-17}$                                         & $-2.05 \times 10^{-18}$                                        & $-1.85 \times 10^{-04}$                                         & $-1.82 \times 10^{-05}$                                        & 0.118                                             & 0.071                                            & 1387.3                                                                                                                   & 502.0                                                                                                                    & 4517.2                                                                                                                   \\
                                                                    &                                                                                                      & LS                                                                    & $1.85 \times 10^{-11}$                                          & $1.71 \times 10^{-11}$                                         & $-7.81 \times 10^{-18}$                                         & $-1.99 \times 10^{-18}$                                        & $-9.09 \times 10^{-06}$                                         & $-2.30 \times 10^{-06}$                                        & 0.127                                             & 0.077                                            & 469.6                                                                                                                    & 506.4                                                                                                                    & 3606.1                                                                                                                   \\
                                                                    & \multirow{-4}{*}{$2^{-6}$}                                                                           & RDF                                                                   & $1.86 \times 10^{-11}$                                          & $1.72 \times 10^{-11}$                                         & $-6.51 \times 10^{-18}$                                         & $-2.09 \times 10^{-18}$                                        & $-5.06 \times 10^{-05}$                                         & $-6.71 \times 10^{-06}$                                        & 0.136                                             & 0.079                                            & 1374.8                                                                                                                   & 506.5                                                                                                                    & 4518.5                                                                                                                   \\ \cline{2-14} 
                                                                    &                                                                                                      & CAG                                                                   & $1.77 \times 10^{-11}$                                          & $1.08 \times 10^{-11}$                                         & $-5.20 \times 10^{-18}$                                         & $-2.42 \times 10^{-18}$                                        & $-5.30 \times 10^{-06}$                                         & $-6.27 \times 10^{-07}$                                        & 0.030                                             & 0.021                                            & 5475.7                                                                                                                   & 7222.0                                                                                                                   & 54614.6                                                                                                                  \\
                                                                    &                                                                                                      & NAG                                                                   & $1.77 \times 10^{-11}$                                          & $1.08 \times 10^{-11}$                                         & $-1.33 \times 10^{-17}$                                         & $-3.44 \times 10^{-18}$                                        & $-3.97 \times 10^{-06}$                                         & $-4.84 \times 10^{-07}$                                        & 0.029                                             & 0.020                                            & 21057.4                                                                                                                  & 7332.8                                                                                                                   & 71855.4                                                                                                                  \\
                                                                    &                                                                                                      & LS                                                                    & $1.77 \times 10^{-11}$                                          & $1.08 \times 10^{-11}$                                         & $-6.75 \times 10^{-18}$                                         & $-3.39 \times 10^{-18}$                                        & $-1.80 \times 10^{-06}$                                         & $-3.47 \times 10^{-07}$                                        & 0.035                                             & 0.023                                            & 4195.2                                                                                                                   & 8492.9                                                                                                                   & 55111.7                                                                                                                  \\
\multirow{-8}{*}{\rotatebox[origin=c]{90}{\shortstack{Structured\\polyhedral}}}                             & \multirow{-4}{*}{$2^{-7}$}                                                                           & RDF                                                                   & $1.77 \times 10^{-11}$                                          & $1.10 \times 10^{-11}$                                         & $-6.22 \times 10^{-18}$                                         & $-3.18 \times 10^{-18}$                                        & $-9.63 \times 10^{-06}$                                         & $-1.84 \times 10^{-06}$                                        & 0.028                                             & 0.019                                            & 11272.1                                                                                                                  & 7313.0                                                                                                                   & 62837.7                                                                                                                  \\ \hline
\end{tabular}%
}
\end{table}

\cref{tab_advection_average_convergence_orders_orientation_schemes} outlines the average convergence orders of shape errors, which are defined by:
\begin{equation}
    \bar{\mathcal{O}}\left(E_{s}\right) = \frac{1}{12 \ln{2}} \sum_{t_i}{\ln{\frac{E_{s,\Delta s=2^{-6}}(t_i)}{E_{s,\Delta s=2^{-7}}(t_i)}}},
\end{equation}
where $t_i\,(i=1,2,\cdots,12)$ represent the sampling time instants and determined as $t_i = 0.25 i$. In tetrahedral and hexahedral meshes, the convergence rate for the CAG scheme is approximately first order, whereas the other three methods exhibit markedly better performance. Notably, in hexahedral meshes, the RDF scheme achieves a convergence rate as high as 1.95. In both general and structured polyhedral meshes, the four orientation schemes demonstrate similar convergence rates, varying between 1.41 and 1.75. Furthermore, the NAG and LS schemes consistently achieve convergence rates ranging from 1.47 to 1.74 across all mesh types. The RDF method maintains roughly second-order convergence in all mesh types except for general polyhedral meshes, where its convergence rate is observed to be 1.41.

\setlength\MAX{3.25mm}
\renewcommand*\Chart[1]{#1~\rlap{\textcolor{blue!20}{\rule{1.95\MAX}{1.5ex}}}\textcolor{blue!70}{\rule{#1\MAX}{1.5ex}}}

\begin{table}[htbp]
\footnotesize
\centering
\caption{Average convergence orders $\bar{\mathcal{O}}\left(E_{s}\right)$ of all meshes.}
\label{tab_advection_average_convergence_orders_orientation_schemes}
\renewcommand{\arraystretch}{1.15}
\begin{tabular}{c|m{0.07\textwidth}|m{0.07\textwidth}|m{0.07\textwidth}|m{0.07\textwidth}}
\rowcolor[HTML]{EFEFEF}
\hline
Meshes &
  \multicolumn{1}{c|}{\cellcolor[HTML]{EFEFEF}CAG} &
  \multicolumn{1}{c|}{\cellcolor[HTML]{EFEFEF}NAG} &
  \multicolumn{1}{c|}{\cellcolor[HTML]{EFEFEF}LS} &
  \multicolumn{1}{c}{\cellcolor[HTML]{EFEFEF}RDF} \\ \hline
Tetrahedral           & \Chart{0.72} & \Chart{1.54} & \Chart{1.74} & \Chart{1.66} \\ \hline
Hexahedral            & \Chart{1.08} & \Chart{1.69} & \Chart{1.48} & \Chart{1.95} \\ \hline
General polyhedral    & \Chart{1.46} & \Chart{1.47} & \Chart{1.50} & \Chart{1.41} \\ \hline
Structured polyhedral & \Chart{1.57} & \Chart{1.55} & \Chart{1.51} & \Chart{1.75} \\ \hline
\end{tabular}
\end{table}

\subsubsection{Comparisons with official-released PLIC-VOF methods in \OF}

Finally, the proposed SimPLIC method is compared with the officially released PLIC-VOF methods in \OF\,v2312. Within these official PLIC-VOF methods, the LS and RDF orientation schemes are utilized, referred to as "isoAdvector-plicLS"\footnote{\href{https://www.openfoam.com/documentation/guides/latest/api/classFoam_1_1reconstruction_1_1gradAlpha.html}{https://www.openfoam.com/documentation/guides/latest/api/classFoam\_1\_1reconstruction\_1\_1gradAlpha.html}} and "isoAdvector-plicRDF"\footnote{\href{https://www.openfoam.com/documentation/guides/latest/api/classFoam_1_1reconstruction_1_1plicRDF.html}{https://www.openfoam.com/documentation/guides/latest/api/classFoam\_1\_1reconstruction\_1\_1plicRDF.html}}, respectively. The LS and RDF schemes in SimPLIC are implemented in exact accordance with the detailed procedures in the isoAdvector PLIC library. The reconstructed interface planes generated by the solvers are illustrated in \cref{tab_advection_plic_surfraces_various_solvers}. These solvers produce virtually identical interface shapes at the same resolution across all four mesh types.

\begin{table}[htbp]
\footnotesize
\centering
\caption{Reconstructed interfaces at $t=1.5\,s$ (\textit{gray}) and $t=3\,s$ (\textit{red}) with various PLIC-VOF solvers.}
\label{tab_advection_plic_surfraces_various_solvers}
\renewcommand{\arraystretch}{1.15}
\begin{tabular}{c|c|c|c|c|c}
\hline
\rowcolor[HTML]{EFEFEF} 
$\Delta s\,[m]$ &
  Mesh &
  isoAdvector-plicLS &
  isoAdvector-plicRDF &
  SimPLIC-LS &
  SimPLIC-RDF \\ \hline
 &
  \rotatebox[origin=c]{90}{Tetrahedral} &
  \includegraphics[height=0.16\textwidth,valign=c]{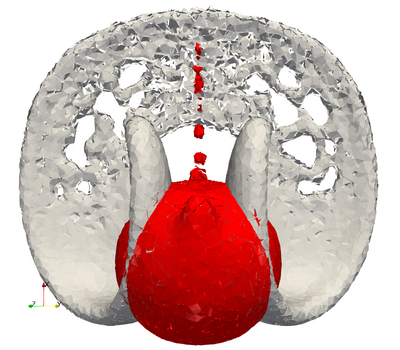} &
  \includegraphics[height=0.16\textwidth,valign=c]{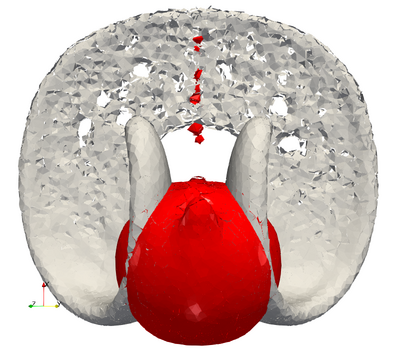} &
  \includegraphics[height=0.16\textwidth,valign=c]{interface_tetMesh_plicVof_LS_mesh_3.png} &
  \includegraphics[height=0.16\textwidth,valign=c]{interface_tetMesh_plicVof_RDF_mesh_3.png} \\
 &
  \rotatebox[origin=c]{90}{Hexahedral} &
  \includegraphics[height=0.16\textwidth,valign=c]{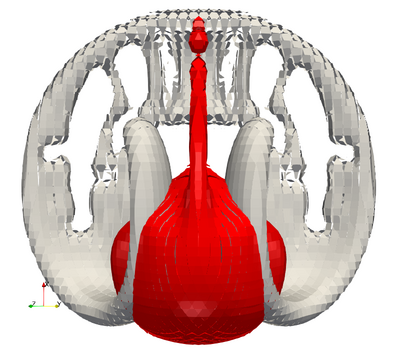} &
  \includegraphics[height=0.16\textwidth,valign=c]{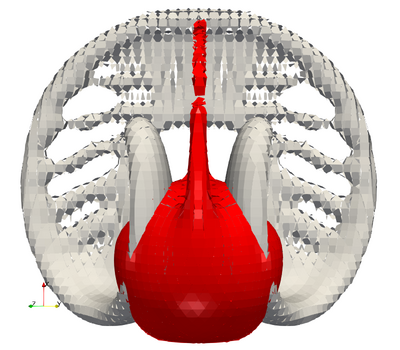} &
  \includegraphics[height=0.16\textwidth,valign=c]{interface_hexMesh_plicVof_LS_mesh_3.png} &
  \includegraphics[height=0.16\textwidth,valign=c]{interface_hexMesh_plicVof_RDF_mesh_3.png} \\
 &
  \rotatebox[origin=c]{90}{\shortstack{General\\polyhedral}} &
  \includegraphics[height=0.16\textwidth,valign=c]{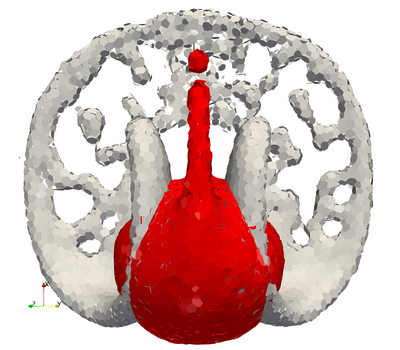} &
  \includegraphics[height=0.16\textwidth,valign=c]{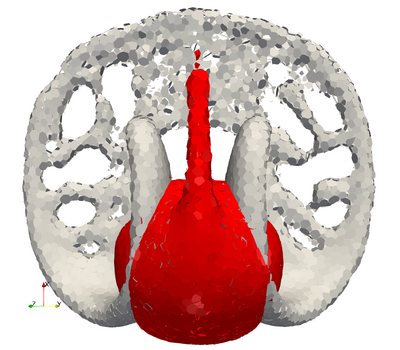} &
  \includegraphics[height=0.16\textwidth,valign=c]{interface_polyMesh_plicVof_LS_mesh_3.png} &
  \includegraphics[height=0.16\textwidth,valign=c]{interface_polyMesh_plicVof_RDF_mesh_3.png} \\
\multirow{-4}{*}[13em]{$2^{-6}$} &
  \rotatebox[origin=c]{90}{\shortstack{Structured\\polyhedral}} &
  \includegraphics[height=0.16\textwidth,valign=c]{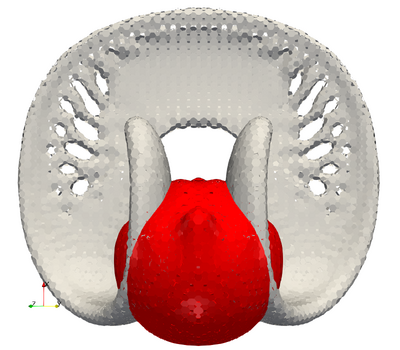} &
  \includegraphics[height=0.16\textwidth,valign=c]{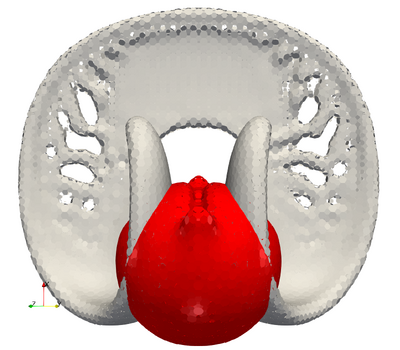} &
  \includegraphics[height=0.16\textwidth,valign=c]{interface_blockPolyMesh_plicVof_LS_mesh_3.png} &
  \includegraphics[height=0.16\textwidth,valign=c]{interface_blockPolyMesh_plicVof_RDF_mesh_3.png} \\ \hline
 &
  \rotatebox[origin=c]{90}{Tetrahedral} &
  \includegraphics[height=0.16\textwidth,valign=c]{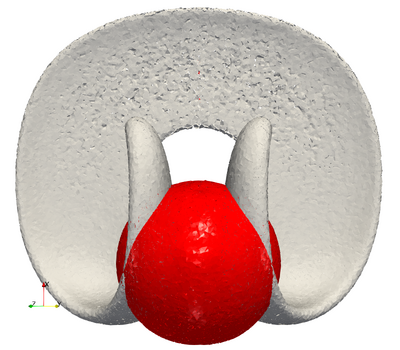} &
  \includegraphics[height=0.16\textwidth,valign=c]{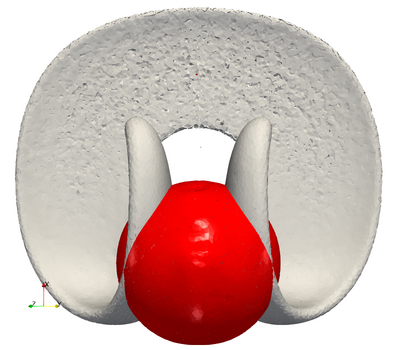} &
  \includegraphics[height=0.16\textwidth,valign=c]{interface_tetMesh_plicVof_LS_mesh_4.png} &
  \includegraphics[height=0.16\textwidth,valign=c]{interface_tetMesh_plicVof_RDF_mesh_4.png} \\
 &
  \rotatebox[origin=c]{90}{Hexahedral} &
  \includegraphics[height=0.16\textwidth,valign=c]{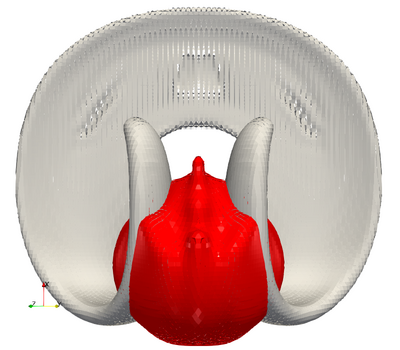} &
  \includegraphics[height=0.16\textwidth,valign=c]{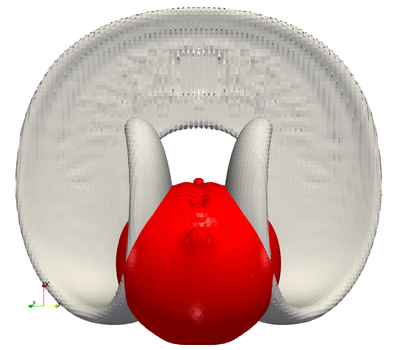} &
  \includegraphics[height=0.16\textwidth,valign=c]{interface_hexMesh_plicVof_LS_mesh_4.png} &
  \includegraphics[height=0.16\textwidth,valign=c]{interface_hexMesh_plicVof_RDF_mesh_4.png} \\
 &
  \rotatebox[origin=c]{90}{\shortstack{General\\polyhedral}} &
  \includegraphics[height=0.16\textwidth,valign=c]{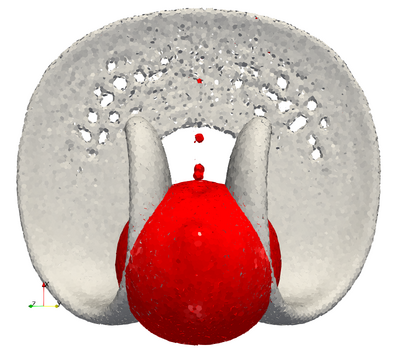} &
  \includegraphics[height=0.16\textwidth,valign=c]{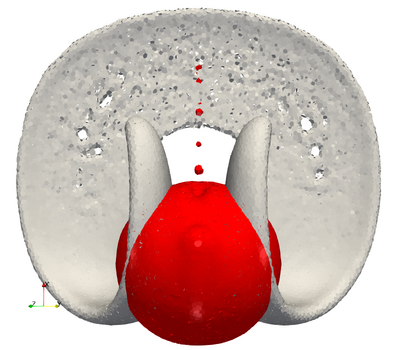} &
  \includegraphics[height=0.16\textwidth,valign=c]{interface_polyMesh_plicVof_LS_mesh_4.png} &
  \includegraphics[height=0.16\textwidth,valign=c]{interface_polyMesh_plicVof_RDF_mesh_4.png} \\
\multirow{-4}{*}[13em]{$2^{-7}$} &
  \rotatebox[origin=c]{90}{\shortstack{Structured\\polyhedral}} &
  \includegraphics[height=0.16\textwidth,valign=c]{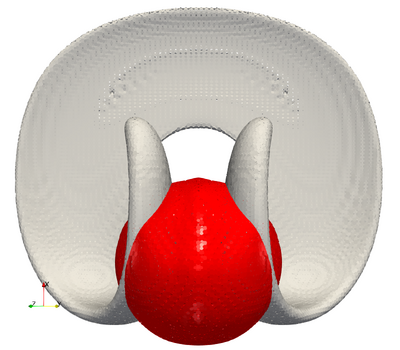} &
  \includegraphics[height=0.16\textwidth,valign=c]{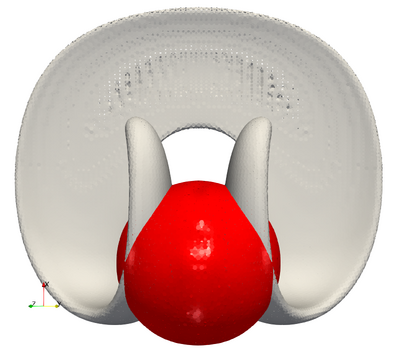} &
  \includegraphics[height=0.16\textwidth,valign=c]{interface_blockPolyMesh_plicVof_LS_mesh_4.png} &
  \includegraphics[height=0.16\textwidth,valign=c]{interface_blockPolyMesh_plicVof_RDF_mesh_4.png} \\ \hline
\end{tabular}%
\end{table}

The errors and execution times for the various PLIC-VOF solvers are detailed in \cref{tab_advection_performance_data_various_solvers}. The SimPLIC-LS and SimPLIC-RDF solvers exhibit volume conservation errors that are almost equivalent to those of isoAdvector-plicLS and isoAdvector-plicRDF, respectively, at the same mesh size for different mesh types. Regarding the lower bound of the fraction field, all solvers achieve values within machine tolerance in general and structured polyhedral meshes, while the minimum fractions increase in tetrahedral and hexahedral meshes for all solvers. Concerning the upper fraction limit, all solvers perform similarly, except SimPLIC-RDF, which shows relatively higher errors compared to isoAdvector-plicRDF in coarser hexahedral and all general polyhedral meshes. In terms of shape errors, all solvers yield identical results in tetrahedral, hexahedral, and structured polyhedral meshes. However, in general polyhedral meshes with mesh resolutions of $\Delta s = 2^{-6}\,m$ and $\Delta s = 2^{-7}\,m$, the SimPLIC-LS and SimPLIC-RDF solvers result in $5.9\% - 10.6\%$ and $38.3\% - 58.1\%$ higher shape errors, respectively, compared to isoAdvector-plicLS and isoAdvector-plicRDF.

\begin{table}[htbp]
\footnotesize
\centering
\caption{Errors and execution times at $t=3\,s$ with various PLIC-VOF solvers.}
\label{tab_advection_performance_data_various_solvers}
\resizebox{\textwidth}{!}{%
\begin{tabular}{c|c|c|c|c|c|c|c|c|c|c|r|r|r}
\hline
\rowcolor[HTML]{EFEFEF} 
\multicolumn{1}{c|}{\cellcolor[HTML]{EFEFEF}}                       & \cellcolor[HTML]{EFEFEF}                                                                             & \multicolumn{1}{c|}{\cellcolor[HTML]{EFEFEF}}                         & \multicolumn{2}{c|}{\cellcolor[HTML]{EFEFEF}$E_v(t)$}                                                & \multicolumn{2}{c|}{\cellcolor[HTML]{EFEFEF}$\alpha_{min}(t)$}                                       & \multicolumn{2}{c|}{\cellcolor[HTML]{EFEFEF}1-$\alpha_{max}(t)$}                                     & \multicolumn{2}{c|}{\cellcolor[HTML]{EFEFEF}$E_s(t)$}                                                & \multicolumn{1}{c|}{\cellcolor[HTML]{EFEFEF}}                                                                            & \multicolumn{1}{c|}{\cellcolor[HTML]{EFEFEF}}                                                                            & \multicolumn{1}{c}{\cellcolor[HTML]{EFEFEF}}                                                                             \\ \cline{4-11}
\rowcolor[HTML]{EFEFEF} 
\multicolumn{1}{c|}{\multirow{-2}{*}{\cellcolor[HTML]{EFEFEF}Mesh}} & \multirow{-2}{*}{\cellcolor[HTML]{EFEFEF}\begin{tabular}[c]{@{}c@{}}$\Delta s$\\ $[m]$\end{tabular}} & \multicolumn{1}{c|}{\multirow{-2}{*}{\cellcolor[HTML]{EFEFEF}Solver}} & \multicolumn{1}{c|}{\cellcolor[HTML]{EFEFEF}amax} & \multicolumn{1}{c|}{\cellcolor[HTML]{EFEFEF}avg} & \multicolumn{1}{c|}{\cellcolor[HTML]{EFEFEF}amax} & \multicolumn{1}{c|}{\cellcolor[HTML]{EFEFEF}avg} & \multicolumn{1}{c|}{\cellcolor[HTML]{EFEFEF}amax} & \multicolumn{1}{c|}{\cellcolor[HTML]{EFEFEF}avg} & \multicolumn{1}{c|}{\cellcolor[HTML]{EFEFEF}amax} & \multicolumn{1}{c|}{\cellcolor[HTML]{EFEFEF}avg} & \multicolumn{1}{c|}{\multirow{-2}{*}{\cellcolor[HTML]{EFEFEF}\begin{tabular}[c]{@{}c@{}}$T_{rec}$\\ $[s]$\end{tabular}}} & \multicolumn{1}{c|}{\multirow{-2}{*}{\cellcolor[HTML]{EFEFEF}\begin{tabular}[c]{@{}c@{}}$T_{adv}$\\ $[s]$\end{tabular}}} & \multicolumn{1}{c}{\multirow{-2}{*}{\cellcolor[HTML]{EFEFEF}\begin{tabular}[c]{@{}c@{}}$T_{calc}$\\ $[s]$\end{tabular}}} \\ \hline
                                                                    &                                                                                                      & isoAdvector-plicLS                                                    & $1.98 \times 10^{-11}$                                          & $1.91 \times 10^{-11}$                                         & $-7.14 \times 10^{-06}$                                         & $-2.46 \times 10^{-06}$                                        & $-7.22 \times 10^{-03}$                                         & $-3.58 \times 10^{-03}$                                        & 0.168                                             & 0.096                                            & 375.1                                                                                                                    & 308.5                                                                                                                    & 2854.3                                                                                                                   \\
                                                                    &                                                                                                      & isoAdvector-plicRDF                                                   & $1.97 \times 10^{-11}$                                          & $1.90 \times 10^{-11}$                                         & $-8.78 \times 10^{-06}$                                         & $-9.10 \times 10^{-07}$                                        & $-7.19 \times 10^{-03}$                                         & $-3.19 \times 10^{-03}$                                        & 0.151                                             & 0.086                                            & 2466.2                                                                                                                   & 305.6                                                                                                                    & 4940.0                                                                                                                   \\
                                                                    &                                                                                                      & SimPLIC-LS                                                            & $1.98 \times 10^{-11}$                                          & $1.91 \times 10^{-11}$                                         & $-6.85 \times 10^{-06}$                                         & $-2.48 \times 10^{-06}$                                        & $-7.23 \times 10^{-03}$                                         & $-3.59 \times 10^{-03}$                                        & 0.168                                             & 0.096                                            & 453.3                                                                                                                    & 313.4                                                                                                                    & 2909.0                                                                                                                   \\
                                                                    & \multirow{-4}{*}{$2^{-6}$}                                                                           & SimPLIC-RDF                                                           & $1.98 \times 10^{-11}$                                          & $1.91 \times 10^{-11}$                                         & $-1.73 \times 10^{-05}$                                         & $-4.49 \times 10^{-06}$                                        & $-7.22 \times 10^{-03}$                                         & $-3.24 \times 10^{-03}$                                        & 0.151                                             & 0.086                                            & 1824.1                                                                                                                   & 310.5                                                                                                                    & 4282.8                                                                                                                   \\ \cline{2-14} 
                                                                    &                                                                                                      & isoAdvector-plicLS                                                    & $2.56 \times 10^{-11}$                                          & $2.13 \times 10^{-11}$                                         & $-1.35 \times 10^{-04}$                                         & $-1.51 \times 10^{-05}$                                        & $-3.02 \times 10^{-03}$                                         & $-1.60 \times 10^{-03}$                                        & 0.032                                             & 0.023                                            & 3613.9                                                                                                                   & 4659.4                                                                                                                   & 44035.8                                                                                                                  \\
                                                                    &                                                                                                      & isoAdvector-plicRDF                                                   & $2.48 \times 10^{-11}$                                          & $2.13 \times 10^{-11}$                                         & $-2.23 \times 10^{-05}$                                         & $-4.19 \times 10^{-06}$                                        & $-2.63 \times 10^{-03}$                                         & $-1.57 \times 10^{-03}$                                        & 0.032                                             & 0.022                                            & 31250.1                                                                                                                  & 4613.6                                                                                                                   & 71563.3                                                                                                                  \\
                                                                    &                                                                                                      & SimPLIC-LS                                                            & $2.53 \times 10^{-11}$                                          & $2.13 \times 10^{-11}$                                         & $-1.38 \times 10^{-04}$                                         & $-1.55 \times 10^{-05}$                                        & $-3.20 \times 10^{-03}$                                         & $-1.61 \times 10^{-03}$                                        & 0.032                                             & 0.023                                            & 4375.6                                                                                                                   & 4533.4                                                                                                                   & 44372.3                                                                                                                  \\
\multirow{-8}{*}{\rotatebox[origin=c]{90}{Tetrahedral}}                                       & \multirow{-4}{*}{$2^{-7}$}                                                 & SimPLIC-RDF                                                           & $2.40 \times 10^{-11}$                                          & $2.09 \times 10^{-11}$                                         & $-1.36 \times 10^{-05}$                                         & $-1.91 \times 10^{-06}$                                        & $-2.51 \times 10^{-03}$                                         & $-1.55 \times 10^{-03}$                                        & 0.032                                             & 0.022                                            & 16918.4                                                                                                                  & 4533.3                                                                                                                   & 56880.6                                                                                                                  \\ \hline
                                                                    &                                                                                                      & isoAdvector-plicLS                                                    & $1.90 \times 10^{-11}$                                          & $1.88 \times 10^{-11}$                                         & $-1.91 \times 10^{-05}$                                         & $-1.62 \times 10^{-06}$                                        & $-1.08 \times 10^{-07}$                                         & $-1.69 \times 10^{-08}$                                        & 0.305                                             & 0.200                                            & 14.9                                                                                                                     & 14.4                                                                                                                     & 114.6                                                                                                                    \\
                                                                    &                                                                                                      & isoAdvector-plicRDF                                                   & $1.91 \times 10^{-11}$                                          & $1.88 \times 10^{-11}$                                         & $-1.11 \times 10^{-03}$                                         & $-9.37 \times 10^{-05}$                                        & $-1.00 \times 10^{-05}$                                         & $-8.34 \times 10^{-07}$                                        & 0.275                                             & 0.172                                            & 42.9                                                                                                                     & 15.7                                                                                                                     & 144.7                                                                                                                    \\
                                                                    &                                                                                                      & SimPLIC-LS                                                            & $1.90 \times 10^{-11}$                                          & $1.88 \times 10^{-11}$                                         & $-1.91 \times 10^{-05}$                                         & $-1.62 \times 10^{-06}$                                        & $-1.08 \times 10^{-07}$                                         & $-1.68 \times 10^{-08}$                                        & 0.305                                             & 0.200                                            & 19.8                                                                                                                     & 14.7                                                                                                                     & 119.2                                                                                                                    \\
                                                                    & \multirow{-4}{*}{$2^{-6}$}                                                                           & SimPLIC-RDF                                                           & $1.90 \times 10^{-11}$                                          & $1.88 \times 10^{-11}$                                         & $-1.03 \times 10^{-03}$                                         & $-8.75 \times 10^{-05}$                                        & $-1.52 \times 10^{-04}$                                         & $-1.54 \times 10^{-05}$                                        & 0.275                                             & 0.172                                            & 61.8                                                                                                                     & 15.2                                                                                                                     & 161.8                                                                                                                    \\ \cline{2-14} 
                                                                    &                                                                                                      & isoAdvector-plicLS                                                    & $2.13 \times 10^{-11}$                                          & $1.94 \times 10^{-11}$                                         & $-2.61 \times 10^{-05}$                                         & $-4.11 \times 10^{-06}$                                        & $-4.24 \times 10^{-05}$                                         & $-4.41 \times 10^{-06}$                                        & 0.106                                             & 0.068                                            & 127.6                                                                                                                    & 208.4                                                                                                                    & 1749.0                                                                                                                   \\
                                                                    &                                                                                                      & isoAdvector-plicRDF                                                   & $2.11 \times 10^{-11}$                                          & $1.93 \times 10^{-11}$                                         & $-2.29 \times 10^{-05}$                                         & $-5.92 \times 10^{-06}$                                        & $-4.54 \times 10^{-06}$                                         & $-4.21 \times 10^{-07}$                                        & 0.081                                             & 0.046                                            & 356.1                                                                                                                    & 209.1                                                                                                                    & 1976.2                                                                                                                   \\
                                                                    &                                                                                                      & SimPLIC-LS                                                            & $2.13 \times 10^{-11}$                                          & $1.94 \times 10^{-11}$                                         & $-2.61 \times 10^{-05}$                                         & $-4.11 \times 10^{-06}$                                        & $-4.24 \times 10^{-05}$                                         & $-4.41 \times 10^{-06}$                                        & 0.106                                             & 0.068                                            & 169.9                                                                                                                    & 210.6                                                                                                                    & 1783.1                                                                                                                   \\
\multirow{-8}{*}{\rotatebox[origin=c]{90}{Hexahedral}}                                        & \multirow{-4}{*}{$2^{-7}$}                                                 & SimPLIC-RDF                                                           & $2.11 \times 10^{-11}$                                          & $1.95 \times 10^{-11}$                                         & $-4.80 \times 10^{-04}$                                         & $-5.97 \times 10^{-05}$                                        & $-1.70 \times 10^{-07}$                                         & $-2.66 \times 10^{-08}$                                        & 0.079                                             & 0.046                                            & 515.4                                                                                                                    & 215.0                                                                                                                    & 2136.8                                                                                                                   \\ \hline
                                                                    &                                                                                                      & isoAdvector-plicLS                                                    & $1.87 \times 10^{-11}$                                          & $1.85 \times 10^{-11}$                                         & $-9.10 \times 10^{-18}$                                         & $-1.44 \times 10^{-18}$                                        & $-1.14 \times 10^{-02}$                                         & $-5.09 \times 10^{-03}$                                        & 0.307                                             & 0.203                                            & 109.8                                                                                                                    & 80.3                                                                                                                     & 588.2                                                                                                                    \\
                                                                    &                                                                                                      & isoAdvector-plicRDF                                                   & $1.88 \times 10^{-11}$                                          & $1.85 \times 10^{-11}$                                         & $-5.42 \times 10^{-19}$                                         & $-1.30 \times 10^{-19}$                                        & $-1.13 \times 10^{-02}$                                         & $-4.68 \times 10^{-03}$                                        & 0.294                                             & 0.189                                            & 390.4                                                                                                                    & 81.8                                                                                                                     & 873.1                                                                                                                    \\
                                                                    &                                                                                                      & SimPLIC-LS                                                            & $1.87 \times 10^{-11}$                                          & $1.85 \times 10^{-11}$                                         & $-4.61 \times 10^{-18}$                                         & $-5.29 \times 10^{-19}$                                        & $-3.80 \times 10^{-02}$                                         & $-2.05 \times 10^{-02}$                                        & 0.311                                             & 0.215                                            & 108.4                                                                                                                    & 98.3                                                                                                                     & 603.1                                                                                                                    \\
                                                                    & \multirow{-4}{*}{$2^{-6}$}                                                                           & SimPLIC-RDF                                                           & $1.87 \times 10^{-11}$                                          & $1.85 \times 10^{-11}$                                         & $-1.06 \times 10^{-18}$                                         & $-2.07 \times 10^{-19}$                                        & $-3.18 \times 10^{-02}$                                         & $-1.83 \times 10^{-02}$                                        & 0.305                                             & 0.209                                            & 325.4                                                                                                                    & 98.3                                                                                                                     & 818.5                                                                                                                    \\ \cline{2-14} 
                                                                    &                                                                                                      & isoAdvector-plicLS                                                    & $1.84 \times 10^{-11}$                                          & $1.70 \times 10^{-11}$                                         & $-1.78 \times 10^{-18}$                                         & $-5.15 \times 10^{-19}$                                        & $-1.56 \times 10^{-02}$                                         & $-1.04 \times 10^{-02}$                                        & 0.074                                             & 0.047                                            & 1099.2                                                                                                                   & 1493.9                                                                                                                   & 10674.2                                                                                                                  \\
                                                                    &                                                                                                      & isoAdvector-plicRDF                                                   & $1.84 \times 10^{-11}$                                          & $1.70 \times 10^{-11}$                                         & $-5.83 \times 10^{-19}$                                         & $-2.20 \times 10^{-19}$                                        & $-1.75 \times 10^{-02}$                                         & $-9.32 \times 10^{-03}$                                        & 0.070                                             & 0.043                                            & 4963.3                                                                                                                   & 1498.3                                                                                                                   & 14564.8                                                                                                                  \\
                                                                    &                                                                                                      & SimPLIC-LS                                                            & $1.84 \times 10^{-11}$                                          & $1.70 \times 10^{-11}$                                         & $-1.13 \times 10^{-18}$                                         & $-3.37 \times 10^{-19}$                                        & $-9.58 \times 10^{-02}$                                         & $-4.97 \times 10^{-02}$                                        & 0.091                                             & 0.065                                            & 1173.0                                                                                                                   & 1663.1                                                                                                                   & 10906.4                                                                                                                  \\
\multirow{-8}{*}{\rotatebox[origin=c]{90}{\shortstack{General\\polyhedral}}}                                & \multirow{-4}{*}{$2^{-7}$}                                   & SimPLIC-RDF                                                           & $1.84 \times 10^{-11}$                                          & $1.70 \times 10^{-11}$                                         & $-1.79 \times 10^{-18}$                                         & $-4.33 \times 10^{-19}$                                        & $-8.08 \times 10^{-02}$                                         & $-4.18 \times 10^{-02}$                                        & 0.098                                             & 0.068                                            & 3615.6                                                                                                                   & 1696.4                                                                                                                   & 13377.9                                                                                                                  \\ \hline
                                                                    &                                                                                                      & isoAdvector-plicLS                                                    & $1.85 \times 10^{-11}$                                          & $1.71 \times 10^{-11}$                                         & $-6.30 \times 10^{-14}$                                         & $-5.25 \times 10^{-15}$                                        & $-9.86 \times 10^{-05}$                                         & $-9.25 \times 10^{-06}$                                        & 0.127                                             & 0.077                                            & 313.2                                                                                                                    & 500.0                                                                                                                    & 3448.3                                                                                                                   \\
                                                                    &                                                                                                      & isoAdvector-plicRDF                                                   & $1.85 \times 10^{-11}$                                          & $1.71 \times 10^{-11}$                                         & $-1.39 \times 10^{-17}$                                         & $-2.52 \times 10^{-18}$                                        & $-7.29 \times 10^{-05}$                                         & $-1.35 \times 10^{-05}$                                        & 0.134                                             & 0.077                                            & 1379.4                                                                                                                   & 503.7                                                                                                                    & 4534.2                                                                                                                   \\
                                                                    &                                                                                                      & SimPLIC-LS                                                            & $1.85 \times 10^{-11}$                                          & $1.71 \times 10^{-11}$                                         & $-7.81 \times 10^{-18}$                                         & $-1.99 \times 10^{-18}$                                        & $-9.09 \times 10^{-06}$                                         & $-2.30 \times 10^{-06}$                                        & 0.127                                             & 0.077                                            & 469.6                                                                                                                    & 506.4                                                                                                                    & 3606.1                                                                                                                   \\
                                                                    & \multirow{-4}{*}{$2^{-6}$}                                                                           & SimPLIC-RDF                                                           & $1.86 \times 10^{-11}$                                          & $1.72 \times 10^{-11}$                                         & $-6.51 \times 10^{-18}$                                         & $-2.09 \times 10^{-18}$                                        & $-5.06 \times 10^{-05}$                                         & $-6.71 \times 10^{-06}$                                        & 0.136                                             & 0.079                                            & 1374.8                                                                                                                   & 506.5                                                                                                                    & 4518.5                                                                                                                   \\ \cline{2-14} 
                                                                    &                                                                                                      & isoAdvector-plicLS                                                    & $1.77 \times 10^{-11}$                                          & $1.11 \times 10^{-11}$                                         & $-8.67 \times 10^{-18}$                                         & $-3.09 \times 10^{-18}$                                        & $-2.86 \times 10^{-05}$                                         & $-2.53 \times 10^{-06}$                                        & 0.035                                             & 0.023                                            & 2655.8                                                                                                                   & 7404.4                                                                                                                   & 58430.2                                                                                                                  \\
                                                                    &                                                                                                      & isoAdvector-plicRDF                                                   & $1.77 \times 10^{-11}$                                          & $1.14 \times 10^{-11}$                                         & $-9.97 \times 10^{-18}$                                         & $-3.59 \times 10^{-18}$                                        & $-1.76 \times 10^{-05}$                                         & $-2.58 \times 10^{-06}$                                        & 0.028                                             & 0.019                                            & 13885.9                                                                                                                  & 8456.8                                                                                                                   & 65355.5                                                                                                                  \\
                                                                    &                                                                                                      & SimPLIC-LS                                                            & $1.77 \times 10^{-11}$                                          & $1.08 \times 10^{-11}$                                         & $-6.75 \times 10^{-18}$                                         & $-3.39 \times 10^{-18}$                                        & $-1.80 \times 10^{-06}$                                         & $-3.47 \times 10^{-07}$                                        & 0.035                                             & 0.023                                            & 4195.2                                                                                                                   & 8492.9                                                                                                                   & 55111.7                                                                                                                  \\
\multirow{-8}{*}{\rotatebox[origin=c]{90}{\shortstack{Structured\\polyhedral}}}                             & \multirow{-4}{*}{$2^{-7}$}                                   & SimPLIC-RDF                                                           & $1.77 \times 10^{-11}$                                          & $1.10 \times 10^{-11}$                                         & $-6.22 \times 10^{-18}$                                         & $-3.18 \times 10^{-18}$                                        & $-9.63 \times 10^{-06}$                                         & $-1.84 \times 10^{-06}$                                        & 0.028                                             & 0.019                                            & 11272.1                                                                                                                  & 7313.0                                                                                                                   & 62837.7                                                                                                                  \\ \hline
\end{tabular}%
}
\end{table}

All solvers demonstrate comparable CPU time consumption during the advection step in tetrahedral, hexahedral, and coarser structured polyhedral meshes. In general polyhedral meshes with resolutions of $\Delta s = 2^{-6}\,m$ and $\Delta s = 2^{-7}\,m$, the SimPLIC-LS and SimPLIC-RDF solvers require approximately $21.3\%$ and $12.3\%$ more advection time, respectively, compared to isoAdvector-plicLS and isoAdvector-plicRDF. In the finer structured polyhedral mesh, SimPLIC-LS operates $14.7\%$ slower than isoAdvector-plicLS, while SimPLIC-RDF is $15.6\%$ faster than isoAdvector-plicRDF.

Significant differences are observed in the reconstruction step. The SimPLIC-LS solver is about $21.0\%$, $33.1\%$, and $54.0\%$ less efficient than isoAdvector-plicLS in tetrahedral, hexahedral, and structured polyhedral meshes, respectively. In general polyhedral meshes, the efficiency gap between SimPLIC-LS and isoAdvector-plicLS is negligible. The SimPLIC-RDF solver is approximately $60.0\%$ faster in tetrahedral meshes and $28.7\%$ faster in general polyhedral meshes compared to isoAdvector-plicRDF. However, SimPLIC-RDF is $44.4\%$ slower than isoAdvector-plicRDF in hexahedral meshes. In coarser structured polyhedral mesh, SimPLIC-RDF matches the efficiency of isoAdvector-plicRDF, but it shows a $23.2\%$ improvement in finer structured polyhedral mesh.

\section{Conclusion}\label{s_conclusion}

A novel PLIC-VOF solver, SimPLIC, has been developed for interface flow simulations on arbitrary unstructured meshes. The SimPLIC method approximates the interfaces as three-dimensional planes and integrates the submerged face areas with Simpson’s rule. A classic benchmark problem involving three-dimensional interface advection has been employed to evaluate the proposed method on four different unstructured meshes.

In initial interface reconstructions within general polyhedral meshes, warped face decomposition reduces $E_{sd}$ across all orientation schemes, slightly enhancing \(\mathcal{O}\left(E_{sd}\right)\) for fraction-gradient-based methods and notably for RDF, from around unity to 1.6. However, it doesn't significantly impact the accuracy of interface advections, while it does considerably decrease computational efficiency.

The RDF scheme consistently delivers the most accurate orientation evaluations, reducing $E_{sd}$ with approximately second-order accuracy across all meshes. The NAG scheme generally outperforms CAG in shape preservation, except in structured polyhedral meshes where CAG and LS errors are similar. LS excels in tetrahedral meshes, but its performance gain is less in polyhedral meshes, and it falls behind NAG in hexahedral meshes. CAG, NAG, and LS exhibit roughly first-order accuracy in $E_{sd}$.

In interface advection, CAG has an approximate first-order convergence rate in tetrahedral and hexahedral meshes, while other methods perform better. All four schemes show similar convergence rates, between 1.41 and 1.75, in general and structured polyhedral meshes. NAG and LS consistently achieve 1.47 to 1.74 convergence rates across all mesh types, with RDF maintaining second-order convergence in all except general polyhedral meshes. $T_{adv}$ shows minimal variance across different schemes and mesh types, except for CAG in tetrahedral and hexahedral meshes, where it leads to more mixed cells and higher CPU times. NAG consumes more advection time than CAG in general and structured polyhedral meshes. RDF has the highest $T_{adv}$ in tetrahedral meshes but is faster than NAG in hexahedral and polyhedral meshes at finer resolutions. Across all mesh types, LS is the most efficient.

The SimPLIC-LS and SimPLIC-RDF solvers exhibit a performance close to the isoAdvector-plicLS and isoAdvector-plicRDF ones, particularly in terms of volume conservation. This parity is evident across different mesh types and sizes. When considering the fraction field's boundaries, all the solvers consistently maintain the lower limit within machine tolerance for general and structured polyhedral meshes. However, in tetrahedral and hexahedral meshes, there's an observed increase in the minimum values. An exception in performance is noted with SimPLIC-RDF, which shows relatively higher errors in the upper fraction limit compared to isoAdvector-plicRDF, especially in coarse hexahedral and all general polyhedral meshes. Regarding shape errors, while all solvers yield identical results in tetrahedral, hexahedral, and structured polyhedral meshes, the SimPLIC solvers display notably higher shape errors in general polyhedral meshes.

The efficiency in CPU time usage during the advection step is another critical area of comparison. Here, the solvers perform similarly in tetrahedral, hexahedral, and coarser structured polyhedral meshes. However, in general polyhedral meshes, the SimPLIC solvers require more time, with SimPLIC-LS being about $21.3\%$ slower and SimPLIC-RDF around $12.3\%$ slower than their isoAdvector counterparts. Interestingly, in finer structured polyhedral mesh, SimPLIC-RDF outperforms isoAdvector-plicRDF, showing a $15.6\%$ improvement in speed. The most significant differences are observed in the reconstruction step, where SimPLIC-LS consistently lags behind isoAdvector-plicLS in efficiency across various mesh types. Conversely, SimPLIC-RDF demonstrates superior speed in tetrahedral and general polyhedral meshes compared to isoAdvector-plicRDF, but it falls behind in hexahedral meshes. In structured polyhedral mesh, SimPLIC-RDF matches the efficiency of isoAdvector-plicRDF in the coarser mesh and surpasses it in the finer one.

In summary, while the SimPLIC method aligns with the isoAdvector ones in several aspects, including volume conservation and shape error consistency, it exhibits variations in efficiency, particularly in the reconstruction step and across different mesh types. This highlights the nuanced performance differences that emerge when dealing with complex mesh geometries and varying solver algorithms, underscoring the importance of tailored optimization for specific mesh configurations and solver requirements.

\section*{Data Accessibility}

The SimPLIC code, along with utilities required to replicate the results presented in this paper, is accessible in the repository at \url{https://github.com/daidezhi/geometricVofExt}. This code represents an extension of \OF\,v2312, the source code of which is available for download at \url{https://dl.openfoam.com/source/v2312/}. Additionally, the unstructured meshes in the OpenFOAM \texttt{polyMesh} format, the dynamic tool surface meshes, and the raw data used in this study can be obtained from \url{https://doi.org/10.5061/dryad.9zw3r22nq}.

\section*{Acknowledgments}

This work was funded by the U.S. Department of Energy High-Performance Computing for Energy Innovation (HPC4EI) program, support for this program was provided by the U.S. DOE Office of Science, Office of Fossil Energy, Office of Energy Efficiency \& Renewable Energy.

The authors gratefully acknowledge the computing resources provided on Improv, a high-performance computing cluster operated by the Laboratory Computing Resource Center (LCRC) at Argonne National Laboratory.

\section*{GOVERNMENT LICENSE}
The submitted manuscript has been created by UChicago Argonne, LLC, Operator of Argonne National Laboratory (``Argonne"). Argonne, a U.S. Department of Energy Office of Science laboratory, is operated under Contract No. DE-AC02-06CH11357. The U.S. Government retains for itself, and others acting on its behalf, a paid-up nonexclusive, irrevocable worldwide license in said article to reproduce, prepare derivative works, distribute copies to the public, and perform publicly and display publicly, by or on behalf of the Government. The Department of Energy will provide public access to these results of federally sponsored research in accordance with the DOE Public Access Plan. \url{http://energy.gov/downloads/doe-public-access-plan}.

\bibliographystyle{IEEEtran}
\bibliography{ref}

\begin{thebibliography}{10}
\providecommand{\url}[1]{#1}
\csname url@samestyle\endcsname
\providecommand{\newblock}{\relax}
\providecommand{\bibinfo}[2]{#2}
\providecommand{\BIBentrySTDinterwordspacing}{\spaceskip=0pt\relax}
\providecommand{\BIBentryALTinterwordstretchfactor}{4}
\providecommand{\BIBentryALTinterwordspacing}{\spaceskip=\fontdimen2\font plus
\BIBentryALTinterwordstretchfactor\fontdimen3\font minus
  \fontdimen4\font\relax}
\providecommand{\BIBforeignlanguage}[2]{{%
\expandafter\ifx\csname l@#1\endcsname\relax
\typeout{** WARNING: IEEEtran.bst: No hyphenation pattern has been}%
\typeout{** loaded for the language `#1'. Using the pattern for}%
\typeout{** the default language instead.}%
\else
\language=\csname l@#1\endcsname
\fi
#2}}
\providecommand{\BIBdecl}{\relax}
\BIBdecl

\bibitem{nichols1975methods}
B.~D. Nichols and C.~W. Hirt, ``Methods for calculating multidimensional,
  transient free surface flows past bodies,'' in \emph{Proceedings of the First
  International Conference on Numerical Ship Hydrodynamics}, vol.~20.\hskip 1em
  plus 0.5em minus 0.4em\relax Naval Ship Research and Development Center
  Bethesda, MD, USA, 1975.

\bibitem{hirt1981volume}
C.~W. Hirt and B.~D. Nichols, ``Volume of fluid ({VOF}) method for the dynamics
  of free boundaries,'' \emph{Journal of Computational Physics}, vol.~39,
  no.~1, pp. 201--225, 1981.

\bibitem{roenby2016computational}
J.~Roenby, H.~Bredmose, and H.~Jasak, ``A computational method for sharp
  interface advection,'' \emph{Royal Society open science}, vol.~3, no.~11, p.
  160405, 2016.

\bibitem{maric2020unstructured}
T.~Mari{\'c}, D.~B. Kothe, and D.~Bothe, ``Unstructured un-split geometrical
  {Volume-of-Fluid} methods--{A} review,'' \emph{Journal of Computational
  Physics}, vol. 420, p. 109695, 2020.

\bibitem{deshpande2012evaluating}
S.~S. Deshpande, L.~Anumolu, and M.~F. Trujillo, ``Evaluating the performance
  of the two-phase flow solver interfoam,'' \emph{Computational science \&
  discovery}, vol.~5, no.~1, p. 014016, 2012.

\bibitem{muzaferija1999two}
S.~Muzaferija, ``A two-fluid navier-stokes solver to simulate water entry,'' in
  \emph{Proceedings of 22nd symposium on naval architecture, 1999}.\hskip 1em
  plus 0.5em minus 0.4em\relax National Academy Press, 1999, pp. 638--651.

\bibitem{ubbink1999method}
O.~Ubbink and R.~Issa, ``A method for capturing sharp fluid interfaces on
  arbitrary meshes,'' \emph{Journal of Computational Physics}, vol. 153, no.~1,
  pp. 26--50, 1999.

\bibitem{dai2019analytical}
D.~Dai and A.~Y. Tong, ``Analytical interface reconstruction algorithms in the
  {PLIC-VOF} method for {3D} polyhedral unstructured meshes,''
  \emph{International Journal for Numerical Methods in Fluids}, vol.~91, no.~5,
  pp. 213--227, 2019.

\bibitem{noh2005slic}
W.~F. Noh and P.~Woodward, ``{SLIC} (simple line interface calculation),'' in
  \emph{Proceedings of the fifth international conference on numerical methods
  in fluid dynamics June 28--July 2, 1976 Twente University, Enschede}.\hskip
  1em plus 0.5em minus 0.4em\relax Springer, 2005, pp. 330--340.

\bibitem{youngs1982time}
D.~L. Youngs, ``Time-dependent multi-material flow with large fluid
  distortion,'' \emph{Numerical Methods for Fluid Dynamics}, 1982.

\bibitem{ashgriz1991flair}
N.~Ashgriz and J.~Poo, ``{FLAIR}: Flux line-segment model for advection and
  interface reconstruction,'' \emph{Journal of Computational Physics}, vol.~93,
  no.~2, pp. 449--468, 1991.

\bibitem{rider1998reconstructing}
W.~J. Rider and D.~B. Kothe, ``Reconstructing volume tracking,'' \emph{Journal
  of Computational Physics}, vol. 141, no.~2, pp. 112--152, 1998.

\bibitem{pilliod2004second}
J.~E. Pilliod~Jr and E.~G. Puckett, ``Second-order accurate volume-of-fluid
  algorithms for tracking material interfaces,'' \emph{Journal of Computational
  Physics}, vol. 199, no.~2, pp. 465--502, 2004.

\bibitem{zhang2008new}
Q.~Zhang and P.~L.-F. Liu, ``A new interface tracking method: The polygonal
  area mapping method,'' \emph{Journal of Computational Physics}, vol. 227,
  no.~8, pp. 4063--4088, 2008.

\bibitem{vignesh2013noniterative}
T.~Vignesh and S.~Bakshi, ``Noniterative interface reconstruction algorithms
  for volume of fluid method,'' \emph{International Journal for Numerical
  Methods in Fluids}, vol.~73, no.~1, pp. 1--18, 2013.

\bibitem{yang2006analytic}
X.~Yang and A.~J. James, ``Analytic relations for reconstructing piecewise
  linear interfaces in triangular and tetrahedral grids,'' \emph{Journal of
  Computational Physics}, vol. 214, no.~1, pp. 41--54, 2006.

\bibitem{huang2012piecewise}
M.~Huang, L.~Wu, and B.~Chen, ``A piecewise linear interface-capturing
  volume-of-fluid method based on unstructured grids,'' \emph{Numerical Heat
  Transfer, Part B: Fundamentals}, vol.~61, no.~5, pp. 412--437, 2012.

\bibitem{ito2013volume}
K.~Ito, T.~Kunugi, H.~Ohshima, and T.~Kawamura, ``A volume-conservative {PLIC}
  algorithm on three-dimensional fully unstructured meshes,'' \emph{Computers
  \& Fluids}, vol.~88, pp. 250--261, 2013.

\bibitem{dai2019numerical}
D.~Dai, \emph{A Numerical Study of Cavitating Flows Based on PLIC-VOF Method
  for Arbitrary Unstructured Meshes}.\hskip 1em plus 0.5em minus 0.4em\relax
  The University of Texas at Arlington, 2019.

\bibitem{swartz1989second}
B.~Swartz, ``The second-order sharpening of blurred smooth borders,''
  \emph{Mathematics of Computation}, vol.~52, no. 186, pp. 675--714, 1989.

\bibitem{mosso1996recent}
S.~Mosso, B.~Swartz, D.~Kothe, and S.~Clancy, ``Recent enhancements of volume
  tracking algorithms for irregular grids,'' in \emph{Los Alamos National
  Laboratory, Los Alamos, NM, LA-UR-96-277), presented at the Parallel CFD
  Conference, Capri, Italy, March}, 1996, pp. 20--23.

\bibitem{scardovelli2003interface}
R.~Scardovelli and S.~Zaleski, ``Interface reconstruction with least-square fit
  and split eulerian--lagrangian advection,'' \emph{International Journal for
  Numerical Methods in Fluids}, vol.~41, no.~3, pp. 251--274, 2003.

\bibitem{aulisa2007interface}
E.~Aulisa, S.~Manservisi, R.~Scardovelli, and S.~Zaleski, ``Interface
  reconstruction with least-squares fit and split advection in
  three-dimensional cartesian geometry,'' \emph{Journal of Computational
  Physics}, vol. 225, no.~2, pp. 2301--2319, 2007.

\bibitem{dyadechko2005moment}
V.~Dyadechko and M.~Shashkov, ``Moment-of-fluid interface reconstruction,''
  \emph{Los Alamos Report LA-UR-05-7571}, p.~49, 2005.

\bibitem{lopez2008new}
J.~L{\'o}pez, C.~Zanzi, P.~G{\'o}mez, F.~Faura, and J.~Hern{\'a}ndez, ``A new
  volume of fluid method in three dimensions—part ii: Piecewise-planar
  interface reconstruction with cubic-b{\'e}zier fit,'' \emph{International
  journal for numerical methods in fluids}, vol.~58, no.~8, pp. 923--944, 2008.

\bibitem{cummins2005estimating}
S.~J. Cummins, M.~M. Francois, and D.~B. Kothe, ``Estimating curvature from
  volume fractions,'' \emph{Computers \& structures}, vol.~83, no. 6-7, pp.
  425--434, 2005.

\bibitem{scheufler2019accurate}
H.~Scheufler and J.~Roenby, ``Accurate and efficient surface reconstruction
  from volume fraction data on general meshes,'' \emph{Journal of Computational
  Physics}, vol. 383, pp. 1--23, 2019.

\bibitem{ahn2008geometric}
H.~T. Ahn and M.~Shashkov, ``Geometric algorithms for 3d interface
  reconstruction,'' in \emph{Proceedings of the 16th international meshing
  roundtable}.\hskip 1em plus 0.5em minus 0.4em\relax Springer, 2008, pp.
  405--422.

\bibitem{lopez2008analytical}
J.~L{\'o}pez and J.~Hern{\'a}ndez, ``Analytical and geometrical tools for {3D}
  volume of fluid methods in general grids,'' \emph{Journal of Computational
  Physics}, vol. 227, no.~12, pp. 5939--5948, 2008.

\bibitem{diot2016interface}
S.~Diot and M.~M. Fran{\c{c}}ois, ``An interface reconstruction method based on
  an analytical formula for 3d arbitrary convex cells,'' \emph{Journal of
  Computational Physics}, vol. 305, pp. 63--74, 2016.

\bibitem{lopez2016new}
J.~L{\'o}pez, J.~Hern{\'a}ndez, P.~G{\'o}mez, and F.~Faura, ``A new volume
  conservation enforcement method for {PLIC} reconstruction in general convex
  grids,'' \emph{Journal of Computational Physics}, vol. 316, pp. 338--359,
  2016.

\bibitem{skarysz2018iterative}
M.~Skarysz, A.~Garmory, and M.~Dianat, ``An iterative interface reconstruction
  method for plic in general convex grids as part of a coupled level set volume
  of fluid solver,'' \emph{Journal of Computational Physics}, vol. 368, pp.
  254--276, 2018.

\bibitem{dai2018analytical}
D.~Dai and A.~Y. Tong, ``An analytical interface reconstruction algorithm in
  the {PLIC-VOF} method for {2D} polygonal unstructured meshes,''
  \emph{International Journal for Numerical Methods in Fluids}, vol.~88, no.~6,
  pp. 265--276, 2018.

\bibitem{chen2019predicted}
X.~Chen and X.~Zhang, ``A predicted-newton's method for solving the interface
  positioning equation in the mof method on general polyhedrons,''
  \emph{Journal of Computational Physics}, vol. 384, pp. 60--76, 2019.

\bibitem{xie2017toward}
B.~Xie and F.~Xiao, ``Toward efficient and accurate interface capturing on
  arbitrary hybrid unstructured grids: The {THINC} method with quadratic
  surface representation and {Gaussian} quadrature,'' \emph{Journal of
  Computational Physics}, vol. 349, pp. 415--440, 2017.

\bibitem{burden2001numerical}
R.~L. Burden and J.~D. Faires, ``Numerical analysis (7th),'' \emph{Prindle
  Weber and Schmidt, Boston}, 2001.

\bibitem{dai2022adaptive}
D.~Dai and A.~Y. Tong, ``The adaptive {PLIC-VOF} method in cavitating flow
  simulations,'' \emph{Computational Thermal Sciences: An International
  Journal}, vol.~14, no.~4, 2022.

\bibitem{scheufler2021twophaseflow}
H.~Scheufler and J.~Roenby, ``Twophaseflow: An {OpenFOAM} based framework for
  development of two phase flow solvers,'' \emph{arXiv preprint
  arXiv:2103.00870}, 2021.

\bibitem{of221_2013_release_notes}
``{OpenFOAM} 2.2.1 {Released},'' \url{https://openfoam.org/release/2-2-1/},
  accessed: 2013-07-11.

\bibitem{maric2014openfoam}
T.~Maric, J.~Hopken, and K.~Mooney, ``The {OpenFOAM} technology primer,'' 2014.

\bibitem{leveque1996high}
R.~J. Leveque, ``High-resolution conservative algorithms for advection in
  incompressible flow,'' \emph{SIAM Journal on Numerical Analysis}, vol.~33,
  no.~2, pp. 627--665, 1996.

\bibitem{shin2011local}
S.~Shin, I.~Yoon, and D.~Juric, ``The local front reconstruction method for
  direct simulation of two- and three-dimensional multiphase flows,''
  \emph{Journal of Computational Physics}, vol. 230, no.~17, pp. 6605--6646,
  2011.

\bibitem{liovic20063d}
P.~Liovic, M.~Rudman, J.-L. Liow, D.~Lakehal, and D.~Kothe, ``A {3D}
  unsplit-advection volume tracking algorithm with planarity-preserving
  interface reconstruction,'' \emph{Computers \& Fluids}, vol.~35, no.~10, pp.
  1011--1032, 2006.

\bibitem{enright2005fast}
D.~Enright, F.~Losasso, and R.~Fedkiw, ``A fast and accurate semi-lagrangian
  particle level set method,'' \emph{Computers \& Structures}, vol.~83, no.
  6-7, pp. 479--490, 2005.

\bibitem{maric2013vofoam}
T.~Maric, H.~Marschall, and D.~Bothe, ``{voFoam}-{A} geometrical volume of
  fluid algorithm on arbitrary unstructured meshes with local dynamic adaptive
  mesh refinement using {OpenFOAM},'' \emph{arXiv preprint arXiv:1305.3417},
  2013.

\bibitem{dormand1980family}
J.~R. Dormand and P.~J. Prince, ``A family of embedded {Runge-Kutta}
  formulae,'' \emph{Journal of Computational and Applied Mathematics}, vol.~6,
  no.~1, pp. 19--26, 1980.

\bibitem{hairer1993solving}
E.~Hairer, S.~P. N{\o}rsett, and G.~Wanner, \emph{Solving ordinary differential
  equations. 1, Nonstiff problems}.\hskip 1em plus 0.5em minus 0.4em\relax
  Springer-Vlg, 1993.

\bibitem{ccm}
{Siemens Digital Industries Software}, ``Simcenter {STAR-CCM+}, version 2306,''
  Siemens 2023.

\bibitem{ccmUG}
{Siemens Digital Industries Software}, ``Simcenter \uppercase{STAR-CCM+} {U}ser
  {G}uide v. 2306,'' Siemens 2023.

\bibitem{blockPolyMeshCode2023}
D.~Dai, ``{blockPolyMesh},'' \url{https://github.com/daidezhi/blockPolyMesh},
  2023.

\end{thebibliography}


\end{document}